\documentclass{aa} 

\usepackage{graphicx}
\usepackage{caption}
\usepackage{subcaption}
\usepackage{txfonts}
\usepackage[colorlinks=true,     linkcolor=blue, citecolor=blue, filecolor=blue, urlcolor=blue]{hyperref}
\begin{document}

\title{Identification and characterization of nascent planetary nebulae with OH and H$_{2}$O masers}

   \author{Rold\'an A. Cala \inst{1}
          \and          
          Jos\'e F. G\'omez \inst{1}
          \and
          Luis F. Miranda \inst{1}
          \and
          Hiroshi Imai \inst{2,3}
          \and 
          Guillem Anglada \inst{1}
          \and 
          Itziar de Gregorio-Monsalvo \inst{4}
          \and 
          Keiichi Ohnaka \inst{5}
          \and
          Olga Su\'arez \inst{6}
          \and
          Daniel Tafoya \inst{7}
          \and
          Lucero Uscanga \inst{8}
          }

   \institute{Instituto de Astrof\'{\i}sica de Andaluc\'{\i}a, CSIC, 	Glorieta de la Astronom\'{\i}a s/n, E-18008 Granada, Spain\\
              \email{rcala@iaa.es}
         \and
         Center for General Education, Comprehensive Institute of Education, Kagoshima University, 1-21-35 Korimoto, Kagoshima 890-0065, Japan
         \and
         Amanogawa Galaxy Astronomy Research Center, Graduate School of Science and Engineering, Kagoshima University, 1-21-35 Korimoto, Kagoshima 890-0065, Japan
         \and
         European Southern Observatory, 3107, Alonso de C\'ordova, Santiago de Chile, Chile
         \and
         Instituto de Astrof\'isica, Universidad Andr\'es Bello, Fern\'andez Concha 700, Las Condes, Santiago, Chile
         \and
         Laboratoire Lagrange, UMR7293, Observatoire de la Côte d’Azur, Université Côte d’Azur, Boulevard de l’Observatoire CS 34229 06304 Nice Cedex, France
         \and
         Department of Space, Earth and Environment, Chalmers University of Technology, Onsala Space Observatory, 439 92 Onsala, Sweden
         \and
         Departamento de Astronomía, Universidad de Guanajuato, A.P. 144, 36000 Guanajuato, Gto., Mexico
       }

   \date{Received ; accepted }

 
  \abstract
  {{Stars like the Sun expel their outer layers and form planetary nebulae (PNe) as they evolve into white dwarfs. PNe exhibit diverse morphologies, the origins of which are not fully understood. PNe} with OH (OHPNe) and H$_{2}$O (H$_{2}$OPNe) masers are thought to be nascent PNe. 
  { However, the number of known OHPNe and H$_{2}$OPNe remains small, and only in eight cases the position of the maser emission has been found to coincide with the PN, using the high astrometric accuracy of interferometric observations.}
  In order to identify more OHPNe and H$_{2}$OPNe, we used public databases and our own ATCA/VLA observations to match the positions of OH and H$_{2}$O masers with known PNe and radio continuum emitters, considering radio continuum emission as a possible tracer of the photoionized gas that characterizes PNe. 
  { Here we report the confirmation of positional coincidence of maser emission with one more PN, and 12 PN candidates.}  
{ Moreover, we have confirmed three evolved stars as `water fountains' (WFs) hosting H$_2$O masers.
  These WFs are associated with radio continuum emission, but their possible nature as PNe has not yet been confirmed}. Although a final characterization of maser-emitting PNe as a group still requires confirmation of more objects, their distribution in the infrared color-color diagrams suggests that they are a { heterogeneous} group of PNe. 
 { In particular, the new OHPN IRAS\,07027$-$7934 has been reported to contain a late [WC]-type central star, while the maser emission implies an O-rich envelope. This property is found in only one other known maser-emitting PN,}
  { although we found evidence that other confirmed and candidate OHPNe may also have mixed chemistry, since they show emission from polycyclic aromatic hydrocarbons.} The new WF IRAS\,18443$-$0231 shows radio continuum that is dominated by strong and variable non-thermal emission, as in magnetized outflows. }
  

   \keywords{Astronomical data bases -- planetary nebulae: general -- masers -- infrared, radio continuum: stars -- stars: winds, outflows }

\titlerunning{Identification and characterization of nascent PNe with OH and H$_{2}$O masers}
\authorrunning{R.~A. Cala et al.}
   \maketitle
%

\section{Introduction}

The morphology of low- and intermediate-mass stars with initial masses ($M_{\rm i}$) up to 8\,M$_{\odot}$ changes at the end of their life, when {the} spherical envelopes created during the asymptotic giant branch (AGB) phase somehow transform into the elliptical, bipolar, multipolar, and, in general, non-spherical forms of planetary nebulae \citep[PNe;][]{man96}, {which are nebulae photoionized} by a hot central star that evolves towards a cooling white dwarf. Consequently, these hot central stars are expected to contain essential information for understanding the complete evolution of PNe. For example, the formation of circumstellar envelopes that are rich in oxygen (O) around central stars rich in carbon (C) \citep{zij91, mir21} would be better understood by studying sources at the {onset of this evolutionary phase}. 

Maser emission has been considered as a tracer of nascent PNe \citep{zij89}. In fact, masers of circumstellar nature are widespread in the AGB phase, but they seem to be extremely rare among PNe. During the AGB phase, the circumstellar envelope grows as the star loses mass at a rate of up to $\simeq 10^{-5}$ M$_\odot$\,yr$^{-1}$, depending on the initial mass of the star \citep{dec19}. If the envelope is O-rich, special physical conditions can pump maser emission from molecules of SiO, H$_{2}$O, and OH \citep[e.g.][]{reid81}, but if it is C-rich, {HCN and SiS masers are observed} \citep[e.g.][]{men18}. Spatially resolved observations show that O-bearing masers in spherically symmetric AGB envelopes are stratified, with SiO, H$_{2}$O, and OH masers typically located at approximately 10, 100 and 1000 au, respectively \citep[e.g.][]{reid81, eli92, dia94}. 

{The intense mass-loss during the AGB removes {a significant part of} the mass of the star until the beginning of the post-AGB phase. During this post-AGB phase, the mass-loss rate drops to 10$^{-8}$ M$_\odot$\,yr$^{-1}$ \citep{blo95b}, the envelope gets detached from the star \citep{vas94, mil16} and then cools and expands. Favorable conditions for maser emission start to disappear, persisting up to 10, 10$^{2}$, and 10$^{3}$ yr after the {beginning of the post-AGB phase} for SiO, H$_{2}$O, and OH molecules, respectively \citep{lew89, gom90}.
The central star shrinks and increases its effective temperature ($T_{\rm eff}$), until it reaches approximately $\simeq$ 25000-30000\,K, at which point it can photoionize the previously detached envelope and form a PN, 10$^{2}$-10$^{4}$ yr after the {beginning of the post-AGB phase}, depending on the initial mass of the star \citep{vas94, mil16}. These timescales make circumstellar masers valuable signposts of nascent PNe. Moreover, soon after the beginning of the PN phase, the increasing number of ionizing photons from the hot central star starts to destroy these molecules, which may prevent the detection of maser emission in evolved PNe. On the other hand, dense molecular clumps, or an unknown special shielding mechanism, have been proposed to explain the presence of molecular masers after the onset of photoionization \citep[][]{taf07}. Furthermore, the presence of thermal emission from some of these molecules in PNe, possibly due to dense condensations of material \citep[e.g.,][]{cox98, mar13, man15}, also indicates that they may survive {on} longer timescales \citep[][]{sch19, gol24, ou24}.

Therefore, the identification of maser-emitting PNe is extremely important because they could be {a} key to characterizing the beginning of the PN phase. So far, no PN has been found to host SiO masers \citep{cal24}, and only in eight cases H$_{2}$O  and/or OH maser emission has been confirmed to spatially coincide with the position of sources spectroscopically classified as PNe. These PNe associated with H$_{2}$O (H$_{2}$OPNe) or OH maser emission (OHPNe) may be among the youngest objects in this evolutionary phase. 
Most of them are bipolar nebulae, and the spatial distribution of the masers trace equatorial rings or toroidal structures close to the central star, although with two important exceptions. In K 3-35, the H$_{2}$O masers also trace the tip of bipolar jets \citep{mir01}. In IRAS 15103$-$5754,  the H$_{2}$O maser emission shows high-velocity components tracing a jet-like structure \citep{gom15b}, which indicates that it is a `water fountain' (WF). Similarly, OH masers have been observed tracing bipolar outflows of post-AGB stars \citep[e.g.,][]{zij01b}. The {presence} of masers in bipolar outflows of post-AGB stars and PNe means that this type of emission can arise not only from the remnants of maser-emitting regions from the AGB phase, but also that they can trace non-spherical mass-loss episodes.

The jet traced by H$_2$O masers in IRAS\,15103$-$5754 mentioned above deserves further attention, since most WFs seem to be binary or multiple systems hosting a post-AGB star \citep[][]{gom17, kho21}, whereas this source is the only WF that seems to host a PN \citep[][]{gom15b}. WFs can be defined, in general terms, as evolved stars whose H$_{2}$O maser emission traces collimated ejections. They have been classically identified by their large velocity range covered by the emission \citep[up to hundreds of km\,s$^{-1}$; e.g.,][]{ima02, gom11, usc23, ima23}, larger than the $\ga 50$ km\,s$^{-1}$ ranges expected by the spherical mass-loss, which builds up large-scale ($\geq$10$^{4}$ au) slowly expanding envelopes (expansion velocities $\la$$25$\,km\,s$^{-1}$, typically traced by OH masers) during the AGB phase \citep[][]{tel91}. The spatial distribution and proper motions of the maser emission clearly show the presence of jets \citep[e.g.][]{sua09, oro19}. In addition to these `classical' WFs, there are other objects in which the velocity range of the H$_{2}$O maser spectra is relatively narrow $\la 50$ km\,s$^{-1}$, but some components fall outside the range covered by the OH maser emission, suggesting the presence of non-spherical mass loss \citep[][]{gom94, eng02}. When observed with high angular resolution, the H$_{2}$O maser emission in these objects seem also to trace jets \citep[e.g.,][]{bob05}, so they can be considered as `low-velocity' WFs. While the difference between classical and low-velocity WFs has been sometimes interpreted as a projection effect \citep{bob05,vle14}, there is some evidence of evolution in the jet velocities of WFs, with slower ones being present closer to the beginning to the post-AGB phase \citep{yung13}. This possible evolutionary effect make WFs relevant also in the context of nascent PNe. Considering that the kinematical ages of WFs \citep[$\sim$10$^{2}$ yr; e.g.,][]{day10, yung11, oro19} are similar to the timescales of disappearance of H$_{2}$O maser emission after the AGB phase, a potential group of PNe with WF characteristics should be extremely rare and probably at the onset of photoionization. 

The central stars of H$_{2}$OPNe and OHPNe are heavily obscured at optical wavelengths, and only in two cases a stellar classification has been proposed. In IRAS\,18061$-$2505, the central star shows broad carbon emission lines and has
been classified as [WC8] \citep[see][and references therein]{mir21}, implying that it is C-rich. Similarly, a [WC11] classification has been suggested for the central star of Vo\,1 \citep[see][and references therein]{zij91}. In Vy\,2-2, \cite{ark17} suggested the presence of weak stellar emission lines (\textit{wels}) similar to those observed in other central stars \citep[see][]{tyl93, mar03, wei20}, although \textit{wels} may not be of real spectral type \citep[see][]{par98, wei15}. With these few cases, it is impossible to assert whether maser-emitting PNe are associated with a particular type of central star or whether they are a common phase of different types of central stars at the beginning of the PN phase. Therefore, it is essential to identify more H$_{2}$OPNe and OHPNe to understand the nature and evolution of these objects.

\section{Procedure to search for new maser-emitting PNe}
\label{sec:procedure}

In order to find new maser-emitting PNe, there are two different aspects to tackle. The first is to ensure that the maser emission is actually associated with a prospective PN. This is the main focus of this paper. The second is to determine whether the maser-emitting source is actually a PN, rather than another type of object.

Regarding the spatial location of the maser emission, many searches for maser emission toward evolved stars are carried out with single-dish telescopes, which typically have beam sizes of several arcminutes. For example, the Green Bank Telescope, one of the largest radio telescopes in the world (100\,m diameter), has a half-power beam size of $\simeq 30''$ at 22 GHz (for H$_{2}$O masers) and $\simeq 8'$ at 1.6 GHz (for OH masers). When the spectrum is obtained with a single-dish antenna only at the catalog position of the source, the detected maser emission can arise from any source within that beam { and, if the contaminant maser is strong enough, the emission can even spill through the antenna sidelobes \citep[cf. the case of  IRAS 19071+0857 in][]{gom15a}, significantly farther away than the nominal beam width at half maximum}. It is possible to improve the positional accuracy of a single-dish observation, by taking several spectra offset from the nominal position of the source and fitting the decrease of emission assuming that the beam has a Gaussian profile. The astrometric error in this case is determined by the antenna rms pointing accuracy (which is typically several arcseconds) and the Gaussian profile fitting error ($1\sigma$ error $\simeq \theta/(2\times {\rm snr})$, where $\theta$ is the telescope beam, and snr is the signal-to-noise ratio of the maser spectrum), which can also amount to several arcseconds, even with a good signal-to-noise ratio. The resulting theoretical astrometric accuracy of tens of arcseconds in single-dish observations may not even be enough to ascertain whether the emission arises from our target of interest, especially in crowded fields, such as the Galactic Bulge. Moreover, observing a maser source with different angular offsets is rarely done in maser surveys, since it is very time-consuming and, therefore, the positional uncertainty for detections in these surveys is of the order of arcminutes. {Hence}, single-dish reports of maser emission are usually not accurate enough to determine the location of the emitting source. In some particular cases, the probability that the emission seen with a single-dish telescope arises from a contaminant source within the beam may be low, specially when observing at high galactic latitudes, or if the velocity of the maser emission is close to the central velocity of the target source obtained with line observations at other frequencies. { In the case of OH masers, a strong infrared source is required for their excitation. While H$_2$O masers do not necessarily require a strong infrared source, H$_2$OPNe often exhibit significant infrared excess due to their thick circumstellar envelopes. Therefore, if there is only a single strong infrared source within a single-dish beam, and the maser velocity is consistent with that of the star, there is little doubt on the identification of the maser-emitting source, especially in the case of OH emission}. However, in general, radio interferometric observations with much higher positional accuracy are necessary for a definitive confirmation of the association of the maser emission with a PN \citep[e.g.][]{mir01, cal24}.

In practice, our general strategy for finding new candidate maser-emitting PNe is to match the position of H$_{2}$O and OH masers and radio continuum emission, considering the latter as a tracer of the photoionized gas that characterizes PNe. 
Both maser and radio continuum emission can be observed with radio interferometers, which provide much improved positional accuracy over single-dish telescopes. The absolute astrometric accuracy in interferometers is determined by different factors, such as phase errors during the observations or how accurately the position of the antennas and the coordinates of the calibrators used for phase reference are known. The astrometric errors due to these factors amount to a fraction (roughly 10\%) of the size of synthesized beam of the interferometers, so, in principle, it is relatively easy to achieve accuracies better than one arcsecond.

The best results can be achieved if the continuum observations are close in frequency to the maser lines, so they can be carried out simultaneously with the same telescope, sharing the same amplitude and phase calibration. In this case, the absolute astrometry is not so important for a spatial match between maser and continuum, but rather the relative positional errors between them, which is only determined by the signal-to-noise ratio of the emission (uncertainty $\simeq \theta/(2\times {\rm snr})$).
These relative positional errors can reach values much smaller than the absolute astrometry of the images for images with a high signal-to-noise ratio, and it is possible to spatially match maser and radio continuum emission with relative accuracies of a few milliarcseconds using interferometers with synthesized beams of $\simeq 1''$ .

{ These extremely high relative accuracies between maser and continuum emission can provide further valuable information, such as the location of masers close to the central star, or on the nebular lobes \citep[e.g.,][]{mir01}. Moreover, these accuracies also apply to the individual maser components, so it is possible to determine their spatial distribution, tracing structures such as toroids or jets.}

After identifying prospective PN candidates by spatially matching radio continuum and maser emission, our second task is to confirm their nature as true PNe, since maser and radio continuum emission can both be present in objects that are not PNe, such as compact and ultracompact H\,{\sc ii} regions, post-AGB stars, galaxies, young stellar objects, symbiotic stars, supernova remnants and red nova remnants \citep[e.g.][and references therein]{hof96, deg05, cho10, qia16b, qia18, qia20, ort20}. The confirmation of an object as a PN requires additional observations, such as multiwavelength imaging and spectroscopy.

For this paper, we started by carrying out an extensive literature search, to identify 
new H$_{2}$O and OH maser-emitting PN candidates. To this end, we cross-matched sources in radio continuum surveys \citep[e.g.][]{zoo90, hel92, bec94, con98, whi05, urq09, mur10, pur13, wan18, ira18, med19, gor21, hale21, ira23, dzi23} with maser databases \citep[e.g.][]{fc89, ben90, enbu15, lad19}. We also matched the same maser databases with objects labeled as true, possible, and likely PNe in the Hong Kong/AAO/Strasbourg H$\alpha$ (HASH) PN database \citep{par16}.

The initial source list compiled after this search was subsequently purged, first by getting rid of bona fide PNe and PN candidates that have already been interferometrically confirmed in the literature as maser emitters \citep[see][and references therein]{cal22}, and of sources whose radio continuum emission is known to be associated with objects that are not evolved stars. 
 In Table\,\ref{tab:sample} we present the final sample of 29 confirmed or candidate PNe that may be associated with maser emission, resulting from our catalog and search of the literature.

\begin{table*}
    \setlength{\tabcolsep}{3.7pt}
      \caption[]{Initial sample of confirmed or possible maser-emitting PNe}
      \label{tab:sample}
      \centering                          
\begin{tabular}{llllcccccc}     
\hline\hline                 
 & & \multicolumn{2}{c}{Coordinates$^a$} & \multicolumn{2}{c}{Maser emission$^b$} & & \multicolumn{3}{c}{References$^c$} \\ 
Name	   & Alt. name & R.A(J2000)	& DEC(J2000) & 	H$_{2}$O & OH & Status$^d$ & H$_{2}$O & OH & Status \\
\hline     
IRAS 07027$-$7934 & Vo 1& 06:59:26.41 & $-$79:38:47.0 & ND & D & SP & 1 & 2 & 3, 4 \\  
IRAS 14079$-$6402 & WRAY 16-146 & 14:11:46.27 & $-$64:16:24.0 & -- & D & SP & -- & 5 & 6\\
IRAS 14086$-$6907 & & 14:12:50.43 & $-$69:21:09.0 & -- & D & SC & -- & 5 & 7 \\
IRAS 15559$-$5546 & WRAY 16-198 & 15:59:57.63 & $-$55:55:33.0 & -- & D & SP & -- & 5 & 8 \\
IRAS 16029$-$5055 & {SCHD 68} & 16:06:40.63 & $-$51:03:57.2 & -- & D & IC & -- & 9 & 10 \\
IRAS 16280$-$4008 &  NGC 6153 & 16:31:30.57 & $-$40:15:13.1 & -- & D & SP & -- & 11 & 12  \\
IRAS 17180$-$2708 & PN M 3-39 & 17:21:11.49  & $-$27:11:37.9 & -- & D & SP & -- & 11 & 8  \\
IRAS 17253$-$2824 & & 17:28:30.68 & $-$28:26:54.0 & -- & D & SC & -- & 5 & 13\\
IRAS 17374$-$2700  & & 17:40:33.15 & $-$27:02:25.3 & -- & D & SC & -- & 5 & 13 \\
IRAS 17385$-$2211 & PN M 3-13 & 17:41:36.60 & $-$22:13:02.9 & -- & D & SP & -- & 5 & 13  \\
PN H 2-18       & & 17:43:28.75 & $-$21:09:52.0 & -- & D & SP & -- & 5 & 14  \\
IRAS 17487$-$1922 & {OH 008.9+03.7} & 17:51:44.92 & $-$19:23:43.8 & ND & D & SC & 15 & 5 & 16 \\
IRAS 18213$-$1245A & & 18:24:10.40 & $-$12:43:22.8 & D & D & IC & 17 & 18 & 19   \\
IRAS 18243$-$1048 & & 18:27:08.24 & $-$10:46:09.6 & -- & D & SC & -- & 5 & 20 \\
IRAS 18271$-$1014 & & 18:29:56.09 & $-$10:12:12.9 & -- & D & SC	& -- & 21 & 22  \\
IRAS 18295$-$2510 & NGC 6644 & 18:32:34.70 & $-$25:07:45.7 &  -- & D & SP & -- & 5 & 23   \\
OH 25.646$-$0.003 & & 18:38:06.14  & $-$06:28:51.2 & -- & D & IC & -- & 18 & 24  \\
IRAS 18464$-$0140 & {OH 31.21$-$0.18} & 18:48:56.66 & $-$01:36:42.9 & D & D & IC & 25 & 25 & 19  \\
IRAS 18443$-$0231 & RFS 505 & 18:47:00.41  & +00:12:29.8 & D & -- & SP & 26 & -- & 27, 28   \\
IRAS 18480+0008 & & 18:50:36.65 & +00:12:29.8 & -- & D & IC & -- & 18 & 19   \\
IRAS 18508+0047 & {GPSR5 33.906$-$0.044} & 18:53:22.10 & +00:50:49.2 & -- & D & IC & -- & 18 & 19  \\
IRAS 19035+0801 & {PM 1$-$279} & 19:05:56.40 & +08:05:57.9 & -- & D & SC	& -- & 29 & 30\\
IRAS 19123+1139 & & 19:14:40.87 & +11:44:49.3 & -- & D & SC & -- & 31 & 22 \\
IRAS 19127+1717 & SS 438 & 19:14:59.72 & +17:22:46.0 & ND & D & SP & 32 & 33 & 34  \\
IRAS 19176+1251 & {GLMP 898} & 19:19:55.77 & +12:57:37.6 & ND & D & IC & 15 & 18 & 19  \\
IRAS 19194+1548 & & 19:21:40.46 & +15:53:54.6 & -- & D & IC & -- & 18 & 19, 35  \\
IRAS 19200+1035 & & 19:22:26.80 & +10:41:21.2 & ND & D & SC & 36 & 37 & 38  \\
IRAS 19508+2659 & & 19:52:57.86 & +27:07:44.7 & D & D & SC & 39 & 40 & 41  \\
IRAS 20266+3856 & & 20:28:30.67 & +39:07:01.2 & ND & D & SC & 31 & 42 & 41\\
\hline 
\end{tabular}

Notes.$^a$ Position of the radio continuum emission reported in the literature, used as the phase center of our observations (Table\,\ref{tab:obs_atca_vla}). $^b$ D: detected. ND: not detected. --: no observations available. $^c${Reference for the maser emission. If the source is a PN candidate, the status column reports the radio continuum detection. If the source is a confirmed PN, the status column provides the reference for its spectroscopic confirmation as PNe.} : 1. \citet{sua09}. 2. \citet{zij91}. 3. \citet{men90}. 4. \citet{sur02}. 5. \citet{tel91}. 6. \citet{sua06}. 7. \citet{hale21}. 8. \citet{frew13}. 9. \citet{sev97}. 10. \citet{ira23}. 11. \citet{tel96}. 12. \citet{pot86}. 13. \citet{kim01} 14. \citet{ges24}. 15. \citet{gom15a}. 16. \citet{koh01}. 17. \citet{walsh14}. 18. \citet{beu19}. 19. \citet{wan18}. 20. \citet{hel92}. 21. \citet{dav93}. 22. \citet{urq09}. 23. \citet{hsia10}. 24. \citet{whi05}. 
25. \citet{fc89}. 26. \citet{urq11}. 27. \citet{coop13}. 28. \citet{kan15}. 29. \citet{yung13}. 30. \citet{van95}. 31. \citet{yung14}. 32. \citet{yoon14}. 33. \citet{lik89}. 34. \citet{whi86}. 35. \citet{sab14} 36. \citet{ben96}. 37. \citet{ben90}. 38. \citet{zij89}. 39. \citet{ces88}. 40. \citet{eder88}. 41. \citet{zoo90}. 42. \citet{tel89}. $^d$ {Source status, as defined in Section}\,\ref{sec:procedure}.\\
\end{table*}

Our list of sources can be classified, based on whether the association between maser and continuum emission has been confirmed with the high positional accuracy of interferometric observations (sources labeled as ``I''), or if only single-dish maser observations have been reported (labeled as ``S''), and whether the target has been previously confirmed as a PN (label ``P''), or the nature of the radio continuum source is still unknown, and therefore, it is only a PN candidate (label ``C'').
Therefore, we can define four different categories of sources, with the following labels used in the seventh column of Table  \,\ref{tab:sample} 
(status):
\begin{enumerate}
    \item SC: radio continuum sources of unknown nature within the beam of a single-dish maser detection.
    \item SP: Planetary nebulae within the beam of a single-dish maser detection.
    \item IC: radio continuum sources whose association with maser emission has been confirmed interferometrically. In the following, we will refer to these sources as candidate maser-emitting PNe (candidate OHPNe and/or candidate H$_2$OPNe).
    \item IP: planetary nebulae with interferometrically confirmed association with maser emission. We will refer to them as confirmed maser-emitting PNe (confirmed OHPNe and/or confirmed H$_2$OPNe).
\end{enumerate}
Our initial search resulted in 11 SC, 10 SP, and eight IC.

In this paper, we have carried out interferometric observations of maser and radio continuum emission of sources labeled as SC and SP in Table\,\ref{tab:sample}, to confirm whether both types of emission are spatially associated, so that we could {update} their status to IC or IP. We also observed the H$_2$O maser emission in the IC IRAS\,19194+1548, to find out whether it is associated with this emission, in addition to OH maser one. We did not pursue observations of {two ICs} (IRAS\,18213$-$1245A and IRAS\,18464$-$0140), since that spatial association was already confirmed from interferometric data in the literature. However, these two sources are discussed in  Sect.\,\ref{sec:individual}, together with the sources for which our interferometric observations allow us to {update} them as either IP or IC.

Determining the evolutionary stage of an evolved star, particularly in the transition from the AGB to the PN phase, remains a complex and widely debated issue, especially in the absence of optical and infrared imaging and spectroscopy. In this context, establishing whether the sources presented in this paper as candidate PNe are genuine PN (that is, an update from C to P status) is beyond the scope of this work. However we discuss in Sect.\,\ref{sec:individual} some of the available evidence that suggests a possible PN nature for some of our targets.

\section{Observations and data processing}

We observed 27 of the 29 sources (see above) of the initial sample listed in Table 1. Additionally, we observed the previously identified OHPNe candidates IRAS\,17494$-$2645 and IRAS\,18019$-$2216 \citep{cal22} to search for H$_{2}$O maser emission. The observations were carried out with the Australia Telescope Compact Array (ATCA) {of the Australia Telescope National Facility (ATNF)} and the Karl G. Jansky Very Large Array (VLA) of the National Radio Astronomy Observatory (NRAO). Table\,\ref{tab:obs_atca_vla} shows the final list of observed targets, as well as the main parameters of our radio interferometric observations. 

\begin{table*}
      \caption[]{Parameters of the ATCA/VLA observations.}
      \label{tab:obs_atca_vla}
      \centering                          
\begin{tabular}{ccclrrcrr}     
\hline\hline                 
 & & &  & \multicolumn{2}{c}{Synthesized beam$^a$} & rms$^b$ & \\ 
Target name	   & Date & 	Band  & Phase calibrator  & (arcsec)  & (deg) & (mJy\,beam$^{-1}$) & Telescope \\
\hline     
IRAS 07027$-$7934 & 2020/10/19 & L &  PKS 0606-796 & $5.2\times 3.3$ & $+72$ & 13 &  ATCA \\  
IRAS 14079$-$6402 & 2020/10/19  & L & PMN J1355-6326 & $6.1\times 3.0$ & $-22$ & 42 & ATCA \\
IRAS 14086$-$6907 & 2020/10/19 & L & PMN J1355-6326 & $5.9\times 3.0$ & $-22$ & 45 &  ATCA\\
IRAS 15559$-$5546 & 2020/10/19 & L & PMN J1650-5044 & $6.1\times 3.3$ & $-40$ & 34 &  ATCA \\
IRAS 16280$-$4008 & 2020/10/19  & L & PMN J1650-5044 & $8.5\times 3.5$ & $-42$ & 29 &  ATCA \\
IRAS 17180$-$2708 & 2020/10/20 & L & PKS 1657-261 & $8.0\times 3.6$	& $-5$ & 59 &   ATCA \\
 & 2021/04/03 & L & J1700-2610 & $107\times 32$	& $-21$ & 39 &  VLA \\
 & 2021/04/01  & K & J1700-2610 & $6.2\times 2.2$	& $-23$ & 8.7 &  VLA \\
IRAS 17221$-$3038 & 2020/10/20 & L & PMN J1733-3722 & $7.9\times 3.4$ & $-4$ & 38 &  ATCA \\
IRAS 17253$-$2824 & 2020/10/20 & L &  PKS 1657-261 & $7.5\times 3.7$ & $+5$ & 33 &  ATCA \\
 & 2021/04/03 & L & J1700-2610 & $110\times 31$ & $-18$ & 23 &  VLA \\
 & 2021/04/01  & K & J1700-2610 & $6.4\times 2.1$ & $-21$ & 6.7 &  VLA \\
IRAS 17374$-$2700  & 2020/10/20 & L & PKS 1657-261 & $8.6\times 3.4$ & $-8$ & 33 &  ATCA \\
& 2021/04/03 & L & J1700-2610 & $118\times 30$ & $-22$ & 24 &  VLA \\
&  2021/04/01 & K & J1700$-$2610 & $6.0\times 2.1$	& $-20$ & 8.4 &  VLA \\
IRAS 17385$-$2211 & 2020/10/20 & L & 1808-209	& $9.0\times 3.4$ & $-1$ & 16 &  ATCA \\
& 2021/04/03 & L & J1700-2610 & $89\times 32$ & $-20$ & 23 &  VLA \\
& 2021/04/01 & K & J1733-1304 & $4.9\times 2.5$	& $-13$ & 13 &  VLA \\
IRAS 17403$-$2107 & 2020/10/20 & L & 1808-209	& $9.6\times 3.3$ & $-2$ & 40 & ATCA \\
& 2021/04/03 & L & J1700-2610  & $83\times 33$ & $-14$ & 38 &  VLA \\
& 2021/04/01 & K & J1733-1304 & $4.6\times 2.2$	& $-13$ & 11 &  VLA \\
IRAS 17487$-$1922 & 2020/10/20 & L & 1808-209	& $12.6\times 3.3$ & $-2$ & 80 &  ATCA \\
& 2021/04/03 & L & J1700-2610 & $79\times 39$ & $-15$ & 10 &  VLA \\
&  2021/04/01 & K & J1733-1304 & $4.3\times 2.2$	& $-12$ & 12 &  VLA \\
IRAS 17494$-$2645 & 2021/04/01 & K & J1820-2528 & $5.8\times 2.2$ & $-20$ & 8.3 &  VLA \\
IRAS 18019$-$2216 & 2021/04/01 & K & J1820-2528 & $5.1\times 2.2$ & $-20$ & 10 &  VLA \\
IRAS 18243$-$1048 & 2020/10/19 & L & 1819-096	& $18.8\times 3.7$ & $-1$ & 12 &  ATCA \\
IRAS 18271$-$1014 & 2020/10/19 & L & 1819-096	& $20.3\times 3.5$ & $0$ & 11 &  ATCA \\
 & 2021/04/03 & L &  J1700-2610 & $63\times 33$ & $-18$ & 20 &  VLA \\
&  2021/04/01 & K & J1832-1035 & $3.9\times 2.4$	& $-17$ & 8.8 & VLA \\
IRAS 18295$-$2510 & 2020/10/20 & L & 1808-209	& $9.6\times 3.4$ & $-7$ & 39 &  ATCA \\
& 2021/04/03 & L & J1700-2610 & $99\times 31$ & $-19$ & 60 &  VLA \\
& 2021/04/01  & K & J1820-2528 & $5.2\times 2.2$	& $-15$ & 14 &  VLA \\
OH 25.646$-$0.003 & 2021/03/25 & K & J1832-1035 & $3.7\times 2.4$ & $-26$ & 12 &  VLA \\
IRAS 18443$-$0231 & 2021/04/03 & L & J1822-0938& $52\times 35$ & $0$ & 75 &  VLA  \\
 & 2021/03/25 & K & J1851+0035  & $4.2\times 3.3$ & $-14$ & 8.4 &  VLA  \\
IRAS 18480+0008 & 2021/03/25 & K & J1851+0035 & $3.4\times 2.4$ & $-23$ & 7.1 &  VLA  \\
IRAS 18508+0047 & 2021/03/25 & K & J1851+0035 & $3.4\times 2.4$ & $-21$ & 12 &  VLA \\
IRAS 18551+0159 & 2021/04/03 & L & J1822-0938 & $51\times 33$ & $-11$ & 70 &  VLA \\
& 2021/03/25 & K & J1851+0035 & $3.3\times 2.4$ & $-21$ & 14 &  VLA \\
IRAS 19035+0801 & 2020/10/19 & L & 1910+052	& $30.2\times 3.4$ & $0$ & 10 & ATCA \\
& 2021/03/25 & L & J1925+2106	& $49\times 35$ & $-6$ &  27 &  VLA  \\
& 2021/03/25 & K & J1851+0035 & $3.1\times 2.4$	& $-21$ & 13 &  VLA  \\
IRAS 19123+1139 & 2020/10/19 & L & 1910+052	& $30.2\times 3.4$ & $0$ & 29 & ATCA \\
& 2021/03/25 & L & J1925+2106 & $47\times 35$ & $-6$ & 75 &  VLA \\
&  2021/03/25 & K & J1924+1540 & $2.9\times 2.4$ & $-17$ & 12 &  VLA \\
IRAS 19127+1717 & 2020/10/19 & L & 1910+052	& $25.8\times 3.2$ & $+2$ & 20  &  ATCA \\
 & 2021/03/25 & L & J1925+2106 & $46\times 36$ & $-3$ & 30 &  VLA \\
&  2021/03/25 & K & J1924+1540 & $2.8\times 2.4$ & $-13$ & 11 &  VLA \\
IRAS 19194+1548 & 2021/03/25 & K & J1924+1540 & $2.8\times 2.4$ & $-14$ & 11 &  VLA \\
IRAS 19200+1035 & 2020/10/19 & L & 1910+052	& $22.7\times 3.4$ & $0$ & 30 &  ATCA \\
 & 2021/03/25 & L & J1925+2106 & $48\times 35$ & $-7$ & 30  & VLA \\
& 2021/03/25 & K & J1924+1540 & $3.0\times 2.4$	& $-15$ & 12 &  VLA \\
IRAS 19508+2659 & 2021/03/25 & K & J2023+3153 &  $3.5\times 2.6$	& $-71$ & 12 & VLA \\
IRAS 20266+3856 & 2021/04/03 & L & J2025+3343 & $43\times 33$ & $+27$ & 70  & VLA \\
&  2021/03/25 & K & J2015+3710 & $3.7\times 2.5$	& $-83$ & 15 & VLA \\
\hline 
\end{tabular}

Notes.$^a$ Major and minor axes and position angle of the continuum data beam. $^b$ Noise level, measured at the channel map with the weakest maser. If there is no detection, measured at the position and channel map with the $V_{\rm LSR}$ of the reported OH maser from single-dish observations.  \\

\end{table*}

\subsection{ATCA}

ATCA observations (ID: C3390, P.I.: R.\,A. Cala) were carried out in two 12-hour sessions performed on October 19 and 20 2020 using the 6A configuration of the telescope. The CFB 1M-0.5k mode of the Compact Array Broadband Backend (CABB) was configured to observe the ground level OH transitions at L band, with the rest frequencies 1612.2309, 1665.4018, 1667.3590, and 1720.5299 MHz, with a total bandwidth of 1 MHz each, sampled over 6146 channels of 0.5 kHz (total velocity coverage of 557 km\,s$^{-1}$ and velocity resolution of $\sim$90 m\,s$^{-1}$ for each line) and full linear polarization products. The velocity coverage is enough to detect the range of $\sim$120 km\,s$^{-1}$, which is the maximum range detected in OH in any of our target sources from single-dish observations. Also, the channel width is narrower than the reported linewidths. In both sessions, we observed continuum emission at an intermediate frequency centered at 2.1 GHz, with full linear polarization and a bandwidth of 2 GHz sampled into 2048 broadband channels. A second intermediate frequency replicates the same frequency configuration. The flux and bandpass calibrator in both observing sessions was PKS 1934-638. Variations in phase and amplitude were tracked by interleaving observations of our targets with nearby phase calibrators (Table \ref{tab:obs_atca_vla}).

Manual flagging of bad data, as well as calibration of visibilities were performed with the MIRIAD software following standard data reduction procedures. For the continuum data, further flagging of the channels containing the frequencies of the ground-state OH maser transitions, as well as imaging of the calibrated visibilities, were done in the Common Astronomy Software Applications \citep[CASA;][]{cas22} using tasks \textit{plotms} and \textit{tclean}, respectively. For the line data, spectral data cubes were obtained with the Astronomical Image Processing System ( AIPS), since it proved to be significantly faster than CASA. First, we corrected the Doppler shift in the spectral line data due to the observatory motion to align the observations of each source with the same kinematical local standard rest (LSR) velocity ($V_{\rm LSR}$), using task \textit{cvel} of AIPS. Imaging was performed using task \textit{imagr}. Self-calibration was attempted on the channel with brightest OH maser emission, but we did not obtain a significant improvement in the signal-to-noise ratio (S/N) of any of the sources. We present data without applying self-calibration. In the detected sources we also tried to split the band of 2 GHz width into several subbands, to better sample the frequency dependence of flux density. However, {the resulting emission either was not detected or had a very low S/N ratio.} Thus, we report their continuum emission over the full bandwidth.

\subsection{VLA}
\label{sec:radio_vla}

VLA observations (ID: 21A-138, PI: R.\,A. Cala) were carried out in the D configuration of the array, in four observing blocks (OBs) performed in 2021 March and April (see Table\,\ref{tab:obs_atca_vla}). Two OBs were dedicated to L-Band (1--2\,GHz) observations and the other two OBs to K-Band (20--24\,GHz) observations. For the L band, the VLA correlator was configured with 8-bit samplers. This allowed us to observe the 1612.2309 MHz {OH} transition within a 4 MHz subband, sampled over 4098 channels (velocity coverage and resolution of $\sim$743 and $\sim$0.18 km\,s$^{-1}$, respectively), and the 1665.4018, 1667.3590, and 1720.5299 MHz {OH} lines, each with a 2-MHz wide subband sampled over 2048 channels (velocity coverage and resolution of $\sim$360 and $\sim$0.18 km\,s$^{-1}$, respectively). Sixteen additional subbands were set up to cover the whole 1 GHz provided by the 8-bit sampler, for continuum detection. For K band, the 8-bit and 3-bit samplers were mixed, in order to efficiently observe spectral line and radio continuum emission simultaneously. The 8-bit sampler was used to observe the 22235.08\,GHz H$_{2}$O maser transition using a 32 MHz subband, sampled over 2048 channels (velocity coverage and resolution of $\sim$421 and $\sim$0.21 km\,s$^{-1}$, respectively). The velocity coverage is enough to detect the range of $\sim$68 km\,s$^{-1}$, which is the maximum velocity range detected of H$_{2}$O masers in any of our target sources from single-dish observations. The spectral resolution is finer than the reported linewidths. Moreover, 4 subbands of 128 MHz each (512 MHz in total) of the 8-bit sampler and the whole 4 GHz bandwidth of the 3-bit sampler pairs were dedicated for detection of continuum emission. The absolute flux calibrator was 3C286. The quasar J1733$-$1304 (QSO B1730$-$130) was used as a bandpass calibrator. Sources close to our targets in the sky were chosen as phase calibrators (Table \ref{tab:obs_atca_vla}), to correct for phase and amplitude variations. 

The continuum and line emission data of each OB was calibrated independently {using} the VLA pipeline implemented in CASA. After initial imaging with CASA, we tested self-calibration using the channels with brightest maser emission in each detected source. We were only successful in improving the S/N of the H$_{2}$O maser data of IRAS 18443$-$0231 and IRAS 18480+0008. Moreover, comparing our self-calibration results with CASA and AIPS, we noticed that the latter provided S/N ratio a factor of $\sim$1.4 times better than CASA in the channel with the most intense maser emission. Hence, we kept the successful self-calibration solutions obtained in AIPS for this channel, and applied them to the rest of the spectral line data and to all continuum spectral windows at K band for these two sources. Self-calibration of the continuum data of IRAS 17494$-$2645 and IRAS 19194+1548 also significantly improved the S/N of this emission. After self-calibration, subsequent imaging of both continuum and line emission was performed in CASA using the task \textit{tclean}. {As in the case of the ATCA data, to better sample the frequency dependence of flux density, we tried to split the data into different frequency ranges. We started with IRAS\,18443$-$0231, which was the strongest source detected and, therefore, the one with the highest S/N. We split the bandwidth of band L and band K data in two different frequency ranges each. These values will be presented below in the discussion of spectral indices in Section \ref{sec:continuum}. In the remaining sources, we noticed that the S/N of the resulting emission in the L band was very low and, in addition, the small relative frequency coverage of the K band around 22\,GHz would not result in a significant improvement of their spectral indices reported in Section\,\ref{sec:continuum}, at the expense of lowering an already limited S/N. Thus, for all sources except IRAS\,18443$-$0231, we present only one value of flux density per observed band}.

For sources IRAS 18443$-$0231, IRAS 18464$-$0140, and IRAS 18508+0047, we found and processed archival VLA data of continuum emission. For IRAS 18443$-$0231 we found VLA B configuration data from observations performed on January 10 2014 (ID: 13A-334, P.I.: K. Menten) which allowed us to create a continuum map of the emission at 5.7\,GHz. For IRAS 18464$-$0140 and IRAS 18508+0047 we found archival VLA data from observations performed in the A configuration on March 08 2022 (ID: 22A-297, PI: A. Yang) which allowed us to create continuum emission maps at 10\,GHz. All of these continuum data were processed with the VLA {\sc CASA} pipeline, successfully self-calibrated, and imaged in CASA. Moreover, for IRAS 18464$-$0140 we reprocessed archival VLA H$_{2}$O maser data from observations performed {with} the C configuration on 21 June 1984 (ID: AF80, PI: J.~R. Forster). The initial calibration of these data was carried out with CASA. Self-calibration was successfully applied with AIPS using the channel with the brightest maser emission.

\section{Results}
\label{sec:results}

The initial criterion we have used to assess the interferometric association between the maser and the continuum emission is that the distance between the maser and the peak of the continuum emission is smaller than the FWHM synthesized beam of the continuum map, in the case of unresolved sources. This match criterion is the same as that used in previous searches of nascent PNe {\citep[e.g.][]{pay88, zij89, deG04, usc12}}. If the continuum emission of the target is resolved, then the criterion is that the maser emission falls within a four-sigma contour level of this resolved emission. We then carefully examined the sources with and without a positional match following these criteria, to minimize false positives and negatives.

In the case of identified possible matches, we have to rule out the possibility that the continuum and the maser emission may arise from different sources lying close in the sky. Most distances between the maser and continuum emission are less than $2"$, which makes the chance of alignment highly unlikely. However, we inspected optical and infrared images to make sure that there is no obvious contaminant source near the main infrared source from which maser emission is most likely to arise. In some cases where we could determine the distribution of maser emission, this shows a morphology that aligns well with the elongation of the nebula traced by radio continuun emission, which futher reinforces their mutual association. We therefore conclude that in all matches we found, the same object is the source of maser and radio continuum radiation. These sources are discussed in Section \ref{sec:individual}.

In Table\,\ref{tab:match} we show the interferometric positions of the radio continuum and maser emission of the matches we found, as well as the angular distance between the two types of emission ($\delta$). In some cases, these spatial matches were obtained from the positions reported in literature data. Moreover, our interferometric ATCA/VLA observations allowed us to upgrade the status of several of the sources labeled SC or SP in Table\,\ref{tab:sample} to either IC or IP. This updated status is also presented in Table\,\ref{tab:match}, where confirmed maser-emitting (IP) sources are listed as OHPNe and/or H$_2$OPNe in the ``nature'' column, depending on the maser species found, and candidate maser emitting (IC) ones listed with the same labels, but enclosed by parentheses. We could also discard as possible maser-emitting PNe several of the objects labeled as SC and SP in Table \ref{tab:sample}, since the maser emission was not spatially associated with the continuum emission (Table \ref{tab:non_match_discarded} and \ref{tab:non_match}).

\begin{sidewaystable*}
    \setlength{\tabcolsep}{3.5pt}
      \caption[]{Characteristics of the radio continuum, H$_{2}$O and OH maser emission of the maser-emitting PNe and candidates {studied in this paper}.}
      \label{tab:match}
      \centering                          
\begin{tabular}{lcccccccr}     
\hline\hline                 
        & \multicolumn{3}{c}{Radio continuum emission} & \multicolumn{4}{c}{Maser emission}  \\
 & Position & Frequency & Flux density & Position &  & $\delta$$^a$& Flux density  \\    
Source	   &	RA, DEC (J2000) & (GHz) & (mJy) & RA, DEC (J2000) & Transition & ($\arcsec$) & (Jy) & Nature$^b$ \\
\hline 
\multicolumn{9}{c}{From our observations} \\
\hline
IRAS 07027$-$7934 & 06:59:26.327 $-$79:38:47.02 & 2.1 & 5.37$\pm$0.28 & 06:59:26.273 $-$79:38.46.99 & OH\,1612\,MHz &  0.15$\pm$0.04 & 0.602$\pm$0.016 & OHPN \\
IRAS 17487$-$1922 & 17:51:44.868 $-$19:23:42.26 & 2.1 & 0.68$\pm$0.18 & 17:51:44.981 $-$19:23:40.21 & OH\,1612\,MHz &  2.6$\pm$1.8 & 0.44$\pm$0.07 & (OHPN) \\
&  & 1.5 & $<$1.45 & 17:51:44.645 $-$19:23:38.67 & OH\,1612\,MHz & 5$\pm$9  & 0.52$\pm$0.05 &  \\
IRAS 18443$-$0231 & 18:47:00.4017 $-$02:27:52.573 & 22.2 & 226.3$\pm$0.4& 18:47:00.4012 $-$02:27:52.552 & H$_{2}$O\,22\,GHz & 0.022$\pm$0.009 & 9.80$\pm$0.03 &  (H$_{2}$OPN), WF \\
IRAS 19035+0801 & 19:05:56.436 $+$08:06:56.88 & 2.1 & 2.63$\pm$0.22 & 19:05:56.413 $+$08:06:57.86 & OH\,1612\,MHz &  1.0$\pm$0.4 & 0.364$\pm$0.019 & (OHPN) \\
&  & 1.5 & 0.93$\pm$0.09 & 19:05:56.289 $+$08:06:56.17 & OH\,1612\,MHz & 2.3$\pm$5.9 & 0.857$\pm$0.019 &  \\
\hline 
\multicolumn{9}{c}{{From our observations and the literature}} \\
\hline
IRAS 18019$-$2216$^c$ & 18:04:57.359 $-$22:15:50.84  & 22.2 & 0.583$\pm$0.018& 18:04:57.347 $-$22:15:50.91 & H$_{2}$O\,22\,GHz &  0.18$\pm$0.26 & 0.332$\pm$0.010 & (H$_{2}$OPN), WF \\
IRAS 18480+0008$^d$ & 18:50:36.659 +00:12:28.25 & 3.0 & 1.90$\pm$0.22 & 18:50:36.685 +00:12:28.34 & OH\,1665\,MHz &  0.4$\pm$1.9 & 0.116$\pm$0.009 &  (OHPN), (WF) \\
& 18:50:36.659 +00:12:28.13 & 22.2 & 2.020$\pm$0.013 & 18:50:36.6614 +00:12:28.191 & H$_{2}$O\,22\,GHz & 0.071$\pm$0.020 & 1.957$\pm$0.014 & (H$_{2}$OPN) \\
\hline
\multicolumn{9}{c}{From the literature} \\
\hline
IRAS 16029$-$5055 & 16:06:40.60 $-$51:03:57.2 & 5.5 & 3.02$\pm$0.15 & 16:06:40.637 $-$51:03:55.73 & OH\,1612\,MHz & 1.5$\pm$0.9 & 1.31$\pm$0.13  &  (OHPN) \\
IRAS 18213$-$1245A & 18:24:09.722 $-$12:43:22.84 & 4.8 & 3.6$\pm$0.8 & 18:24:09.727 $-$12:43:23.90 & OH\,1665\,MHz & 1.1$\pm$1.0  & 0.139$\pm$0.005 & (OHPN)\\
                & & & & 18:24:09.706 $-$12:43:24.04 & H$_{2}$O\,22\,GHz & 1.22$\pm$1.41 & 3.70$\pm$0.04 &  (H$_{2}$OPN) \\
IRAS 18464$-$0140 & 18:48:56.647 $-$01:36:42.98 & 3.0 & 4.8$\pm$0.4 & 18:48:56.523 $-$01:36:43.32 & OH\,1612\,MHz & 1.9$\pm$1.9 & 1.749$\pm$0.021 & (OHPN)  \\
                & & & & 18:48:56.62 $-$01:36:44.3 & OH\,1665\,MHz & 1.4$\pm$1.7 & 0.500$\pm$0.010   \\
                & & & & 18:48:56.523 $-$01:36:42.99 & OH\,1667\,MHz & 1.9$\pm$1.5 & 1.90$\pm$0.010   \\
                & & & & 18.48.56.65 $-$01.36.43.04 & H$_{2}$O\,22\,GHz & 0.1$\pm$0.4 & 16.11$\pm$0.14 & (H$_{2}$OPN), WF  \\
IRAS 18508+0047 & 18:53:22.196	+00:50:49.03 & 1.4 & 2.58$\pm$0.22 & 18:53:22.29 +00:50:48.46 & OH\,1612\,MHz & 1.5$\pm$1.9 & 0.529$\pm$0.005 &  (OHPN)   \\
IRAS 19176+1251 & 19:19:55.744 +12:57:37.70 & 1.5 & 3.01$\pm$0.28 & 19:19:55.850 +12:57:37.80 & OH\,1720\,MHz & 1.6$\pm$0.4 & 0.115$\pm$0.003 & (OHPN) \\
IRAS 19194+1548 & 19:21:40.374 +15:53:55.16 & 1.5 & 3.23$\pm$0.26 & 19:21:40.514 +15:53:54.74 & OH\,1612\,MHz & 2.1$\pm$1.2 & 0.917$\pm$0.008 & OHPN    \\ 
                & & & & 19:21:40.286 +15:53:46.39 & OH\,6035\,MHz & 8.86$\pm$1.12 & 4.80$\pm$0.11  \\
OH\,25.646$-$0.003 & 18:38:06.242 $-$06:28:51.22 & 5 & 2.21$\pm$0.04 & 18:38:06.165 $-$06:28:54.10 & OH\,1612\,MHz & 3.1$\pm$1.9 & 0.254$\pm$0.005 & (OHPN)  \\

\hline
\end{tabular}

Notes: $^a$ Relative distance between the radio continuum emission peak and the maser spot. $^b$ In parenthesis sources candidates to PN and candidates to WF. WF: `water fountain' characteristics. $^c$ This source is a previously reported OHPN candidate \citep{cal22}. {$^d$ The flux density at 3 GHz was reported in VLASS \citep{gor21}, and the 1665 MHz was reported in THOR \citep{beu19}. }

\end{sidewaystable*}

\begin{figure*}
      \centering
            \includegraphics[width=0.99\hsize]{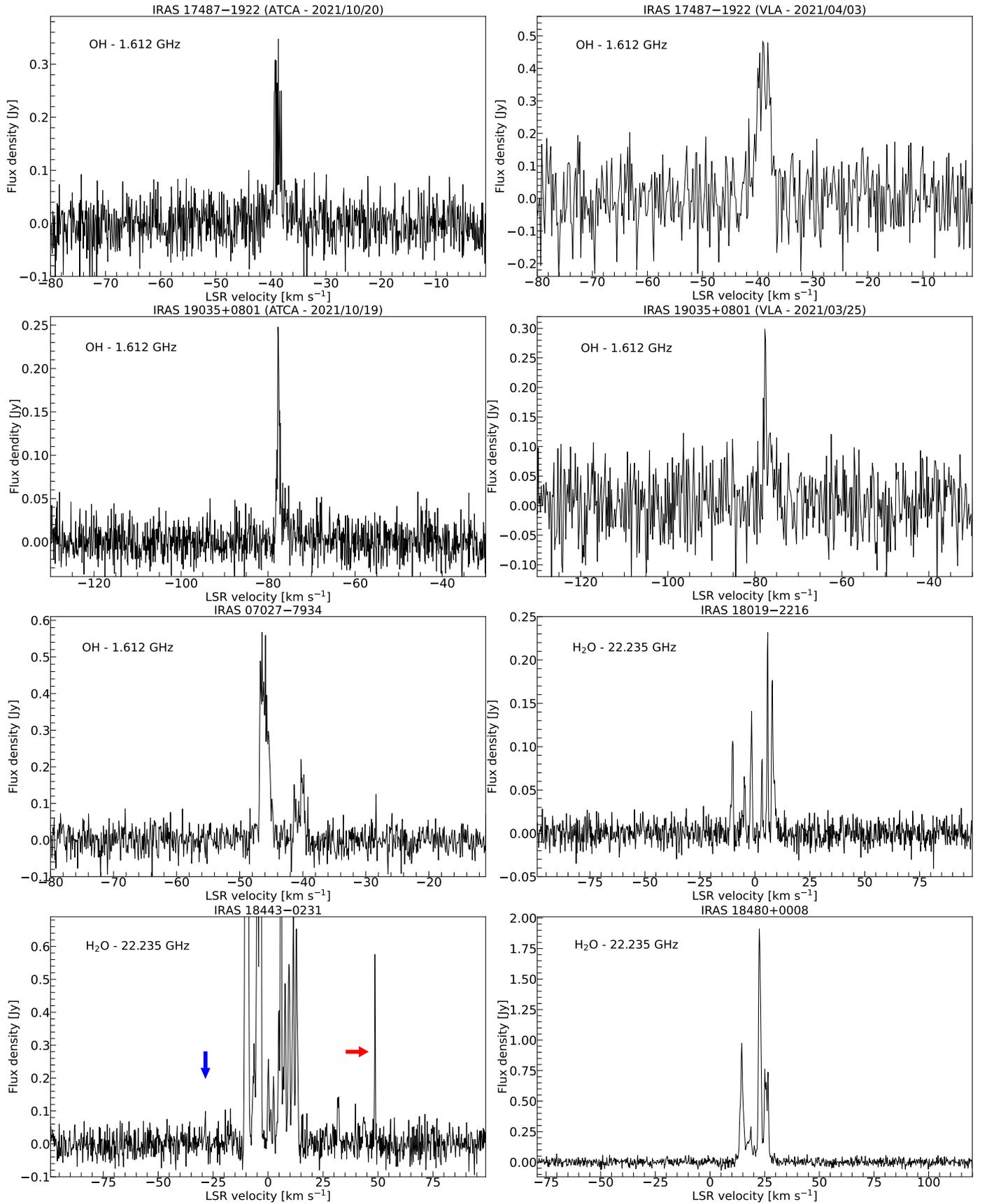}
      \caption{Spectra of OH and H$_{2}$O emission detected in our observations. In IRAS 18443$-$0231 we show {an enlarged view}  to the most blueshifted and redshifted H$_{2}$O masers, which are indicated by a blue and red arrow, respectively.}
      \label{fig:maser_spe}
\end{figure*}

\begin{figure*}
\begin{subfigure}{.5\textwidth}
  \centering
  \includegraphics[width=.92\linewidth]{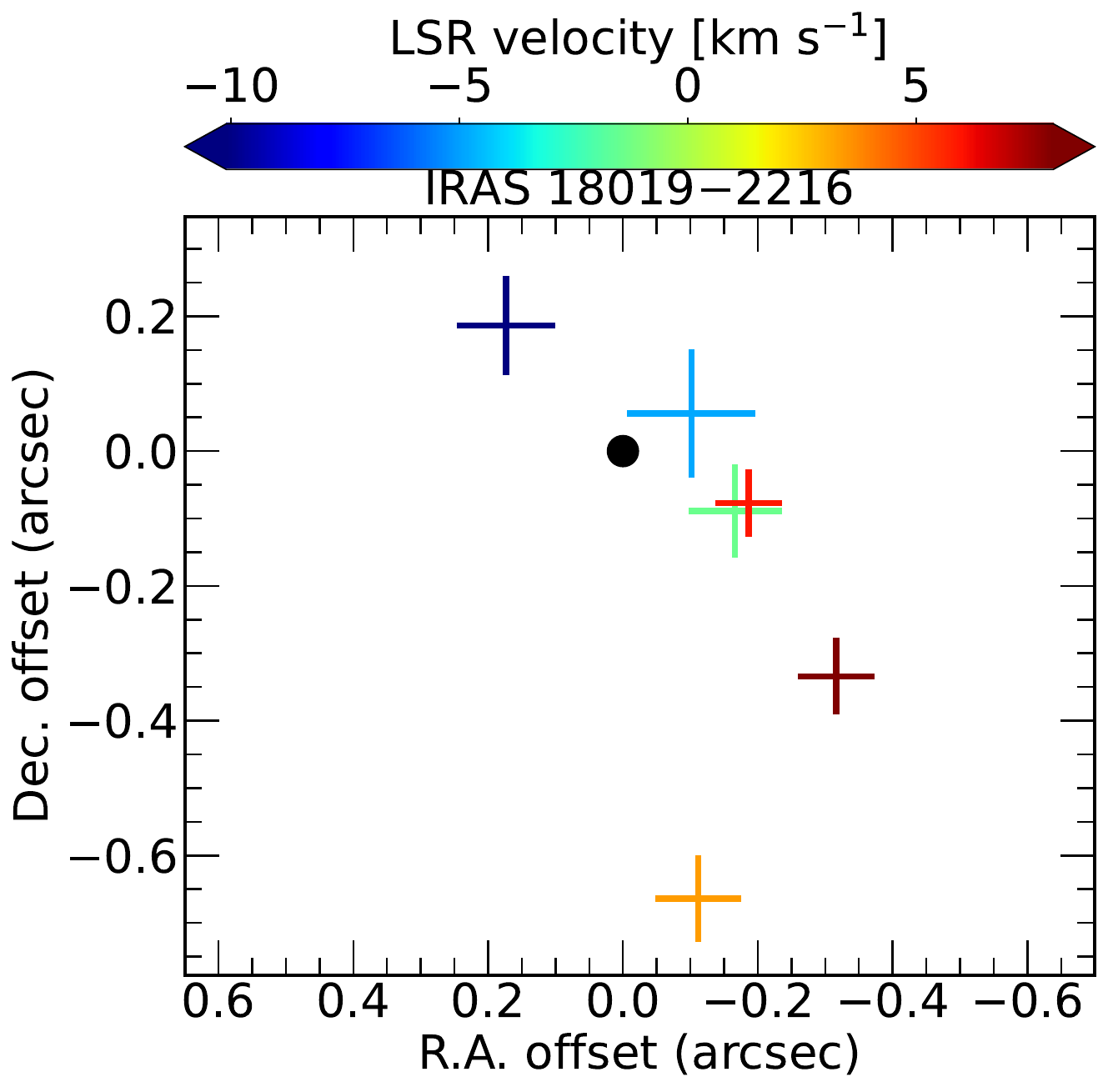}
\end{subfigure}
\begin{subfigure}{.5\textwidth}
  \centering
  \includegraphics[width=.89\linewidth]{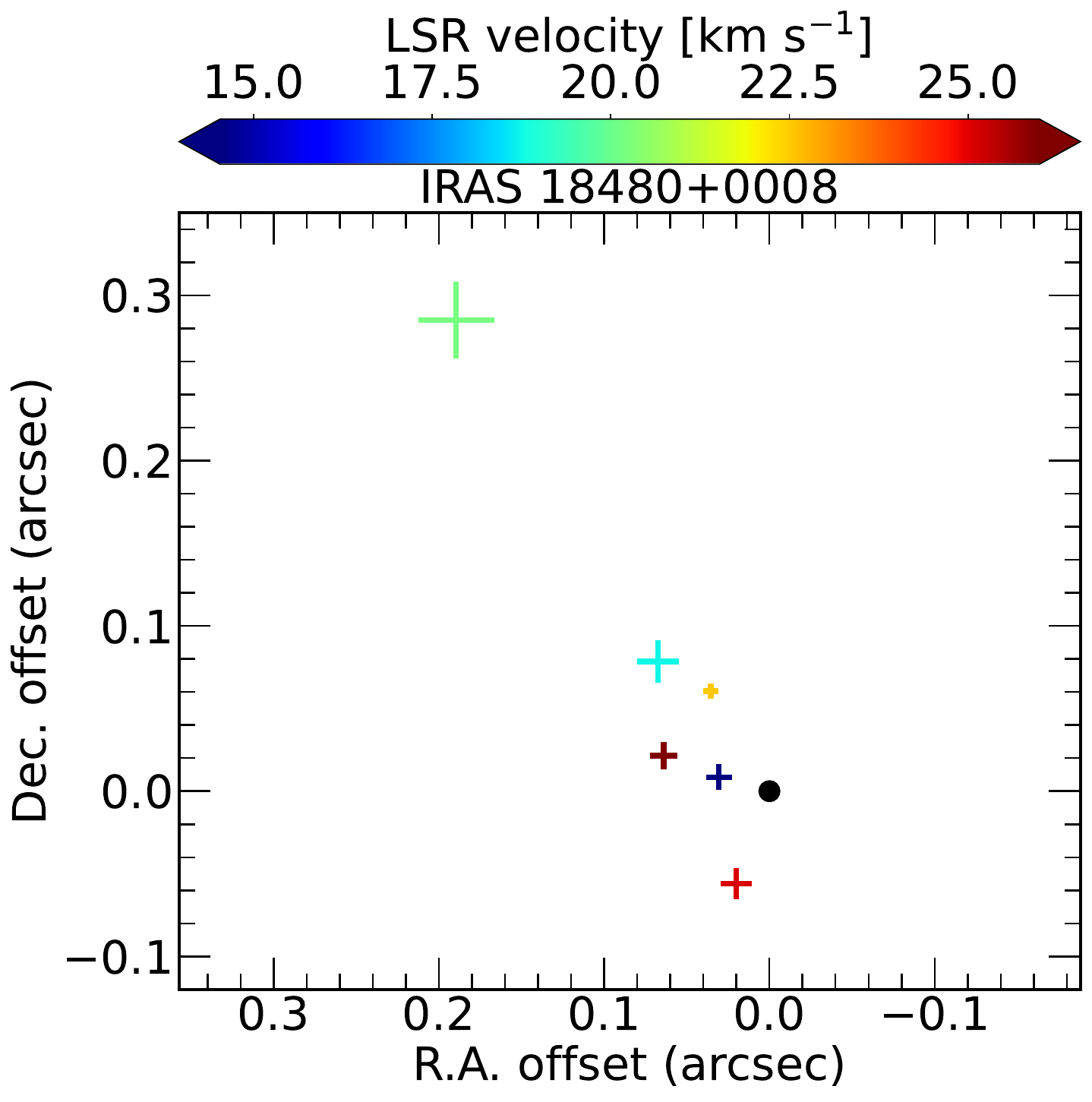}
\end{subfigure}
\begin{subfigure}{.5\textwidth}
  \centering
  \includegraphics[width=.93\linewidth]{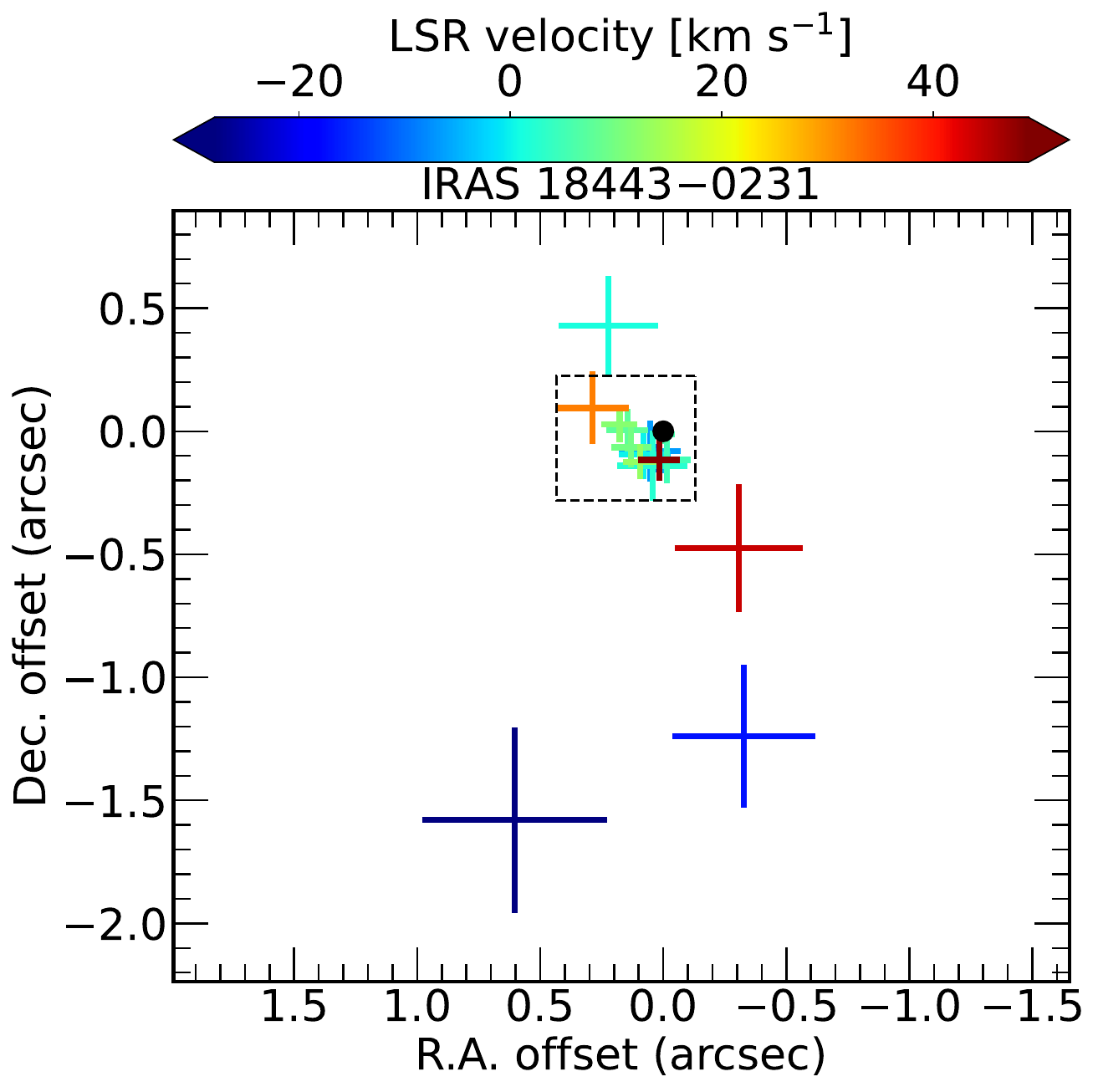}
\end{subfigure}
\begin{subfigure}{.5\textwidth}
  \centering
  \includegraphics[width=.91\linewidth]{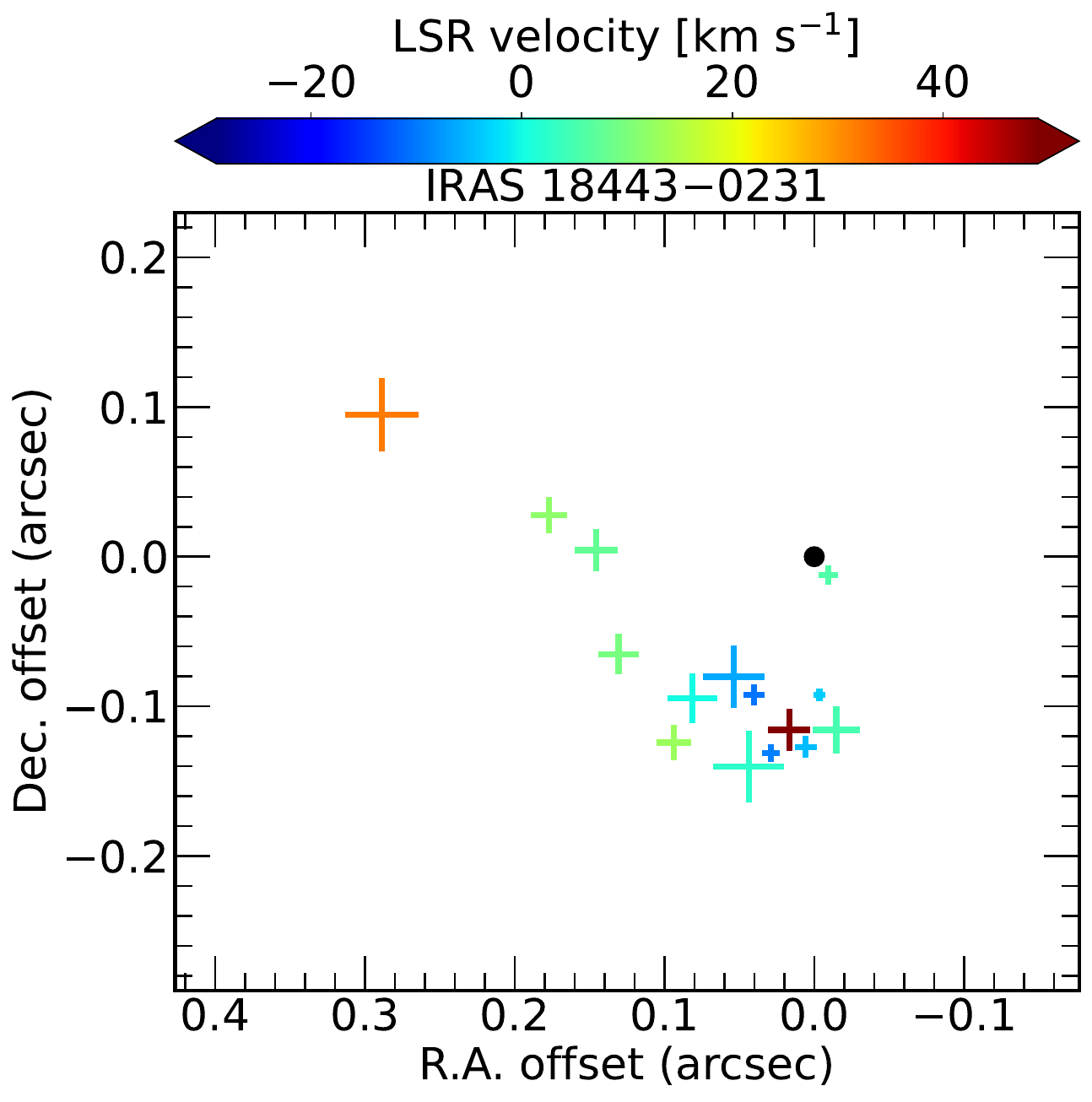}
\end{subfigure}
\caption{Spatial distribution of the H$_{2}$O maser components (crosses) with respect to the position of the radio continuum emission peak (filled circle) at 22.22 GHz. The colorbar represent the LSR velocity of the masers and the {symbols sizes are} their positional uncertainties. The bottom right panel is a close-up of H$_{2}$O maser components in the central area of IRAS\,18443$-$0231 (marked by a black  dashed rectangle in the left panel), close to the continuum emission peak.}
\label{fig:maser_spa}
\end{figure*}

Spectra of the detected maser emission in our data are presented in Fig.\,\ref{fig:maser_spe}. Our criterion for claiming a positive detection of a maser spot is that the emission is present above 3 sigma in at least three consecutive {velocity} channels and at a compatible position, which makes the probability of a false positive virtually negligible.
In three objects, our positional accuracy was enough to determine the spatial distribution of the individual H$_2$O maser components. Maps of these distributions are presented in Fig.\,\ref{fig:maser_spa}. As mentioned in Appendix \ref{app:a}, the positional uncertainties are those obtained with a fit of an elliptical Gaussian to the channel with the peak emission of each maser component. The radio continuum emission appears unresolved in all cases, except for IRAS\,18443$-$0231 (Sect. \ref{sec:i18443}), where we can see some extended emission.  

To minimize the possibility of false negatives, we also inspected images at different wavelengths of the sources for which the maser emission falls outside the beam of the continuum observations (for unresolved sources) or the total extent of the emission (for resolved sources). Our goal was to discard the fact that, despite their apparent misalignment, the maser and continuum emission may arise from different locations from the same source (e.g., the central star or the lobes in a PNe). These sources are discussed in Appendix \ref{app:non_match}, and listed in Table \ref{tab:non_match_discarded}. In the case of sources in which either maser or continuum emission was not detected in our data (Table \ref{tab:non_match}), we cannot confirm or discard a putative association, since variability could bring either emission below our sensitivity threshold.

\section{Source characterization}
\label{sec:character}

{After confirmation of the association between maser and radio continuum emission, we investigated whether the identified objects are consistent with PNe.} Although some of the sources in which we found an interferometric match between maser and continuum emission have been classified spectroscopically as PNe in the literature, no such classification is available for the ones we termed candidate maser-emitting PNe. In this section, we present some further characterization of all sources in Table \ref{tab:match}, which shows that all candidate sources seem to be evolved objects, rather than extragalactic sources or YSOs, source types which could also show maser and continuum emission. Therefore, all our candidate sources are compatible with being PNe, although in several cases further observations are necessary to determine whether the radio continuum emission arises from a photoionized region or from shock-ionized gas, as seen in post-AGB stars \citep[e.g.,][]{ps17}.

\subsection{Spectral energy distribution (SED)}
\label{sec:sed}

We have collected all available archival optical, infrared, and millimeter data, together with the radio data presented in Section \ref{sec:continuum}, to build their SED between $\sim$0.5\,$\mu$m and 40\,cm. Optical and infrared data were obtained from the Two Micron All Sky Survey (2MASS), AKARI, Deep Near Infrared Survey of the Southern Sky (DENIS), Gaia, \textit{Herschel} Space Observatory, Infrared Astronomical Satellite (IRAS), Midcourse Space Experiment (\textit{MSX}), Panoramic Survey Telescope and Rapid Response System (Pan-STARRS), Spitzer Space Telescope, United Kingdom Infrared Telescope (UKIRT), United States Naval Observatory (USNO)-B catalog, Visible and Infrared Survey Telescope for Astronomy variable survey (VVV), and Wide-field Infrared Survey Explorer (all-\textit{WISE} survey). Flux densities at 3 mm were obtained from \cite{gin20}. {These SEDs are presented in Fig.\,\ref{fig:sed}, with the exception of IRAS 17494$-$2645 and IRAS 18019$-$2216}, which were already shown in \cite{cal22}, these SEDs are presented in Fig.\,\ref{fig:sed}. 

\begin{figure*}
      \centering
            \includegraphics[width=0.33\hsize]{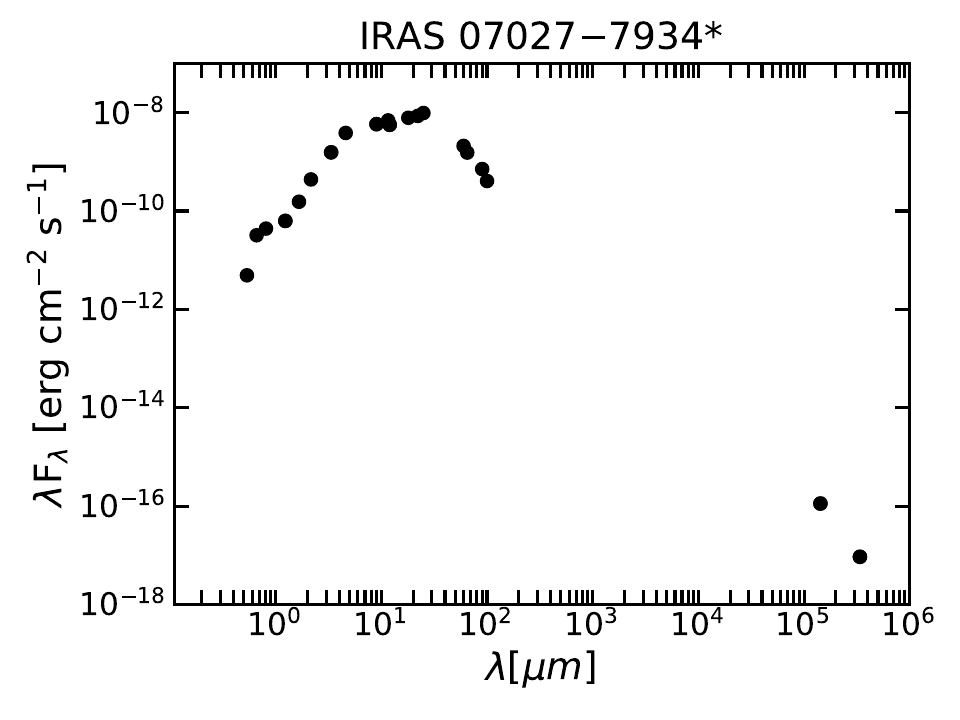}
            \includegraphics[width=0.33\hsize]{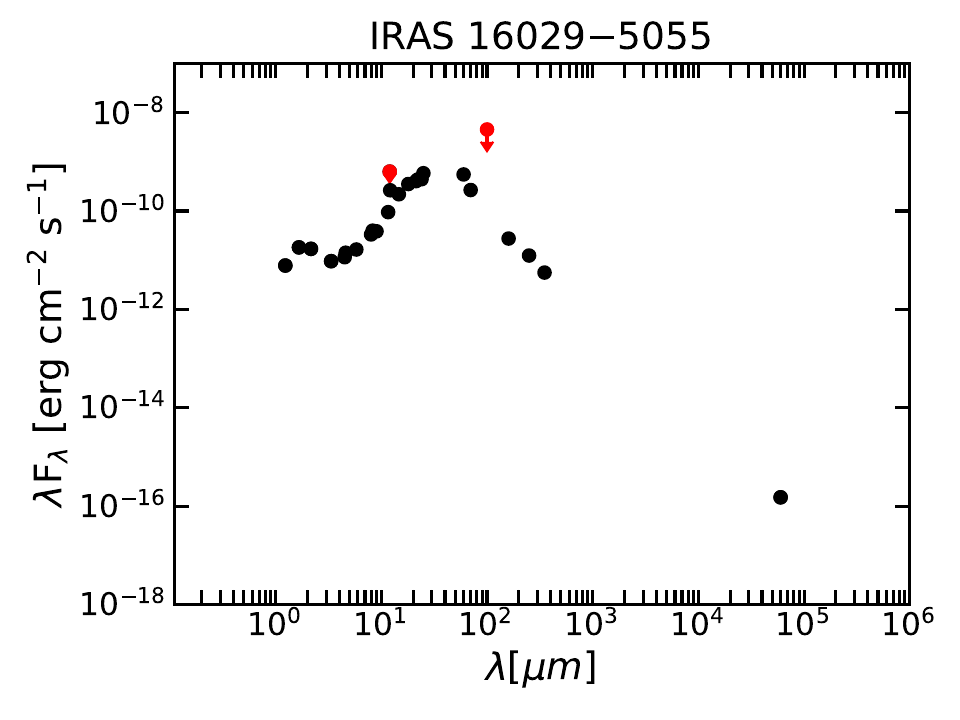}
            \includegraphics[width=0.33\hsize]{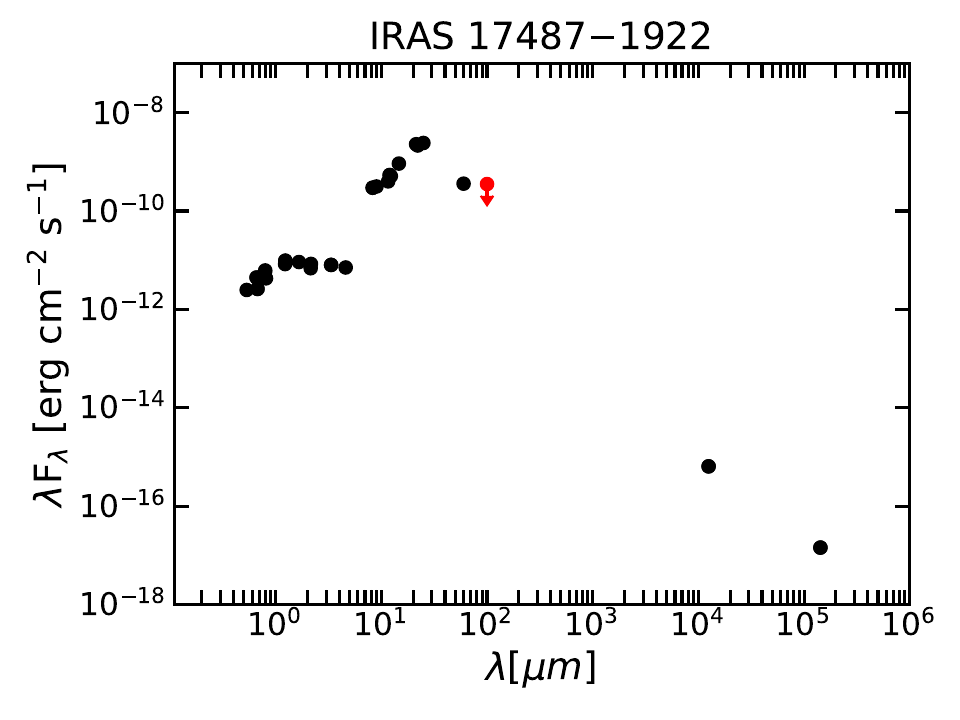}
            \includegraphics[width=0.33\hsize]{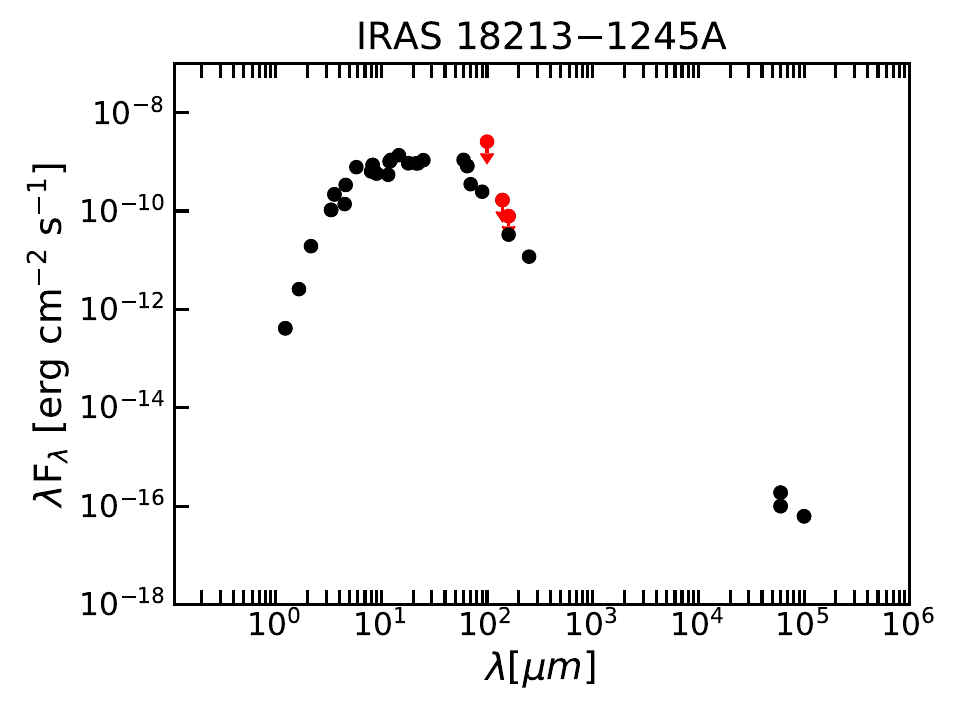}
            \includegraphics[width=0.33\hsize]{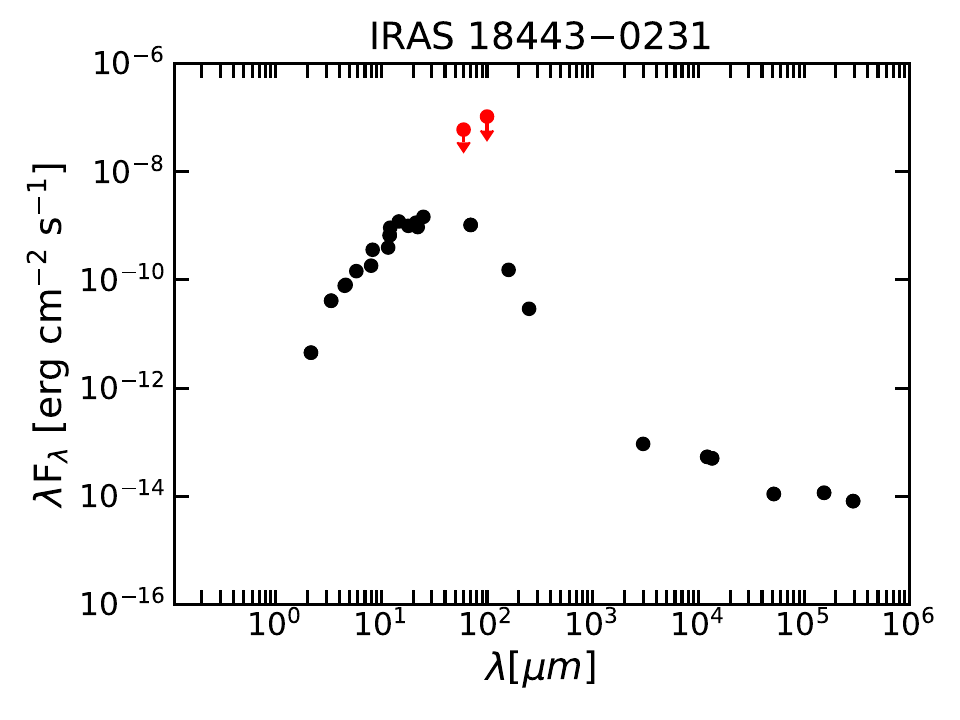}
            \includegraphics[width=0.33\hsize]{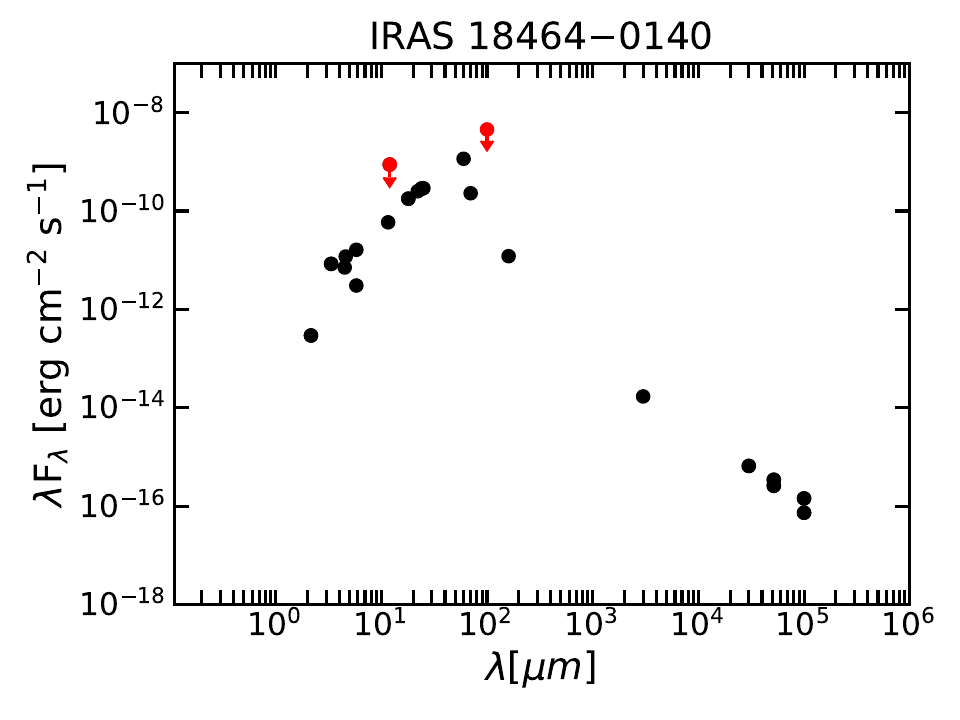}
            \includegraphics[width=0.33\hsize]{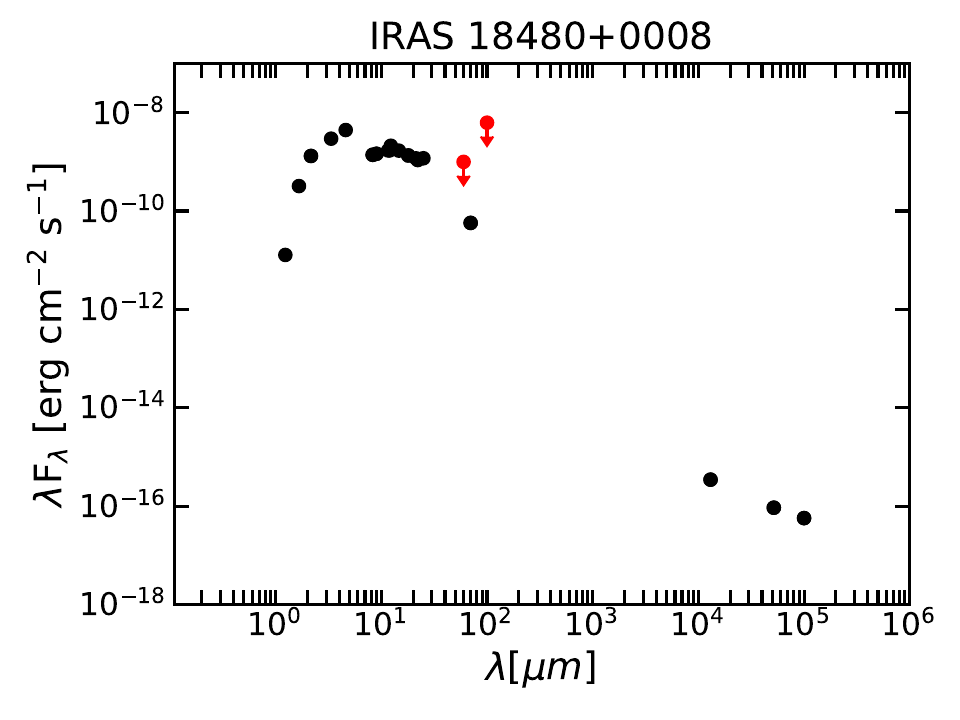}
            \includegraphics[width=0.33\hsize]{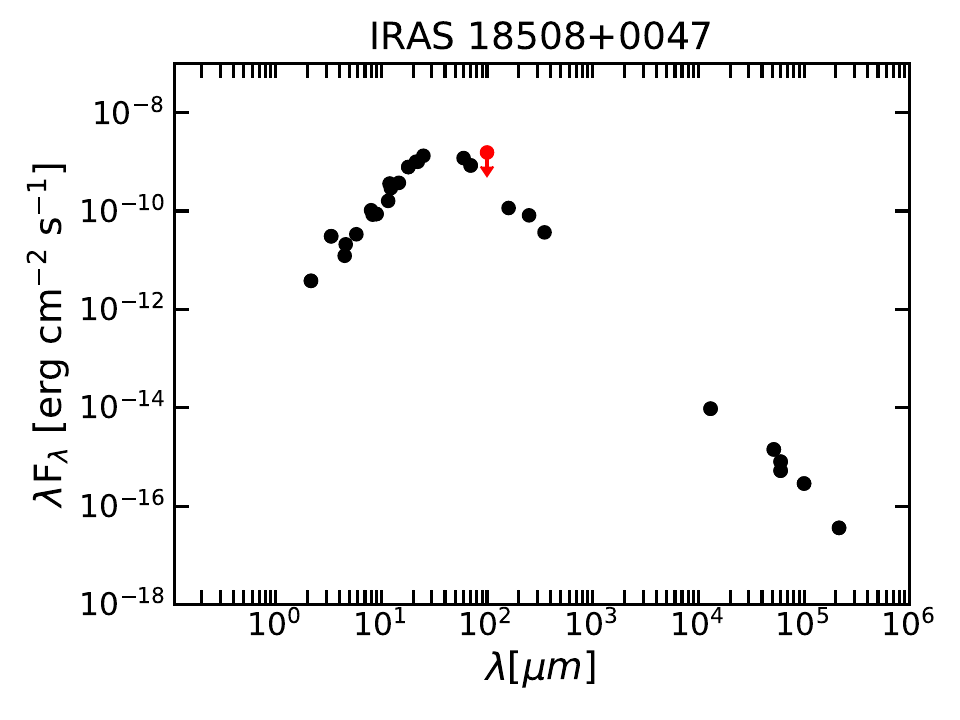}
            \includegraphics[width=0.33\hsize]{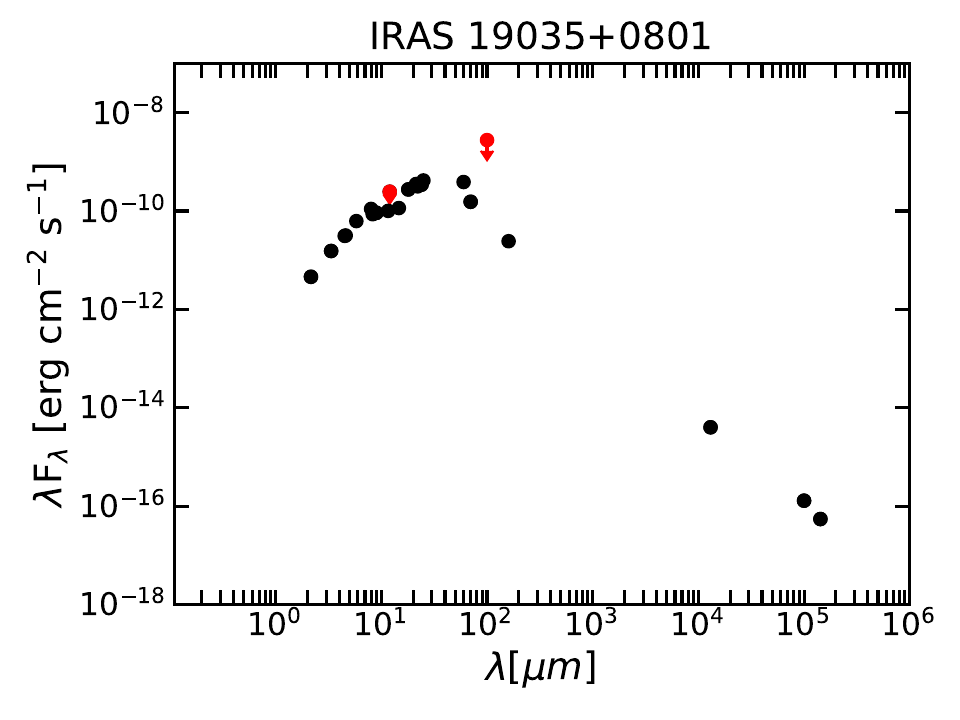}
            \includegraphics[width=0.33\hsize]{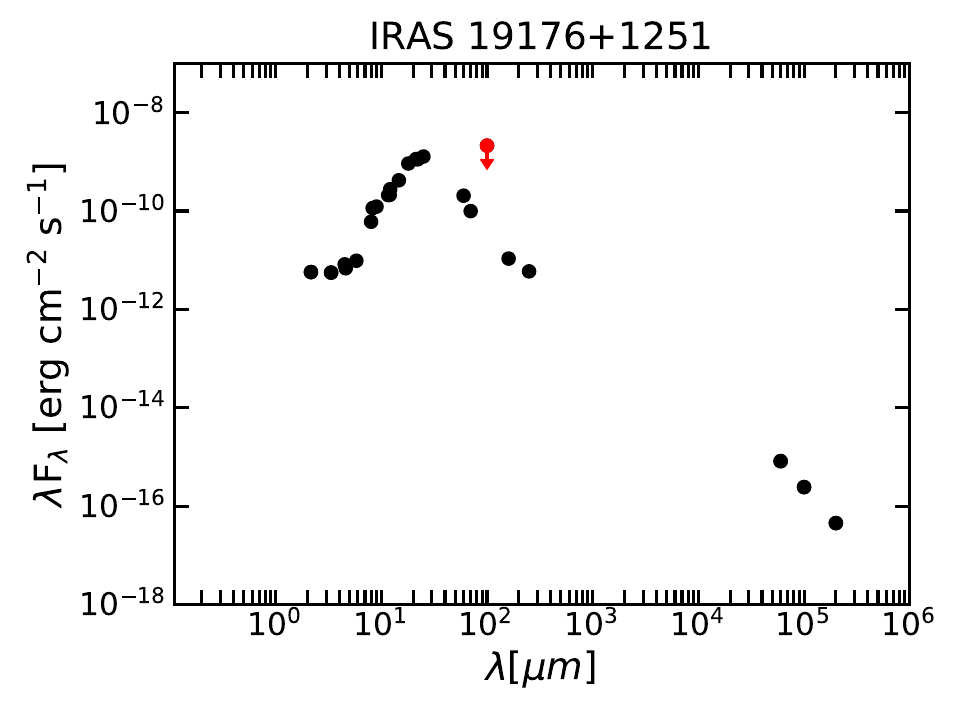}
            \includegraphics[width=0.33\hsize]{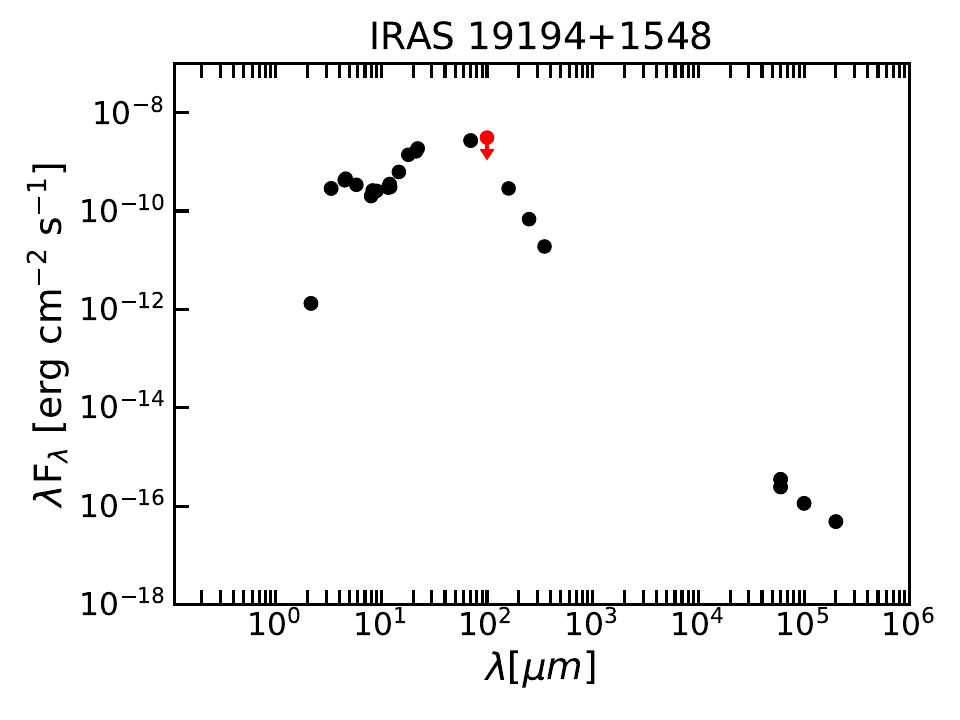}
            \includegraphics[width=0.33\hsize]{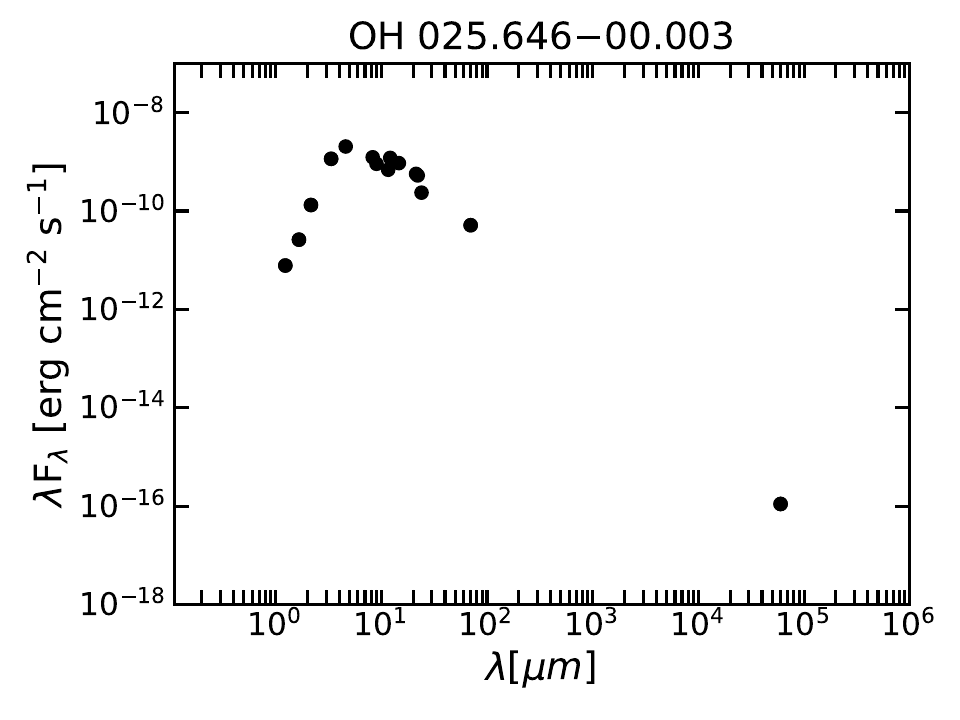}

      \caption{SED of the new maser-emitting PN (indicated with an asterisk next to their name) and candidates (rest of the panels). The circles are the photometric values found for each of the new identified sources. The size of the circles is larger than the errors. Red color stands for upper limits. }
      \label{fig:sed}
\end{figure*}

Although SEDs alone is not sufficient to unambiguously determine the nature of these objects, there are some useful trends that can give support to a prospective classification.
The SEDs of all sources in this paper show significant similarities and, in particular, they present their emission peak at $\lambda$$<$60\,$\mu$m. This feature is also observed in the SEDs of confirmed and candidate H$_{2}$OPNe and OHPNe previously reported \citep{usc12, cal22} and, in general, it is also the case { of compact PNe \citep[][]{zha91}}, and optically obscured post-AGB stars and PNe \citep[e.g.][]{ram09, ram12}. Moreover, the shape of the SED and the wavelength of the peak is significantly different from other types of objects {that present both maser and radio continuum emission, such as YSOs or galaxies} and therefore, could contaminate our group of candidates. For instance,  the SED of YSOs usually peaks at wavelengths around 1\,$\mu$m or longer than 70\,$\mu$m \citep[e.g.][]{rob07}. In the particular case of low-mass YSOs, radio continuum emission from shock-ionized jets is more prominent in the earlier stages (class 0 and class I sources), when the SED peaks at long wavelengths \citep[$>100$\,$\mu$m;][]{dun08,dun13}. In the case of high-mass YSOs, the peak of the SED for H\,{\sc ii} regions (including ultracompact ones) usually lies at wavelengths longer than 70\,$\mu$m \citep{wood89, and12}. 

In galaxies with active {galactic} nuclei, the peak of the SED seems to be at wavelengths shorter than 1\,$\mu$m or longer than 100\,$\mu$m \citep[e.g.][]{dale05, pop11, brown14, lyu22}. Thus, the differences between the SEDs in Fig.\ \ref{fig:sed} and those of typical YSOs and galaxies indicate that our targets are more likely to be evolved objects. Moreover, among these, symbiotic stars (SySts) can also present maser emission \citep[e.g.,][]{cho10}, {although OH ones are extremely rare }\citep[e.g.,][]{nor84,sea95}. The SED peak of SySts usually lies at $\lambda\le 10$ \,$\mu$m \citep{akr19}. This only happens in two of the objects in Fig. \ref{fig:sed}, which will be discussed below. Therefore, the SED of most of the newly identified objects in this paper seems more consistent with post-AGB stars and PNe hosting maser emission. 

\subsection{Infrared morphologies}

We inspected the available mid- and near-infrared images of the new IC sources reported in this work. All seem to be compact sources in the mid-infrared, with the exception of IRAS\,19194+1548, which is compact but marginally resolved. Furthermore, this source (Sec.\,\ref{sec:i19194}) shows a point-symmetric morphology in near-infrared images that is typical of evolved stars, rather than YSOs. The compact and isolated morphology of the sources classified as IC in this paper supports their identification as evolved stars rather than YSOs, which are usually surrounded by diffuse and extended structures from their parental clouds.

\subsection{Mid-infrared colors}

{To further elucidate if the emission properties of the identified sources are more compatible with circumstellar envelopes of evolved stars,} Fig.\,\ref{fig:msx} shows an \textit{MSX} color-color diagram of all known maser-emitting PNe and candidates (collecting the sources presented in this paper, and those previously reported in the literature). This color-color diagram, where each color is defined as $[a]-[b]$=$2.5\,log(S_{a}/S_{b})$, was presented by \cite{sev02} to help identify post-AGB stars among evolved stars with OH maser emission, and later used by \cite{usc12} and \cite{cal22} in the characterization of maser-emitting PNe. We note that this diagram does not provide a clear-cut diagnosis of the nature of each source, but it shows quadrants where it is more likely to find sources of a particular type. The confirmed and candidate maser-emitting PNe cluster mainly in the quadrant of late post-AGB stars, as expected if they are all bona fide PNe. However, some interesting cases are identified in this paper, clearly located in the quadrant of AGB stars (the confirmed PN IRAS\,07027-7934 and the candidate PNe IRAS\,18123-1245A and OH\,025.646-00.003), which deserve further attention. These objects may be rapidly evolving to the PN phase, while still retaining an envelope with AGB characteristics.

\begin{figure}
      \centering
           \includegraphics[width=0.95\hsize]{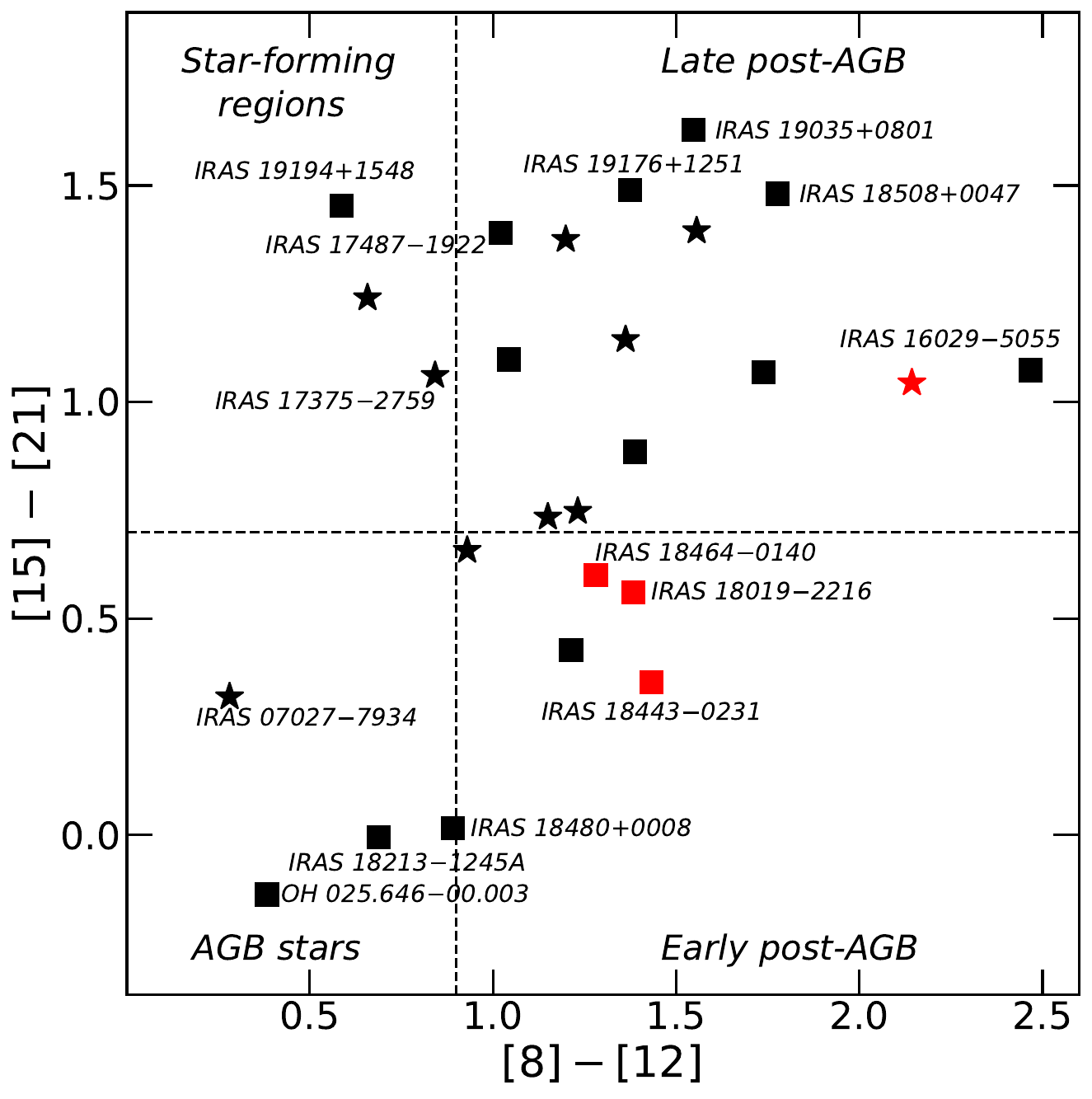}
      \caption{\textit{MSX} color-color diagram, as defined by \cite{sev02}, including all known and new maser-emitting PNe and candidates, and adopted from \cite{cal22}. The vertical and horizontal dashed lines separate the diagram into four quadrants, where different types of sources tend to cluster. The filled star and square symbols without source names represent the previously reported maser-emitting PNe and candidates, respectively. These same symbols with source names are the new maser-emitting PNe and candidates, respectively. IRAS\,16029$-$5055, IRAS\,17375$-$2759 and IRAS\,19176+1251 are included as PNe (see Section\,\ref{sec:spitzer}). {Symbols with red color represent sources with} WF characteristics.}
      \label{fig:msx}
\end{figure}

In Fig.\,\ref{fig:wise} we show a \textit{WISE} color-color diagram of all known and new maser-emitting PNe and candidates, adopted from \cite{cal22}. The diagram was presented in \cite{gom17} for a comparison between the colors of WFs and the optically obscured post-AGB stars. \cite{cal22} suggested that the colors 10.5$\leq$[3.4]$-$[22]$\leq$13.5, and 4.0$\leq$[4.6]$-$[12]$\leq$7.0 could identify new nascent PNe and candidates. Adding to this diagram the new confirmed and candidate maser-emitting PN found in this paper, we can see that these objects tend to cluster in the area delimited by the colors mentioned above. However, these limits are not strict, since some sources lie clearly outside of them. In particular, there are two candidates with $[3.4]-[22]\simeq 5$ (IRAS 18480+0008 and OH 25.646$-$00.003), which seem to correspond to objects within or close to the quadrant of AGB stars in Fig.\,\ref{fig:msx}, suggesting that they may be different from the rest of confirmed or candidate maser-emitting PNe. 

\begin{figure}
      \centering
           \includegraphics[width=0.95\hsize]{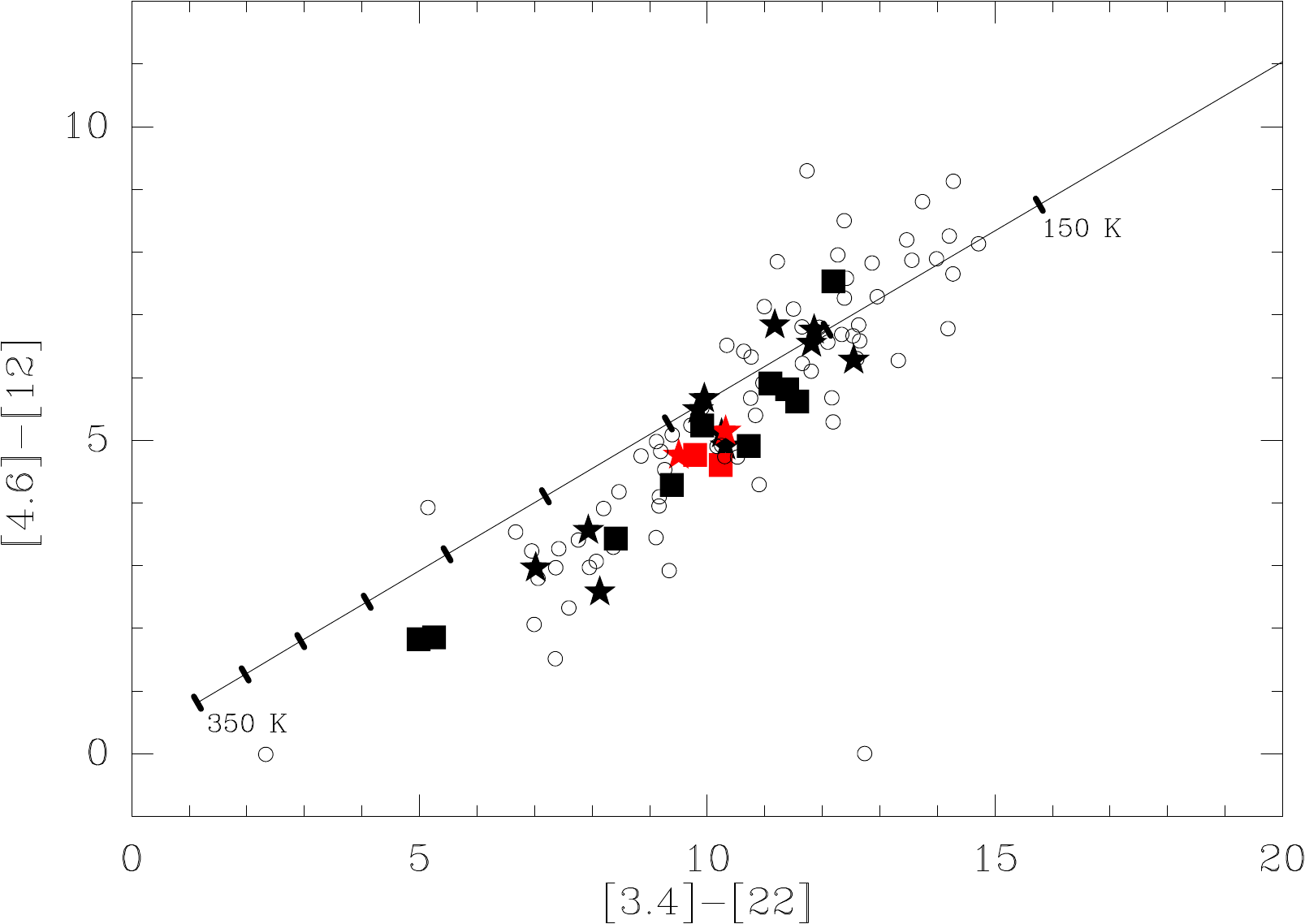}
      \caption{\textit{WISE} color–color diagram, similar to the one presented in \cite{gom17}, including all known and new maser-emitting PNe and candidates, and adopted from \cite{cal22}. The stars and squares represents the confirmed maser-emitting PNe and candidates, respectively, as presented in Fig.\,\ref{fig:msx}. The red color stands for WF nature of the source. The open circles represent obscured post-AGB candidates \citep{ram09, ram12}. The solid line represents the loci of the colors for blackbody brightness distributions. The tick marks in the blackbody line go from 150 to 350\,K at increment steps of 25\,K.}
      \label{fig:wise}
\end{figure}

\subsection{Mid-infrared spectra and diagnostic diagram} \label{sec:spitzer}

We have searched for \textit{Spitzer} spectra of objects observed with the Infrared Spectrograph \citep[IRS;][]{hou04} and enhanced data products in the \textit{Spitzer} Heritage Archive. {We found spectra, obtained with the Short-Low (SL) and Short-High (SH) modules covering the range 5.13-14.29\,$\mu$m and 9.9-19.6\,$\mu$m, respectively, for two of the new OHPN candidates identified in this work, IRAS\,16029$-$5055 (Program ID: 50652; PI:\,D.\,Engels) and IRAS\,19176+1251 (Program ID: 30258; PI:\,P.\,Garcia-Lario), which we show in Fig.\,\ref{fig:spitzer}. We also found \textit{Spitzer} spectra from one previously reported OHPN, IRAS 17393$-$2727 (Program ID: 3633; PI:\,M.\,Bobrowsky) \citep[][]{pot87, gar07}, and one previously reported OHPN candidate, IRAS 17375$-$2759 (Program ID: 3235; PI:\,M.\,Waelkens) \citep[][]{usc12},} which are also shown in those figures, for comparison purposes. In addition, we found two highly processed archive spectra (2-45\,$\mu$m) in the Infrared Space Observatory (ISO) data center\footnote{The spectra presented by \citet{usc12} are available in the ISO data center and the technical report is available at \url{https://nida.esac.esa.int/hpdp/technical_reports/technote52.pdf}} of IRAS\,17347$-$3139 and K\,3-35, two previously reported OHPN and H$_{2}$OPN \citep[][]{mir01, deG04, taf09}, which were shown in \citet{usc12}, and we analyze them below, also for comparison purposes.

\begin{figure*}
      \centering
            \includegraphics[width=0.46\hsize]{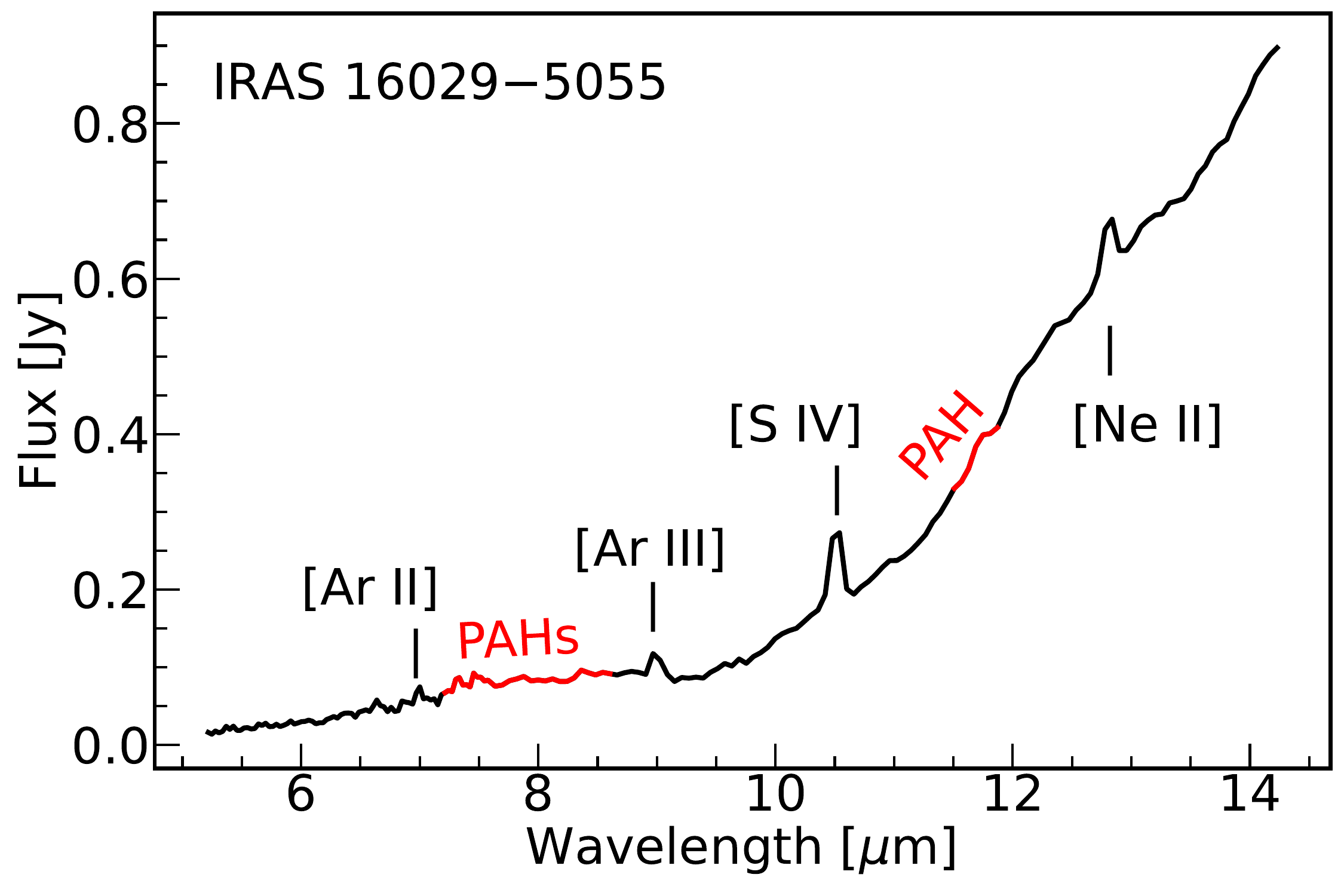}
            \includegraphics[width=0.463\hsize]{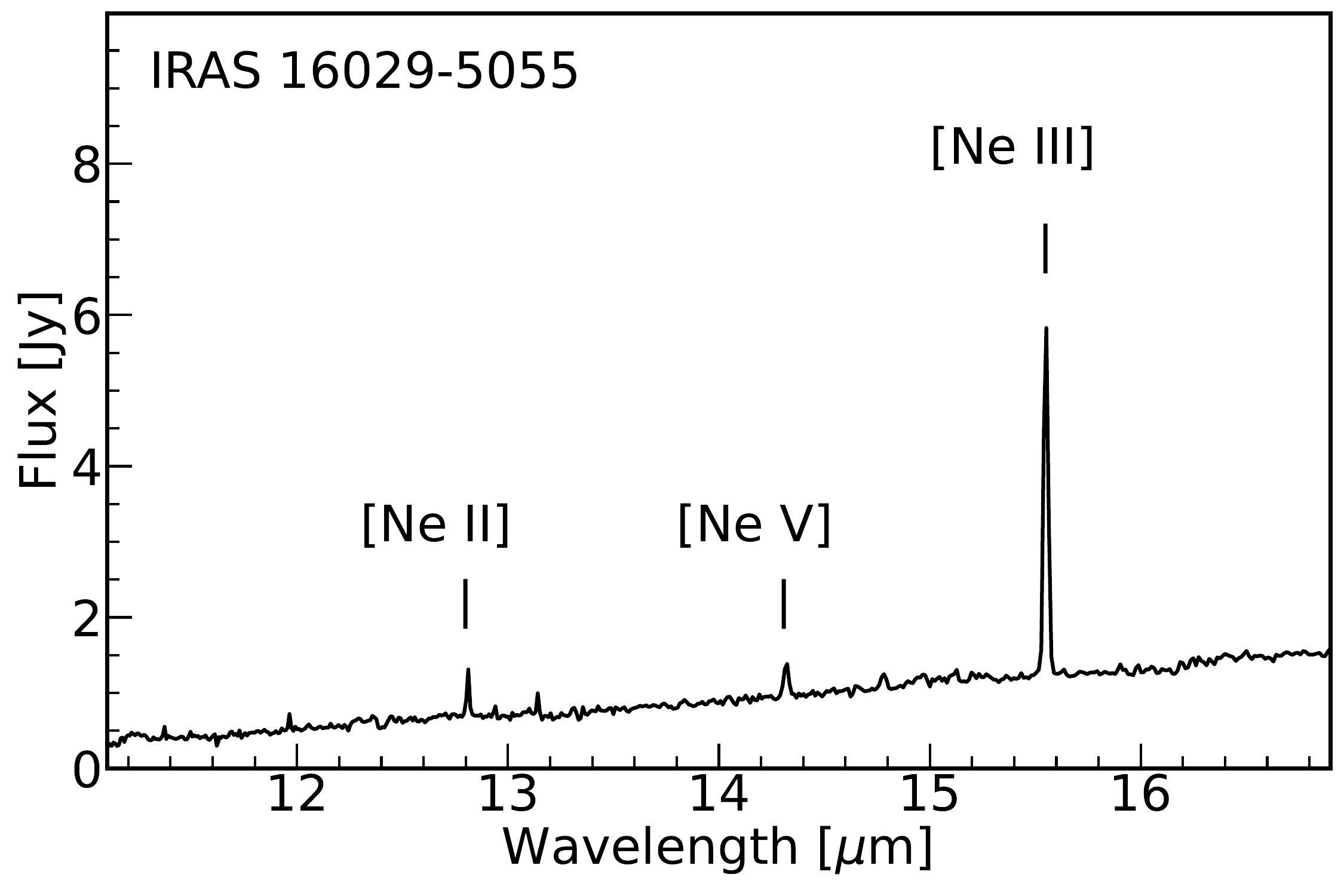}
            \includegraphics[width=0.46\hsize]{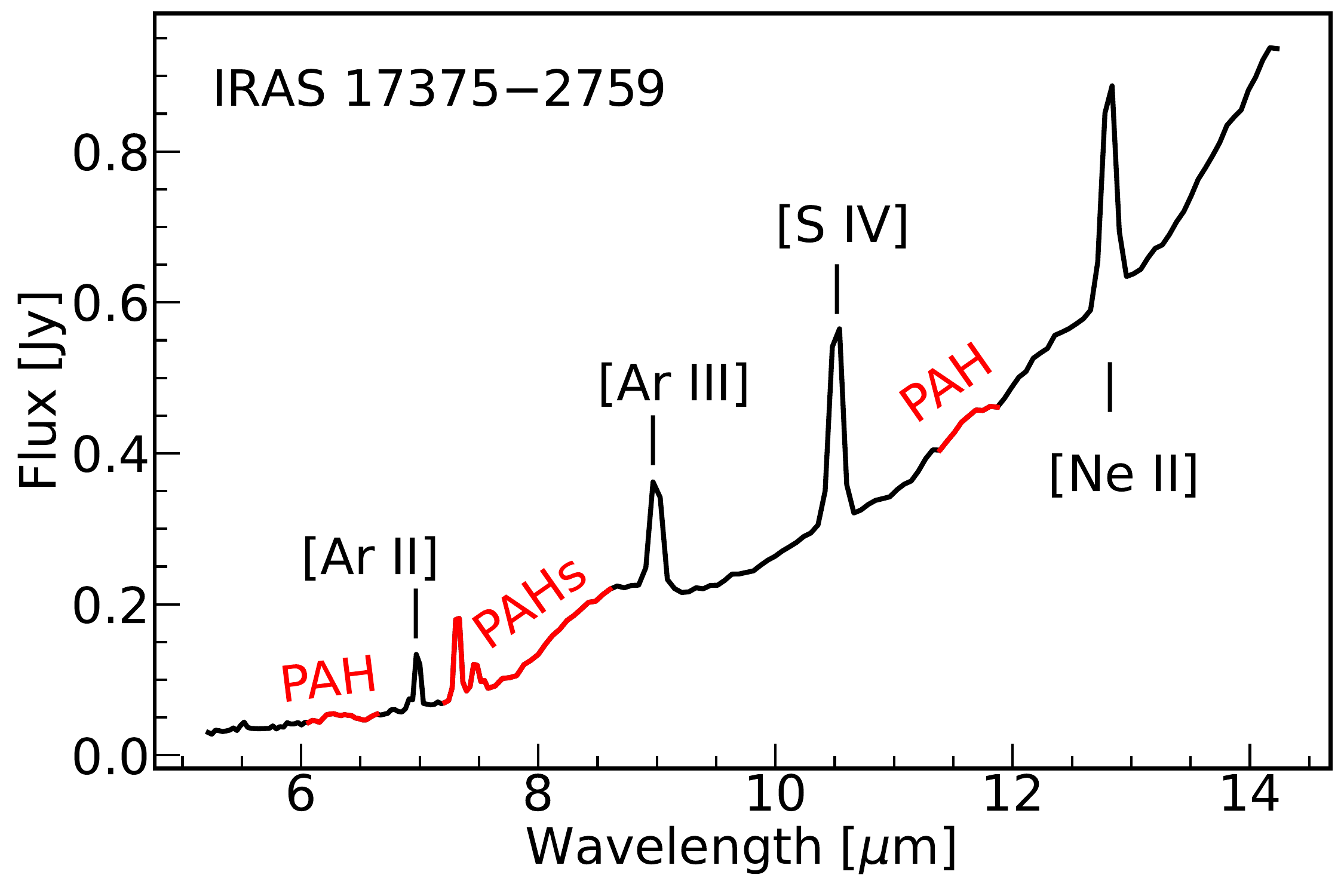}
            \includegraphics[width=0.46\hsize]{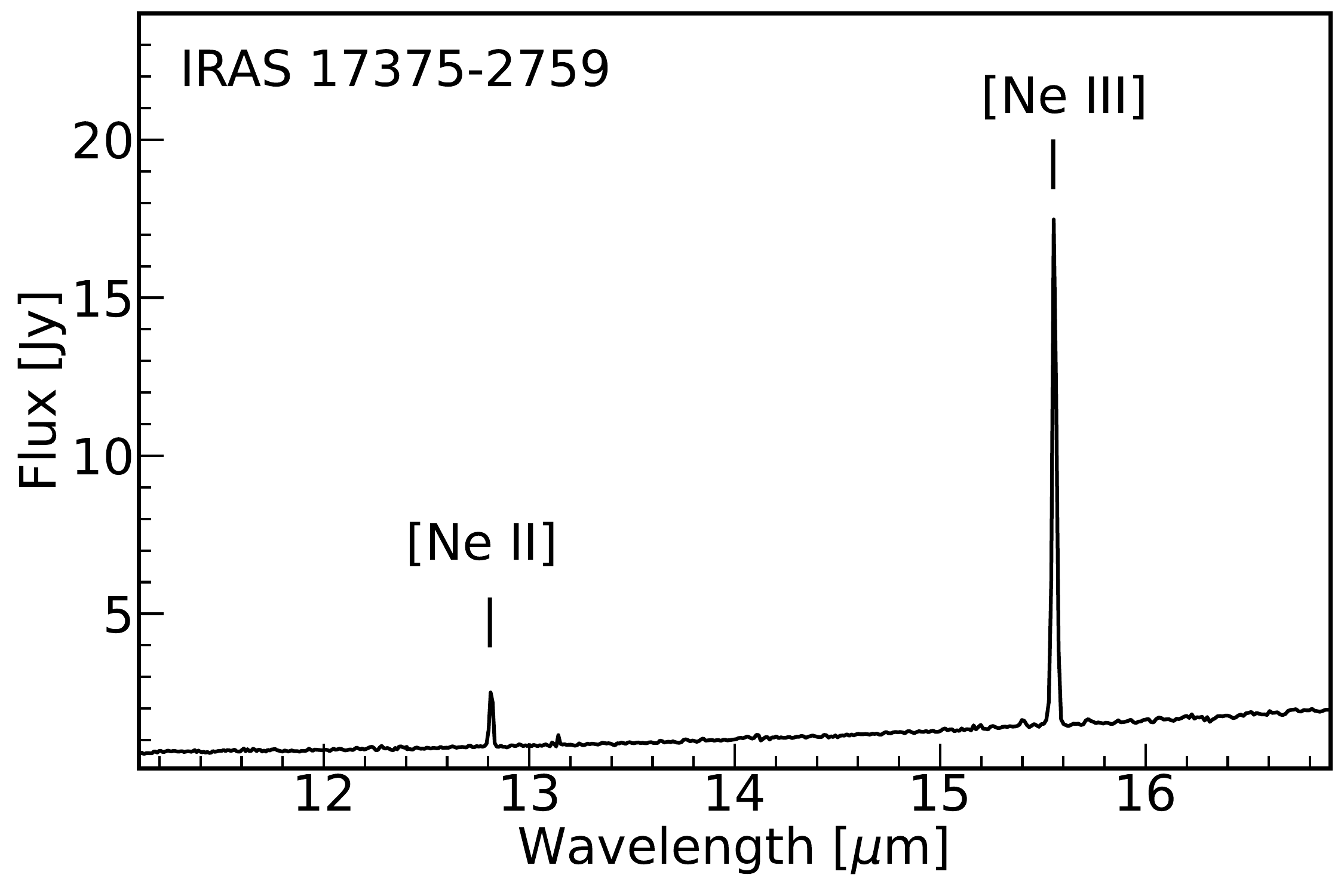}
            \includegraphics[width=0.46\hsize]{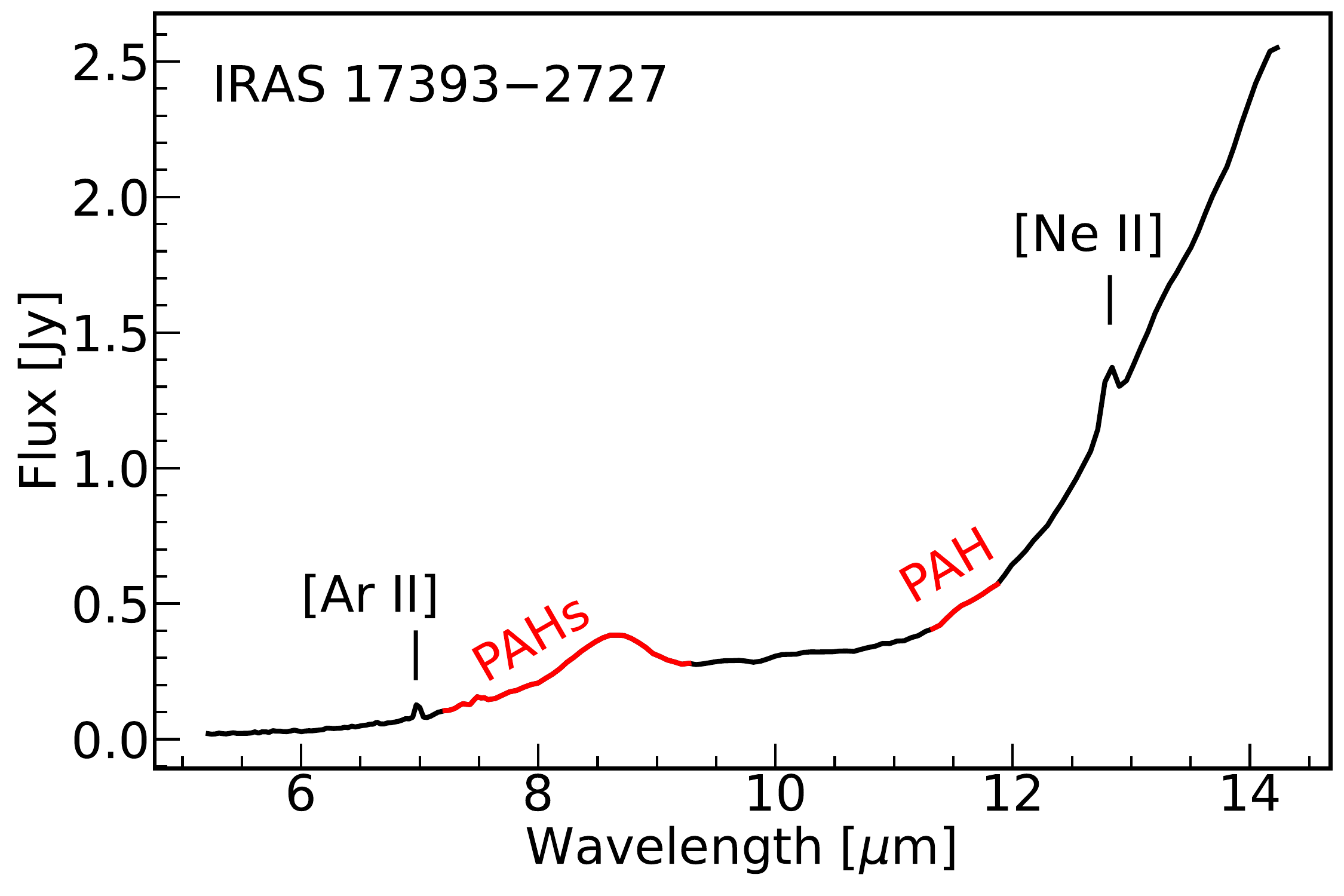}
            \includegraphics[width=0.46\hsize]{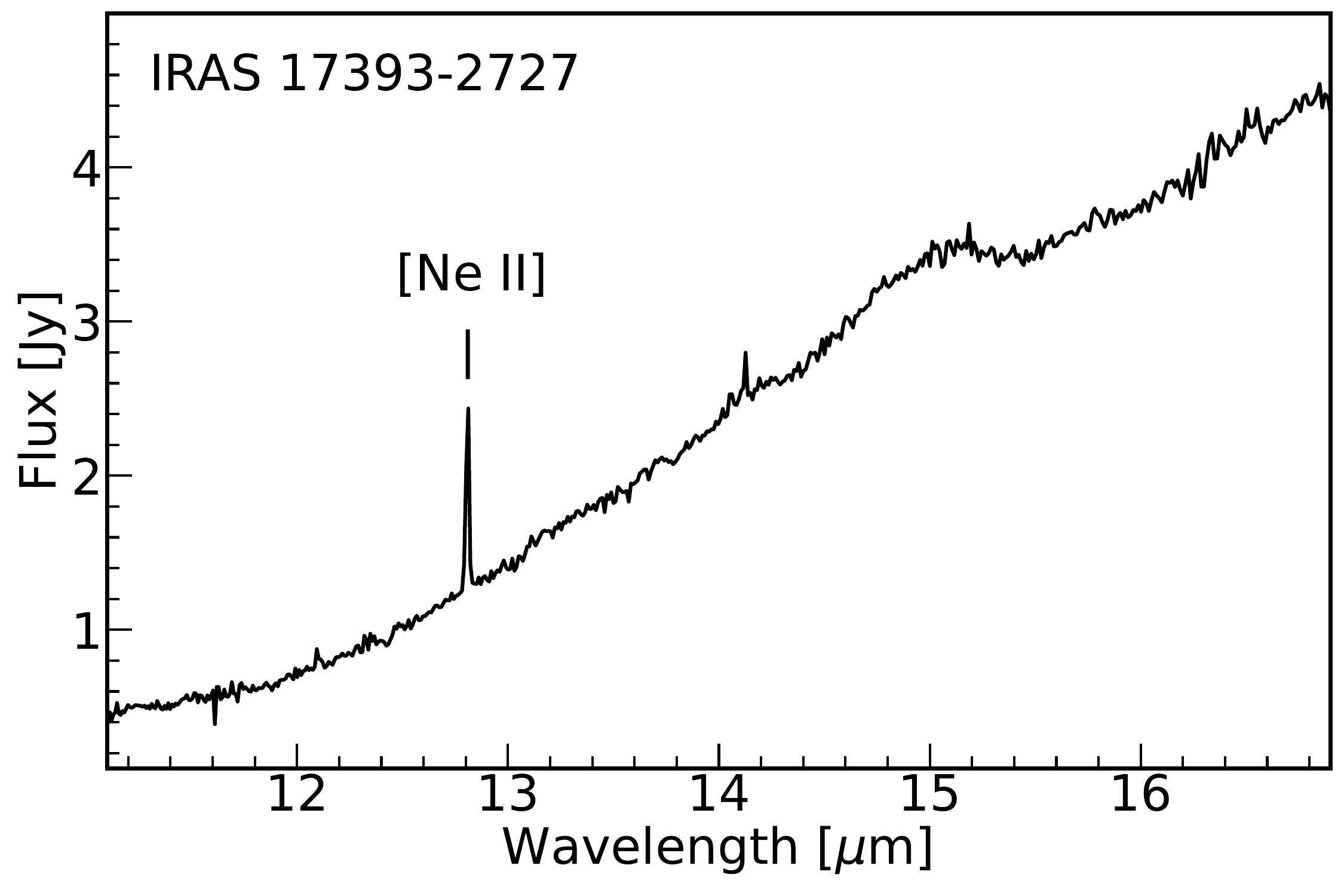}
            \includegraphics[width=0.46\hsize]{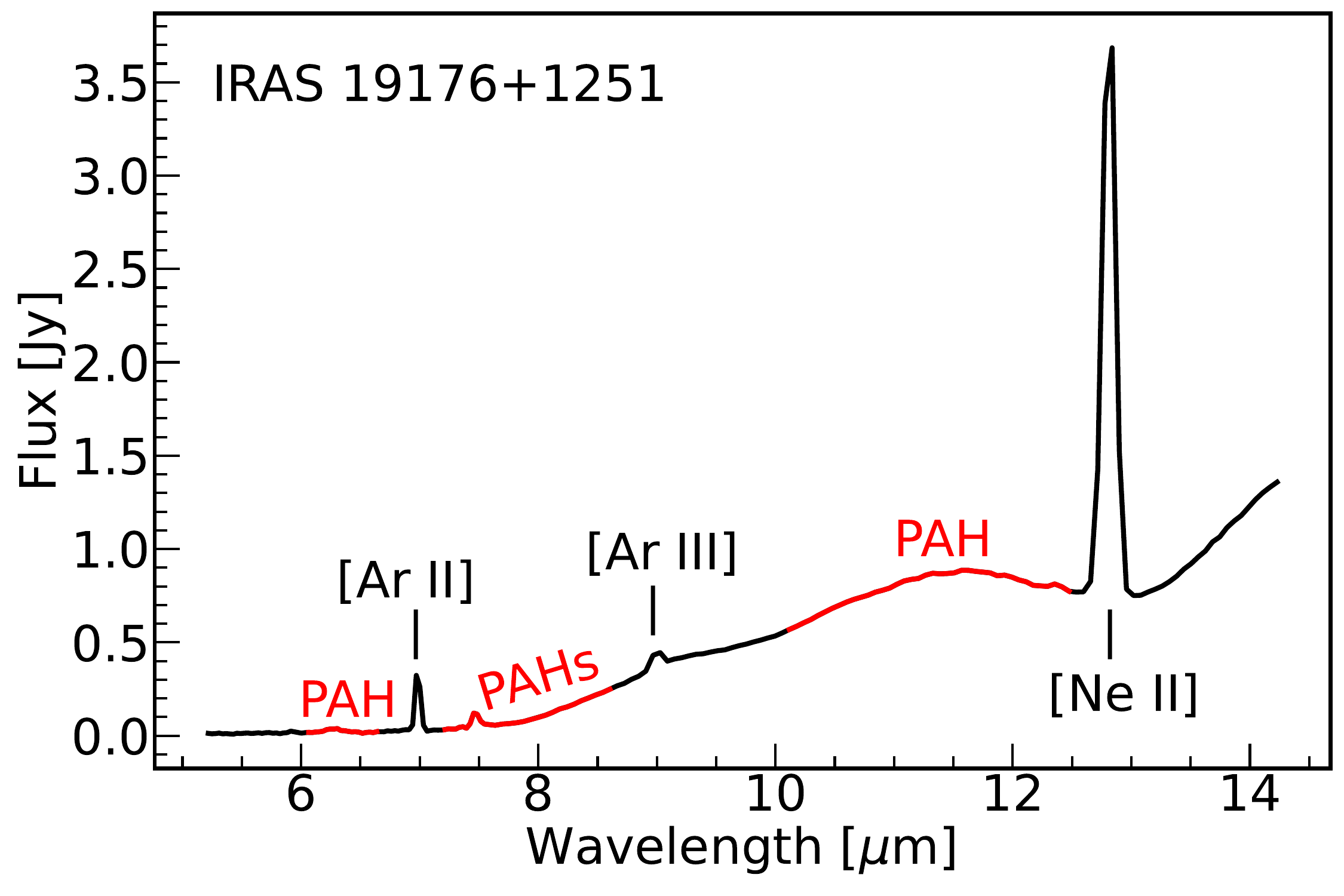}
            \includegraphics[width=0.46\hsize]{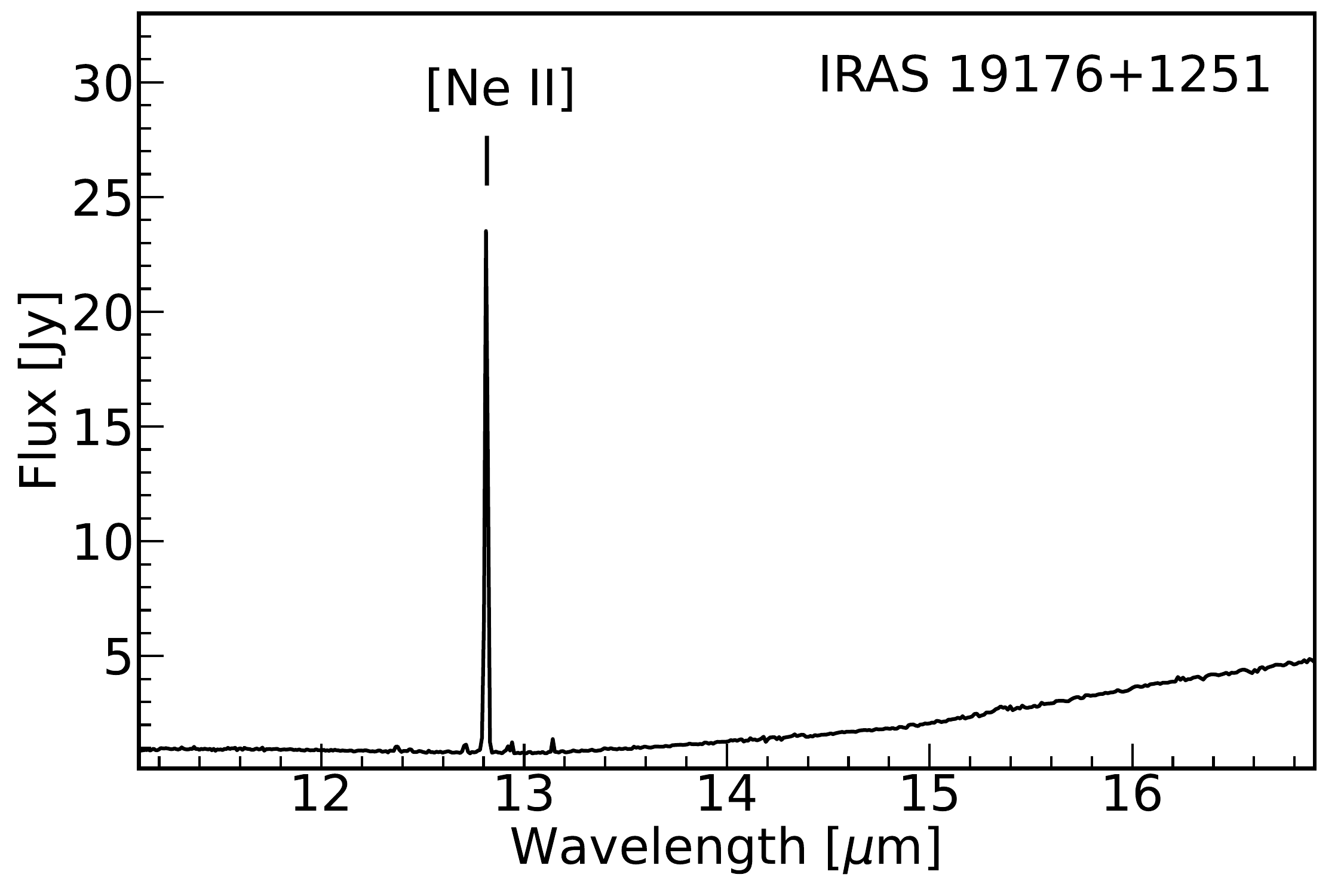}
      \caption{Mid-infrared spectra of IRAS 16029$-$5055, IRAS 17375$-$2759, IRAS 17393$-$2727, and IRAS 19176+1251{, as seen by the IRS SL (left) and SH (right) modules of \textit{Spitzer}}. Narrow and broad emission lines (red color) from ionized gas and dust grains, respectively, are present in their spectra. {The narrow emission lines of [Ar\,{\sc ii}], [Ar\,{\sc iii}], [S\,{\sc iv}], [Ne\,{\sc ii}], [Ne\,{\sc iii}], and [Ne\,{\sc v}] are observed at around 6.98, 8.98, 10.51, 12.84\,$\mu$m, 15.55, and 14.32\,$\mu$m, respectively.} The broad emission lines are usually attributed to emission from PAHs, which present a different chemistry (C-rich) to that of the O-rich maser-emitting regions.}
      \label{fig:spitzer}
\end{figure*}

The \textit{Spitzer} spectra shown in Fig.\,\ref{fig:spitzer} display narrow and wide emission {features} commonly associated with ionized gas and dust grains, respectively. Usual ionized gas tracers are the fine-structure spectral lines of [Ar\,{\sc ii}], [Ar\,{\sc iii}], [Ne\,{\sc ii}], [Ne\,{\sc iii}], [Ne\,{\sc v}], and [S\,{\sc iv}] at 6.98, 8.98, 12.81, 15.55, 14.32, and 10.51\,$\mu$m, respectively, while the broad {features} at 6.3, 8.6, and 11.6 $\mu$m are commonly associated with emission from polycyclic aromatic hydrocarbons (PAHs). In addition, there is spectrally resolved emission of PAHs at 7.4\,$\mu$m. The spectra of galaxies, H\,{\sc ii} regions, and YSOs with low and high mass display an absorption feature at 9.8\,$\mu$m \citep[e.g.,][]{pee04, simp12}, which is absent in the spectra shown here. Moreover, the lines of the ionized gas in this wavelength range are characteristic of PNe, and not of post-AGB stars \citep[see also][for \textit{Spitzer} spectra of post-AGB stars and PNe in the Milky Way and Large Magellanic Cloud]{mat14}, so IRAS\,16029$-$5055, IRAS\,17375$-$2759, and IRAS\,19176+1251 are likely to be PNe. 

In order to better determine if these spectral lines tracing ionized gas in the mid-infrared actually identify these objects as PNe, we used the Mexican Million Models Database \citep[3MdB;][]{mor15, ala19}, which allowed us to analyze whether the line intensities are consistent with shocks or photoionization. We have used the shock models of \citet{sut17} and the photoionization models of \citet{fer17} stored in the 3MdB to create line-ratio diagnostic diagrams. We did not search for exact models to match the observed line ratios. Instead, we used the maximum number of models to define the entire region within these diagrams that is compatible to either origin of ionization. More details on the grid parameters are provided in the Appendix\,\ref{ap:models}. The final diagnostic diagram is shown in Fig.\,\ref{fig:3mdb_dd3}, displaying line ratios compatible with shock ionization and photoionization, which we compared with our data.

\begin{figure*}
      \centering
            \includegraphics[width=0.9\hsize]{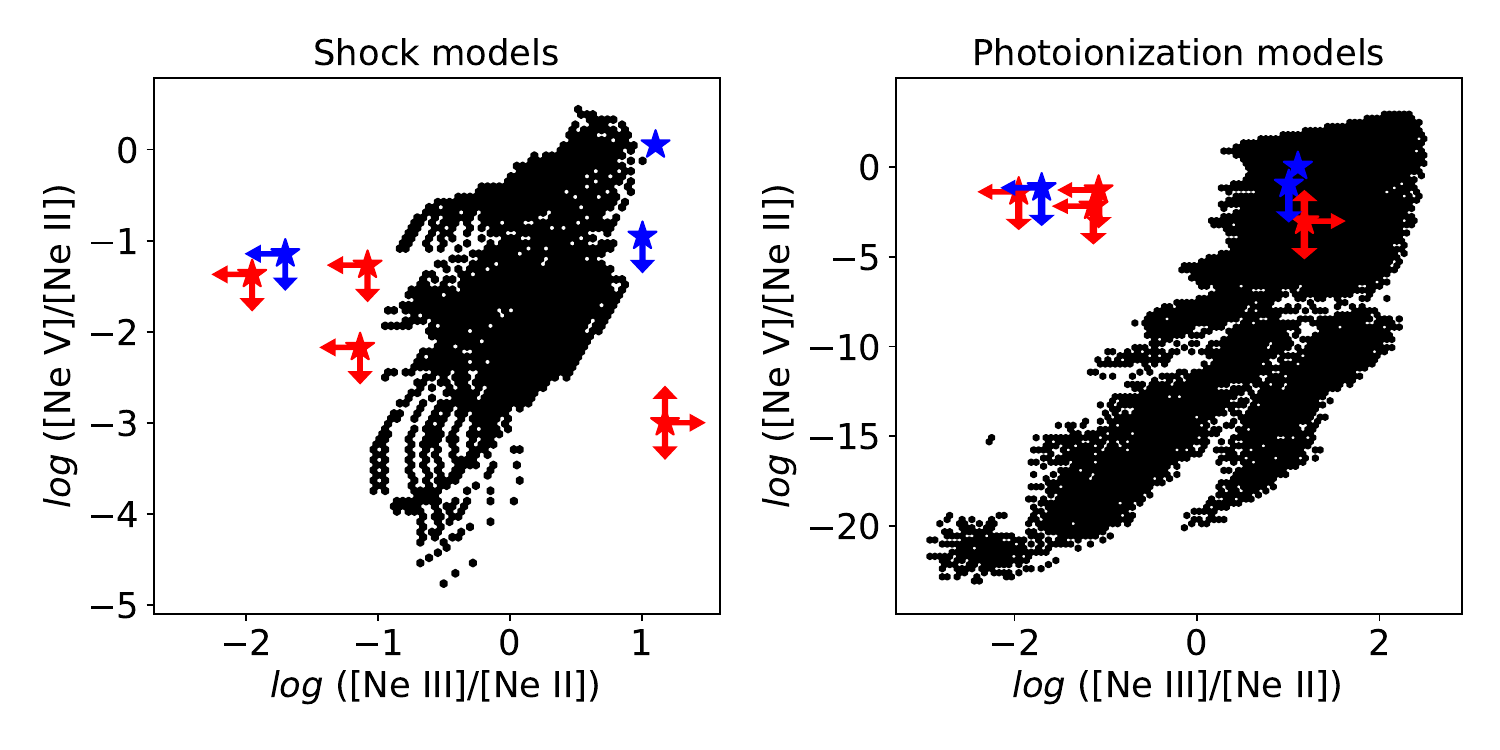}
      \caption{Line ratio diagnostic diagram for the mid-infrared emission lines of [Ne\,{\sc ii}], [Ne\,{\sc iii}], and [Ne\,{\sc v}] at 12.81, 15.55, and 14.32\,$\mu$m, respectively. The left panel shows shock models, and the right panel shows photoionization models. The blue stars mark the new objects in which the presence of photoionized gas is favored by these diagrams (IRAS\,16029$-$5055, IRAS\,17375$-$2759, and IRAS\,19176+1251). The red stars mark previously reported maser-emitting PNe (IRAS\,17347$-$3139, IRAS\,17393$-$2727, and K\,3-35). The sizes of the symbols are larger than their errors. The arrows indicate values of line ratios that are either upper or lower limits, derived from the upper limits for the lines not detected in each case (see Fig.\,\ref{fig:spitzer}). In one case (K\,3-35, red star with three arrows) the value $log$\,([Ne\,{\sc v}]/[Ne\,{\sc ii}]) is undetermined, and it could be located anywhere in the vertical direction. See Appendix\,\ref{ap:models} for more details on the grid parameters.}
      \label{fig:3mdb_dd3}
\end{figure*}

In this comparison, we have not corrected the spectra by extinction. This extinction is relatively low in the mid-infrared, and we have used ratios of lines that are close in wavelength ([Ne\,{\sc ii}], [Ne\,{\sc iii}] and [Ne\,{\sc v}] seen at 12.81, 15.55, and 14.32\,$\mu$m, respectively) so their line intensity ratios are practically insensitive to extinction \citep[e.g.,][]{gen98, mar02, per10, wea10}. In addition, each spectrum has been continuum-subtracted using a polynomial curve fitting, and then each emission line was measured with an individual Gaussian fitting using Astropy \citep{ast13, ast18, ast22}.

Although we can reasonably expect that the origin of ionization in nascent PNe is a combination of both shocks and photoionization, the location of the sources in the line-ratio diagnostic diagrams that separate either process is revealing. The intensity ratios fall outside the region covered by the shock ionization models in all cases, specially in the [Ne\,{\sc iii}]/[Ne\,{\sc ii}] ratio, whereas they can be consistent with the photoionization models. The incompatibility with shocks alone indicates that these fine-structure spectral lines must have a significant photoionization contribution. Therefore, these models favor the presence of photoionized gas in IRAS\,16029$-$5055, IRAS\,17375$-$2759, and IRAS\,19176+1259. For comparison, previously reported maser-emitting PNe IRAS\,17347$-$3139, IRAS\,17393$-$2727, and K\,3-35 also falls outside the region of shock models in Fig.,\ref{fig:3mdb_dd3} (left).

\subsection{Spectral indices and variability of the radio continuum emission}
\label{sec:continuum}

The dependence of the flux density ($S_{\nu}$) of the radio continuum emission with frequency ($\nu$) can give useful hints on the emission processes and therefore on the nature of the sources. Spectral indices ($\alpha$, with $S_\nu\propto \nu^\alpha$) of thermal (free-free) emission of the photoionized gas in PNe must be in the range from $+2$ at low frequencies (optically thick regime) and $-0.1$ at high frequencies (optically thin regime). The observed radio spectra usually show two different ranges \citep[e.g.,][]{tay87,aaq91,gru07}. At low frequencies,  $\alpha$ ranges from $\simeq +0.6$ to +2, indicating (partially) optically thick emission, while at {high} frequencies the slope flattens to values $\alpha\simeq -0.1$. The turn-over frequency separating both regimes can be used to estimate the emission measure (EM) of the nebulae that, in its turn, is thought to be related with age, with younger PN having higher EM \citep{kwo81}. Thus, the turn-over frequencies would tend to decrease with time.
In the particular case of a nascent PN, while the ionization front advances 
{ at supersonic velocities \citep[R-type ionization front,][]{kah54,fra90}, the plasma still briefly maintains}
the density distribution of the circumstellar envelope created by mass-loss in the AGB phase, with a radial dependence of the electron density to be $n_{e}\propto r^{-2}$. In this case, the radio spectral index must be around +0.6 (partially optically thick) in most of the centimeter domain \citep[see e.g.,][]{ol75, pan75}. Spectral indices on this order, and high turn-over frequencies ($> 20$ GHz) are indeed observed in nascent PNe \citep[e.g.,][]{gom05, cal22}. However, we note that the same spectral index is also expected in the case of shock-ionized winds, including biconical jets \citep{rey86}, which could be present before the PN phase. In this case, considerations of the required mass-loss rate to produce the observed flux densities can help us to discern if the emission is compatible with shock ionization or photoionization \citep[see][]{taf09, usc14}. One of such cases is IRAS\,17347$-$3139, which was reported before as PN because the required mass-loss rates are incompatible with shock ionization alone \citep[][]{taf09}, and in this work (Sec.\,\ref{sec:spitzer}) we spectroscopically confirmed the presence of photoionized gas in this object and, therefore, corroborated that it is a PN.

\begin{figure*}[h]
      \centering
            \includegraphics[width=0.32\hsize]{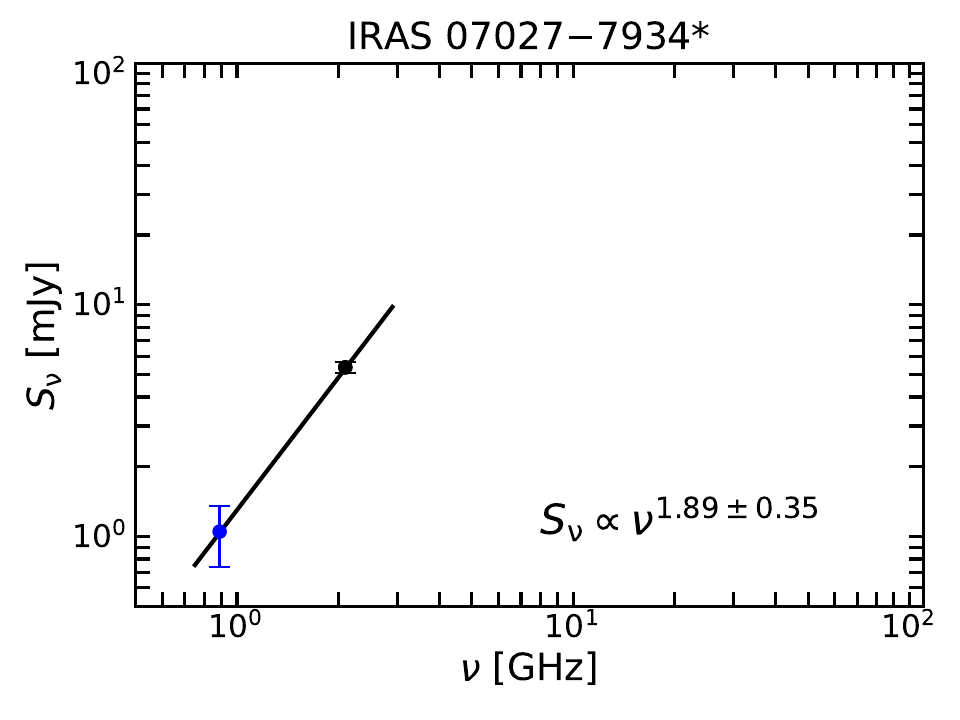}
            \includegraphics[width=0.32\hsize]{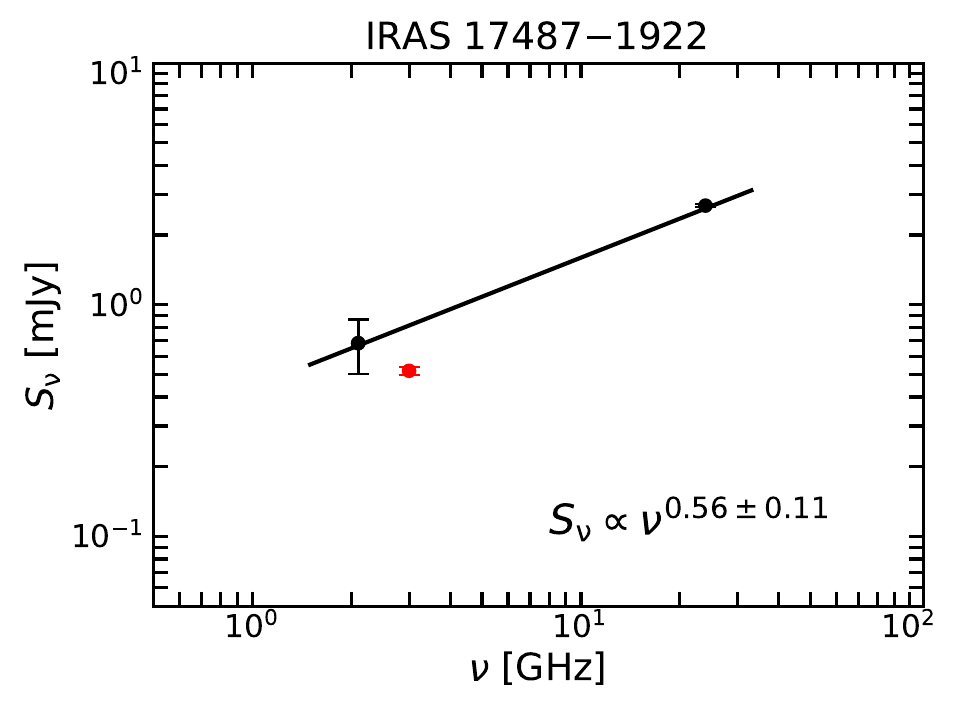}
            \includegraphics[width=0.32\hsize]{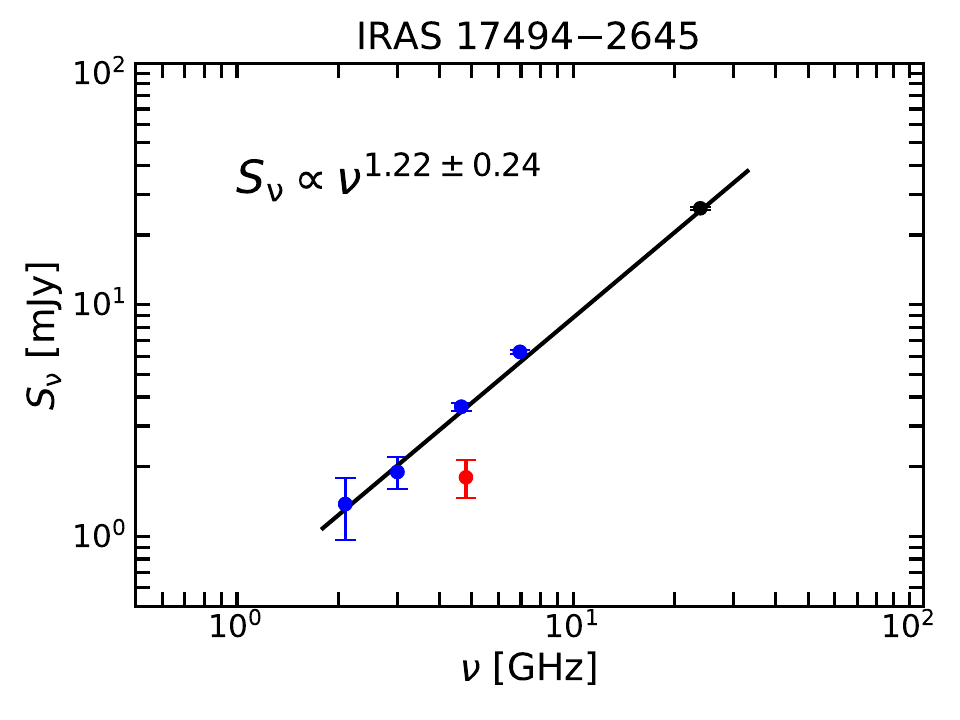}
            \includegraphics[width=0.32\hsize]{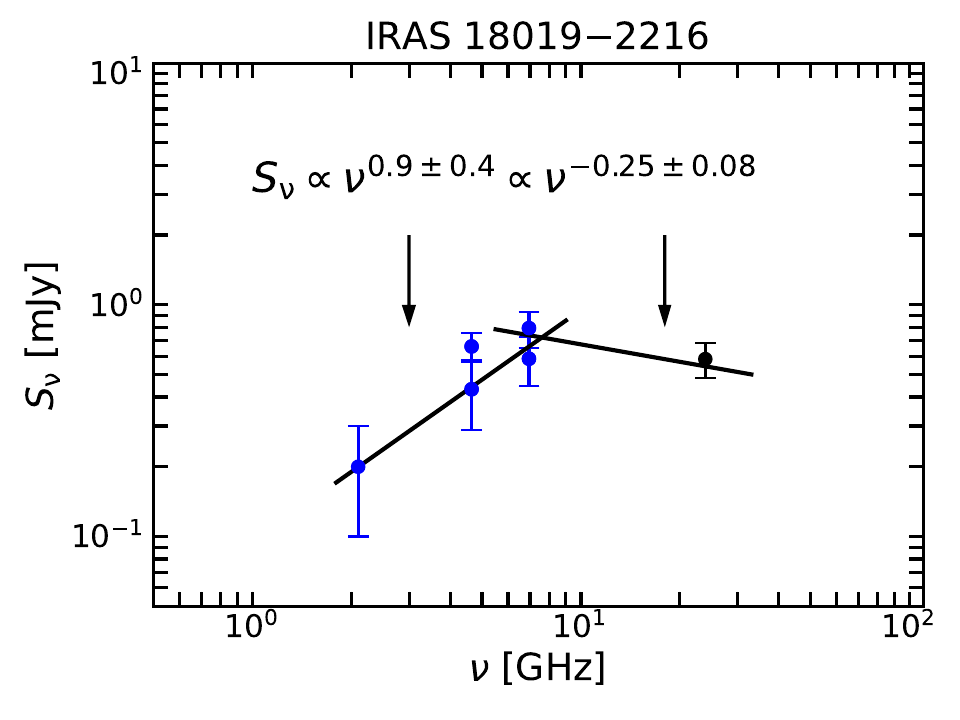}
            \includegraphics[width=0.32\hsize]{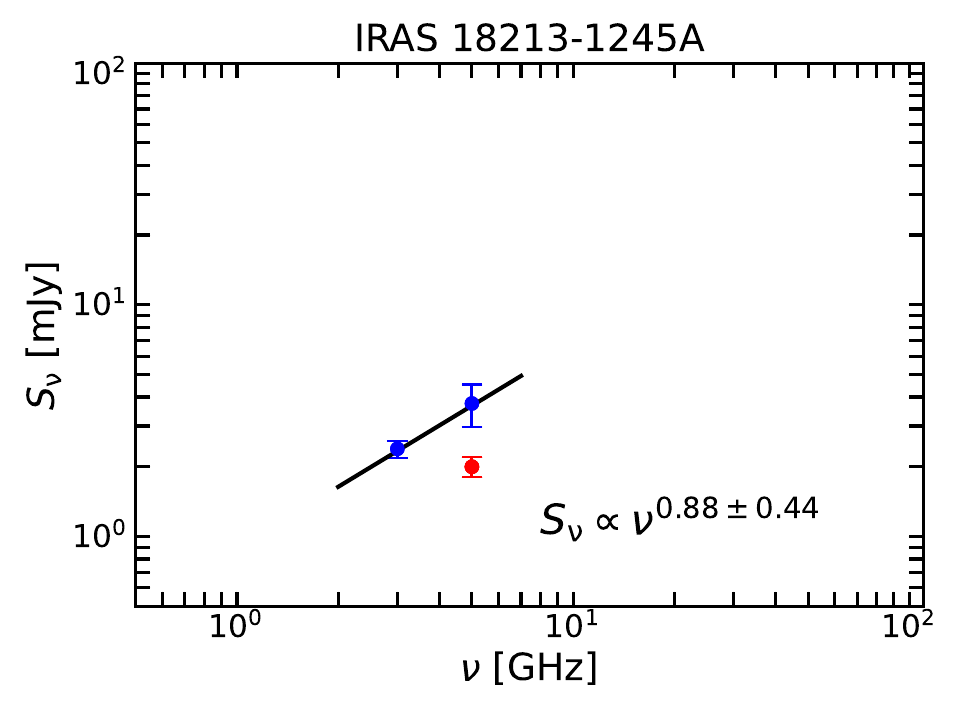}
            \includegraphics[width=0.32\hsize]{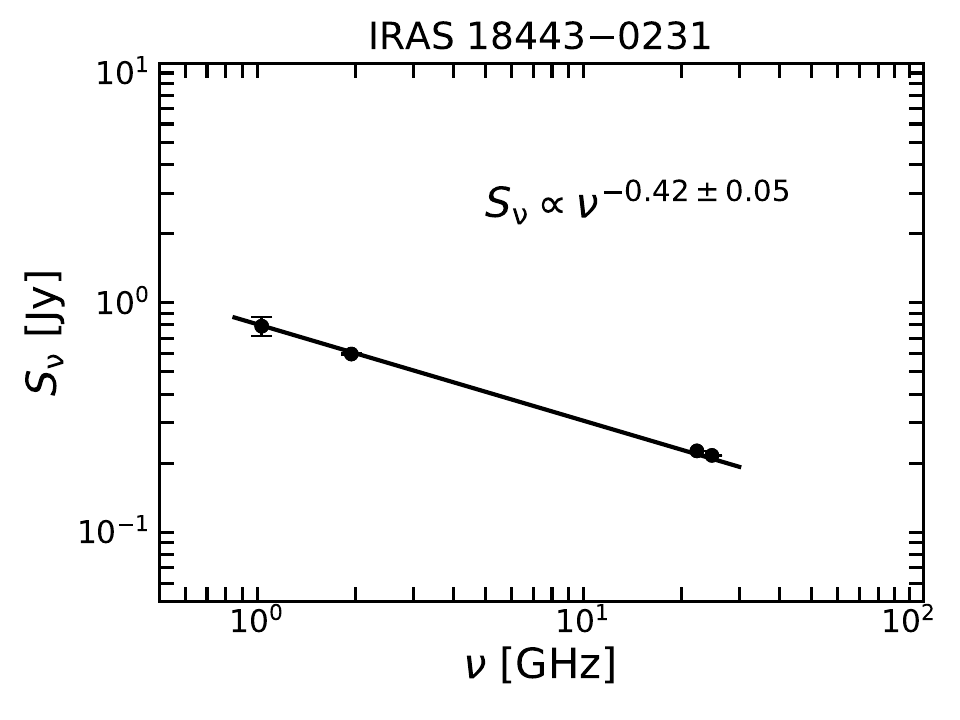}
            \includegraphics[width=0.32\hsize]{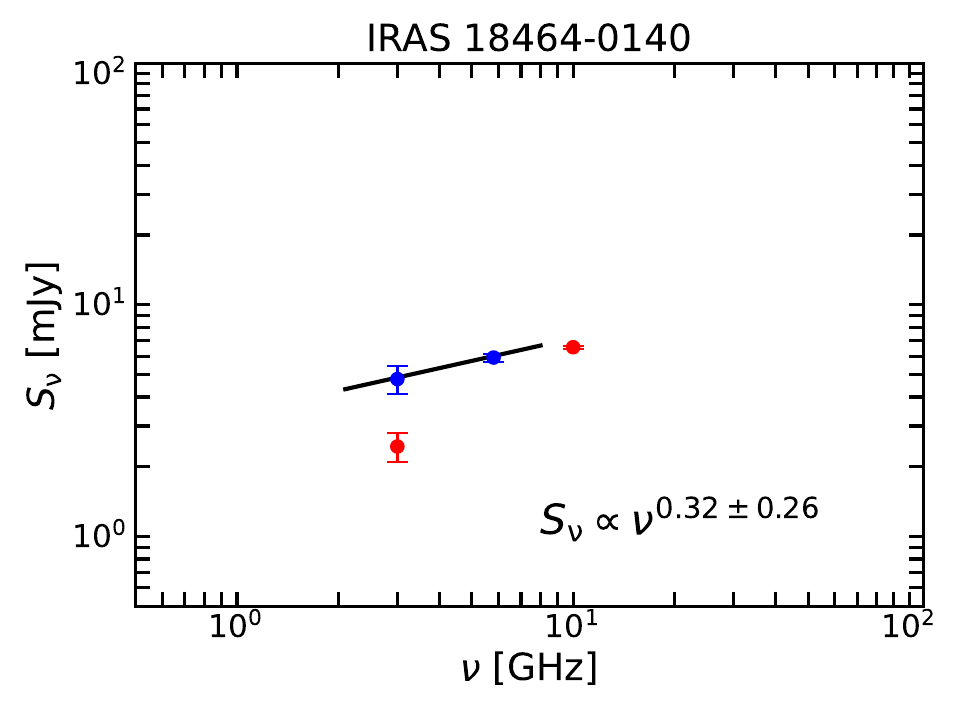}
            \includegraphics[width=0.32\hsize]{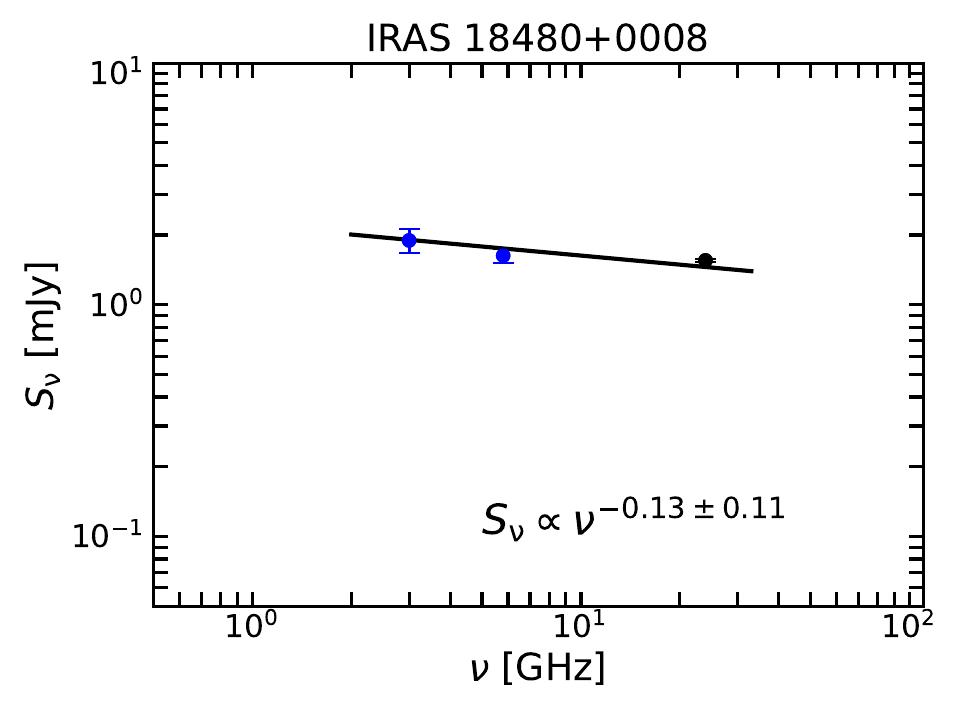}
            \includegraphics[width=0.32\hsize]{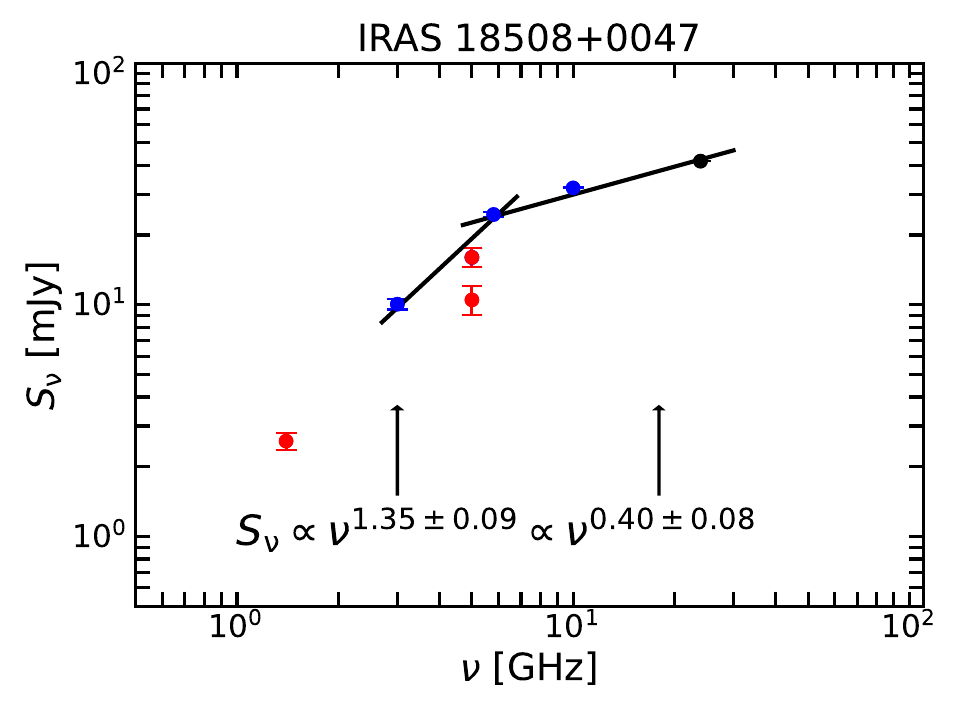}
            \includegraphics[width=0.32\hsize]{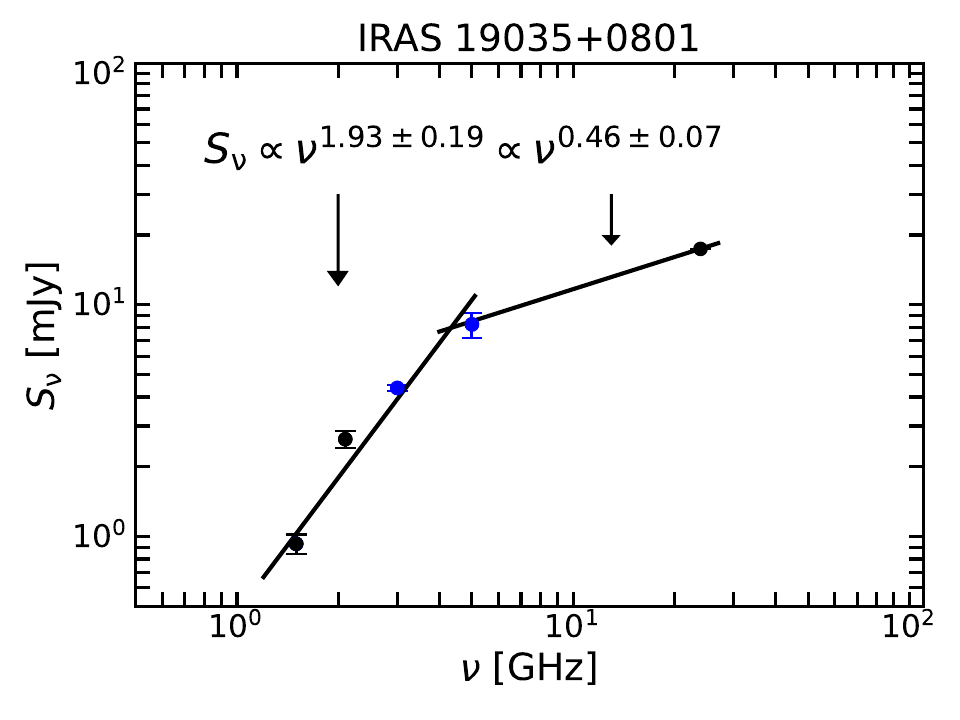}
            \includegraphics[width=0.32\hsize]{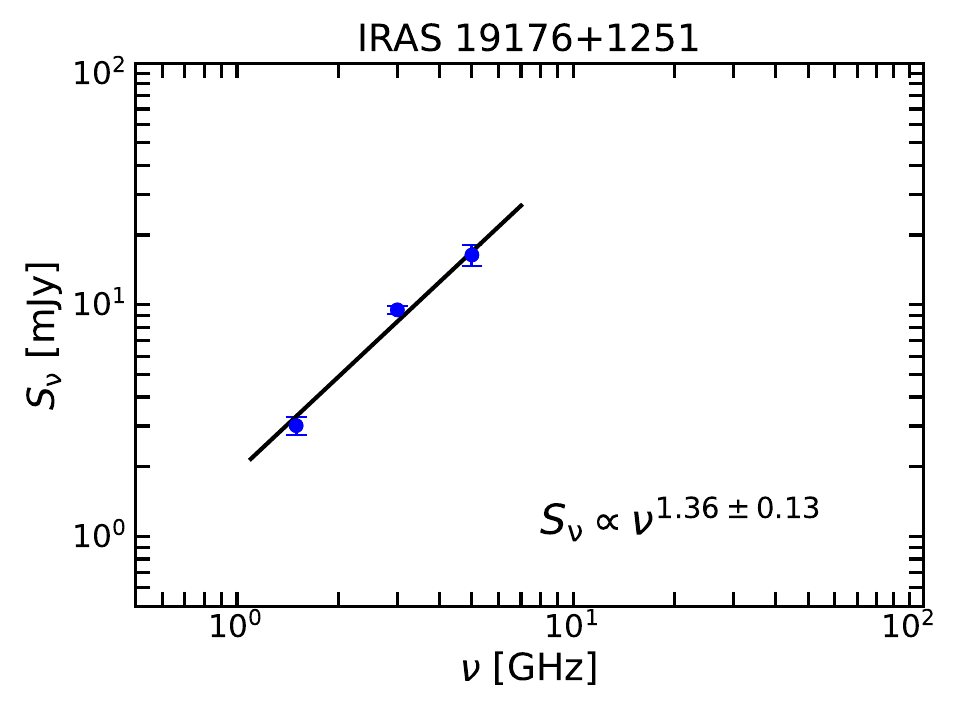}
            \includegraphics[width=0.32\hsize]{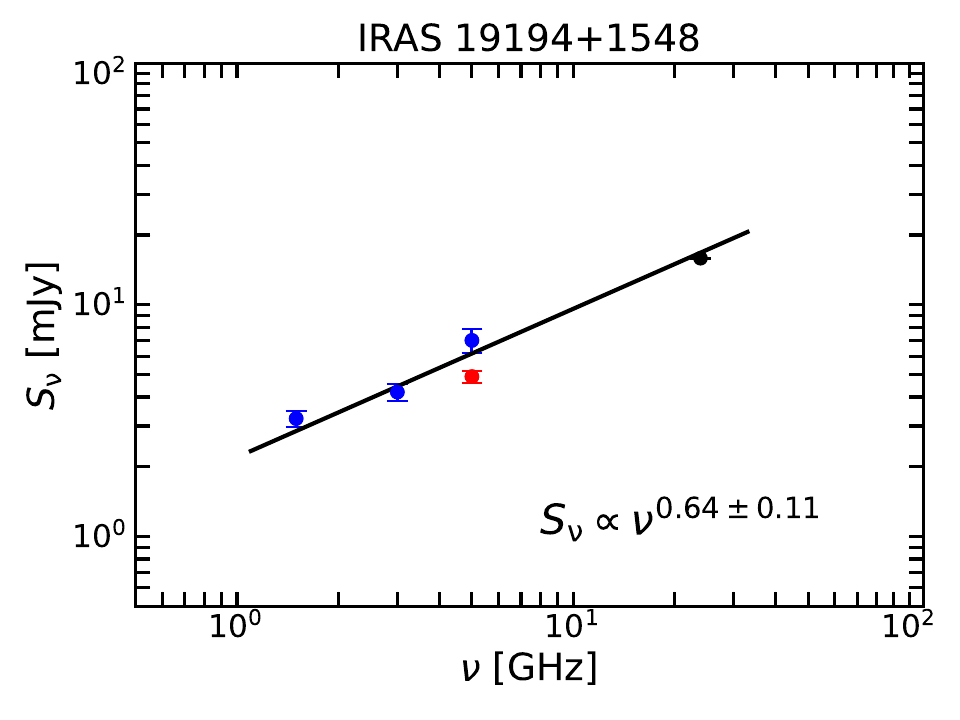}
      \caption{Radio continuum emission of the new maser-emitting PN (indicated with an asterisk next to their name) and candidates (rest of the panels), including the new flux density values at 24.5 GHz from our VLA observations in the previously reported OHPN candidates IRAS\,17494$-$2645 and IRAS\,18019$-$2216 \citep{cal22}. The source names and obtained spectral indices ($\alpha$) are indicated in each panel. Black circles are data from our ATCA/VLA observations. Blue circles are data from the literature which, together with the black circles, were employed to fit the spectral indices. Red circles correspond to values from epochs earlier than the blue and black ones, except for IRAS 18464$-$0140, where the point at 10\,GHz is from a more recent epoch. Thus, these red points were not used in the spectral index fits, since they were obtained at a significantly different epoch. {See the text for more details.}}
      \label{fig:spectral_indices} 
\end{figure*}

Together with the radio continuum emission in our data, we collected publicly available measurements of flux densities in our targets at different frequencies. Recent observational efforts have aimed at updating and creating new all-sky radio continuum surveys such as, for instance, the VLA Sky Survey \citep[VLASS;][]{gor21}, Co-Ordinated Radio `N' Infrared Survey for High-mass star formation \citep[CORNISH;][]{ira18,ira23}, GLObal view on Star formation in the Milky Way \citep[GLOSTAR;][]{med19, dzi23}, Rapid Australian SKA Pathfinder \citep[RASKAP;][]{hale21}, or The HI/OH/Recombination line survey of the inner Milky Way \citep[THOR;][]{wan18}. 

{Maser-emitting PNe and candidates show increasing flux densities in timescales of years and decades \citep{gom05, taf09, cal22}, with the exception of IRAS\,15103$-$5754, which shows a decrease non-thermal radio continuum emission in timescales of a few months \citep{sua15}.} Hence, simultaneous observations across a large part of the centimeter domain are crucial to obtain a precise determination of the radio spectral indices. Using data at different frequencies taken at different epochs can result in unrealistic slopes in the radio spectrum, so special care should be taken in these cases. Obtaining spectral indices from different catalogs or observational projects can provide a useful first approximation, but a definite confirmation of the actual radio spectrum can only be obtained with (quasi-)simultaneous observations.
In our case, while flux densities reported in the surveys mentioned above can be used to obtain preliminary radio spectra of the new identified maser-emitting sources, we can obtain even more solid confirmation of the spectral indices by focusing only on the measured flux density values at different frequencies in our own observations for several targets, which were taken within less than eight days for each individual source (Table \ref{tab:obs_atca_vla}). 

Fig.\,\ref{fig:spectral_indices} shows the flux densities and determined spectral indices of the new maser-emitting PNe and candidates, using the data from our observations (black symbols) and from the surveys mentioned above (red and blue symbols). The black and blue symbols were used in the fits to estimate spectral indices. {In the case of the red points, most of them were obtained from an earlier epoch than the blue ones and/or our data. In particular, the red point of IRAS\,17487$-$1922 was reported by VLASS from observations performed on 30/06/2019, which significantly deviates from the observed tendency from the data points of our own VLA observations; the one of IRAS\,17494$-$2645 was discussed by \citet{cal22}; the one in IRAS\,18213$-$1245A was reported by \citet{whi05}; in IRAS 18464$-$0140 the red point at 5 GHz was reported by \citet{whi05}, and that at 10 GHz is from 08/03/2022 (see Section\,\ref{sec:radio_vla}), which is the only one that is from a more recent epoch; in IRAS\,18508+0047, the red point at 1.4 GHz was reported by \citet{bec94}, while those at 5 GHz were reported by \citet{whi05} and \citet{urq09}; the red point in IRAS 19194+1548 was reported by \citet{urq09}. These red points} were not used in the spectral index fit, since they consistently show lower flux densities {and where obtained at a significantly different time than other data,} but are useful to illustrate the variability of the objects. The new OHPN candidates IRAS 16029$-$5055 and OH 025.646$-$00.003 are not presented in this figure, since we found only one flux density value available in the literature with no other data in the ATCA and VLA archives, and thus we cannot obtain the spectral index {values} of these sources. 

The radio spectrum of most of the sources identified here seems compatible with free-free emission and, therefore, these sources could be PNe. These indices would further suggest that they are not background galaxies or pulsars, which tend to show negative spectral indices, due to the non-thermal nature of their emission \citep[e.g.][]{mar00}. The most noticeable exception is IRAS 18443$-$0231, for which the obtained $\alpha$=$-$0.42$\pm$0.05 from our VLA observations between 1-24 GHz is consistent with non-thermal radio continuum emission. This result is robust, since it was obtained from our observations and taken close in time. This source is described in detail in Section \ref{sec:i18443}, providing strong arguments that it is an evolved star, {despite its non-thermal radio continuum spectrum.}

For some sources, there are indications of a spectral index change along the spectrum, although we cannot always confirm this due to the lack of quasi-simultaneous observations at different frequencies. The study of the turnover frequencies is interesting, since they may reveal some additional information. For example, the turnover between $\alpha$$\simeq$+0.6 and $-0.1$ in a nascent PN could indicate that the electron density in the innermost regions falls well below the $n_{e}\propto r^{-2}$ law. This could happen if there is an inner, non-ionized hole close to the central star \citep[][]{ang18}. Such a central hole is expected, since the PN is the result of ionizing a circumstellar envelope ejected in previous evolutionary phases. The size of such a hole will increase as the PN evolves, causing this turnover frequency to decrease with time. A possible example of such a turnover to optically thin emission ($\alpha$$\simeq$$-$0.1) at high frequencies is IRAS 18019$-$2216. Assuming a turnover frequency of 8\,GHz, the radius of this hole would be of $\simeq$12\,au\,$\times$\,$d$ for a spherical ionized region {surrounding that hole}, where $d$ is the distance in kpc from the object \citep[see Eq.\,10 of][]{ang18}, a small size that would be consistent with a young PN. On the other hand, a turnover between the optically thick regime ($\alpha$$\simeq$+2) and the standard $\alpha$$\simeq$+0.6 expected in nascent PNe may trace the existence of a sharp outer boundary in the ionized nebula. As the ionization front advances in a nascent PN, one could also expect this turnover frequency to decrease over time. Possible sources with this turnover frequency from $\alpha$$\simeq$+2 to +0.6 are IRAS 18508$+$0047 and IRAS 19035$+$0801. However, only in the latter case can our own data, obtained close in time, give more solid support to this change of spectral index.

As mentioned above, the points marked in red in Fig. \ref{fig:spectral_indices} are data from the literature that were not used in the fit of spectral indices because they were obtained at a significantly different epoch than the blue or black ones. In several cases, they clearly depart from the fit. This shows a clear trend of flux density increase with time. This has been found in other confirmed and candidate maser-emitting PNe \citep[e.g.][]{gom05,cal22}, and has been interpreted as tracking the growth of pristine ionizing regions at the beginning of the PN phase.

Therefore, the combined evidence provided by the spectral indices and variability of the radio continuum emission provides further support that the sources identified in this work are compatible with being nascent PNe.

\section{Details on individual sources}
\label{sec:individual}

\subsection{The OHPN IRAS 07027$-$7934: interferometric determination of maser positions in our observations}
\label{sec:i07027}

This source is the spectroscopically confirmed PN Vo\,1 \citep[][]{men90, sur02}. OH observations with the 26-m Hartebeeshoek \citep[beam=0.5$\degr$;][]{zij91} and the 64-m Parkes antennas \citep[beam=$13'$;][]{tel91} show a symmetric double peak at 1612 MHz, with flux densities of $S_{\nu}$\,$\simeq$\,10\,Jy\,beam$^{-1}$ and $V_{\rm LSR}$$\simeq$$-$40 and $-47$\,km\,s$^{-1}$, as well as a single-peaked spectrum at 1667 MHz ($S_{\nu}$\,$\simeq$\,2\,Jy\,beam$^{-1}$, $V_{\rm LSR}$$\simeq$$-$51\,km\,s$^{-1}$). \cite{sua09} did not detect any H$_{2}$O maser emission in this source in spectra taken with the Parkes antenna (rms=0.22\,Jy\,beam$^{-1}$). 
Despite the reported OH detections were carried out with single-dish antenna, \citet{zij91} indicated that the Parkes beam was sufficiently small ($13'$) that the chance of confusion with other sources is negligible, given the high galactic latitude of IRAS 07027$-$7934. { The LSR velocities of the reported OH maser components, ranging between $-$51 and $-$40 \,km\,s$^{-1}$, are blueshifted with respect to the systemic velocity of the source, traced by the CO peak \citep[with $V_{\rm LSR}\simeq -27$ and $-22$ km\,s$^{-1}$;][]{lou90, zij91}, and they are actually close to the blueshifted edge of the observed CO spectrum ($\simeq -40$\,km\,s$^{-1}$). While double-peaked OH spectra in AGB stars have mean velocities coinciding with the stellar one, the blueshifted bias of maser emission seen in this source is a commonly observed trend in maser-emitting PNe \citep[][]{usc12}. }

In our interferometric ATCA observations, we detected maser emission at 1612 MHz (Fig.\,\ref{fig:maser_spa}), but not at 1667 MHz (rms=13\,mJy\,beam$^{-1}$). The LSR velocities of the {observed} maser components (Fig.\,\ref{fig:maser_spe}; Table\,\ref{tab:masers_obs}) are similar to the 1612 MHz OH masers previously reported. However, the symmetric double peak spectrum{ of the 1612\,MHz line} has changed and is now clearly asymmetric, and has faded by a factor of $\simeq 20$. This fading correlates with the one of the maser at 1667 MHz, whose non-detection indicates a weakening of at least a factor of 50. The positional accuracy of our ATCA data does not allow us to determine the spatial distribution of the OH masers. 

Our map of radio continuum emission at 2.1\,GHz shows an unresolved source. Furthermore, the OH maser emission is located only at $0\farcs15\pm0\farcs04$ from the peak of the radio continuum, well within the beam of our ATCA observations (5$\farcs$3$\times$3$\farcs$3). The positional accuracy in this match between maser and continuum emission improves by four orders of magnitude the one provided by previous single dish observations, which was determined by the size of  the Parkes beam ($13'$). { This positional coincidence further reinforces the classification of this source as an OHPN, previously proposed by \citet{zij91} based on single-dish OH observations}. Moreover, the masers arise from regions relatively close to the central star, as previously seen in {other} OHPNe \citep[]{chr98, mir01, taf09, qia16a}, and not from the faint extended structure observed in the H$\alpha$ image, with an angular diameter of $\la 15\arcsec$ \citep{zij91}.  

Vo\,1 has been proposed to be a link between OH/IR stars and C-rich central stars \citep{zij91} since the observed OH masers could have been emitted only by an O-rich star at the AGB phase, and now Vo\,1 harbors a [WC11] central star \citep{men90}. Hence, this OH maser emission, the crystalline silicates and water ice \citep{coh02}, suggest that Vo\,1 is, in general, an O-rich nebula, while PAH emission arises from an elliptical shell surrounding the [WC11] central star over a region of $\sim$1$\farcs$0$\times$0$\farcs$8 \citep{gar06}.

\subsection{New candidate maser-emitting PNe in our observations}

\subsubsection{IRAS 17487$-$1922}

This source has been classified as a post-AGB star \citep{scz07, ram09}. Single-peaked OH maser emission at 1612 MHz was reported at $V_{\rm LSR}$=$-$37.7 km\,s$^{-1}$ based on 64-m Parkes observations \citep{tel91}. We have detected this OH maser emission in our interferometric VLA and ATCA observations, and resolved it into several individual spectral features. In particular, in our ATCA observations, which have a  better spectral resolution, we identified six emission features (Fig.\,\ref{fig:maser_spe}, Table \ref{tab:masers_obs1}). The positional accuracy of our ATCA/VLA data does not allow us to determine the spatial distribution of the OH masers. No H$_{2}$O masers were detected in previous observations with the 100-m Green Bank antenna \citep{gom15a}, nor in our own VLA observations. The radio continuum emission {is unresolved}, and is reported here for the first time, as well as its association with OH masers. Hence, IRAS\,17487$-$1922 is a new OHPN candidate.

\subsubsection{IRAS 18019$-$2216}
\label{sec:i18019}

This source has been classified as an OHPN candidate by means of radio interferometric observations \citep{cal22}. Emission of H$_{2}$O masers was not detected in observations made in 2010 \citep[rms=36\,mJy\,beam$^{-1}$;][]{walsh14}. However, in our observations, made in 2021, we detected 22.235\,GHz H$_{2}$O maser emission, where the brightest maser feature shows a $S_{\nu}$=332\,mJy\,beam$^{-1}$ (Fig.\,\ref{fig:maser_spe} and Table\,\ref{tab:masers_obs2}). Thus the H$_{2}$O masers seem to have emerged or presented significant variability between 2010 and 2021. Moreover, in IRAS\,18019$-$2216, the OH and H$_{2}$O maser-emitting material are kinematically different, since the velocity range of the H$_{2}$O maser emission ({from $-$10 to $+$8 km\,s$^{-1}$}) falls outside that of the OH (from $-$38 to $-$29 km\,s$^{-1}$), as seen in other WFs \citep[e.g.,][]{bob05, yung13, vle14}. Therefore, apart from being a H$_{2}$OPN and OHPN candidate, and taking into account this small spread of its H$_2$O maser emission, IRAS 18019$-$2216 is also a new `low-velocity' WF \citep[see][for this classification]{yung13}.

The spatial distribution of the H$_{2}$O masers in IRAS 18019$-$2216 (Fig.\,\ref{fig:maser_spa}, top left) shows that the blueshifted and redshifted maser components occupy two different regions, (to the NE and SW, respectively), as expected in a bipolar jet. Fig.\,\ref{fig:i18019_pv_6GHz} (left) {shows the radio continuum map of IRAS 18019$-$2216 at 5.7\,GHz obtained by GLOSTAR \citep{dzi23} with the B configuration of the VLA}. An elliptical Gaussian fit to the continuum emission indicates a deconvolved size of ($\simeq$3$\farcs$0$\pm$0$\farcs$53$)\,\times$\,(0$\farcs$90$\pm$0$\farcs$03), and {position angle} $\sim$18$\pm$3$\degr$. This orientation is similar to the direction of the bipolar jet traced by the H$_{2}$O masers{, as it can be seen in Fig.\,\ref{fig:i18019_pv_6GHz}, where we have assume that the peak emission of the radio continuum emission at 22.22 GHz from our VLA observations lies at the same position as that of this map at 5.7 GHz. This} radio continuum emission at 22.22 GHz only shows {an unresolved} source. 

\begin{figure} 
\begin{subfigure}{.5\textwidth}
  \centering
  \includegraphics[width=.90\linewidth]{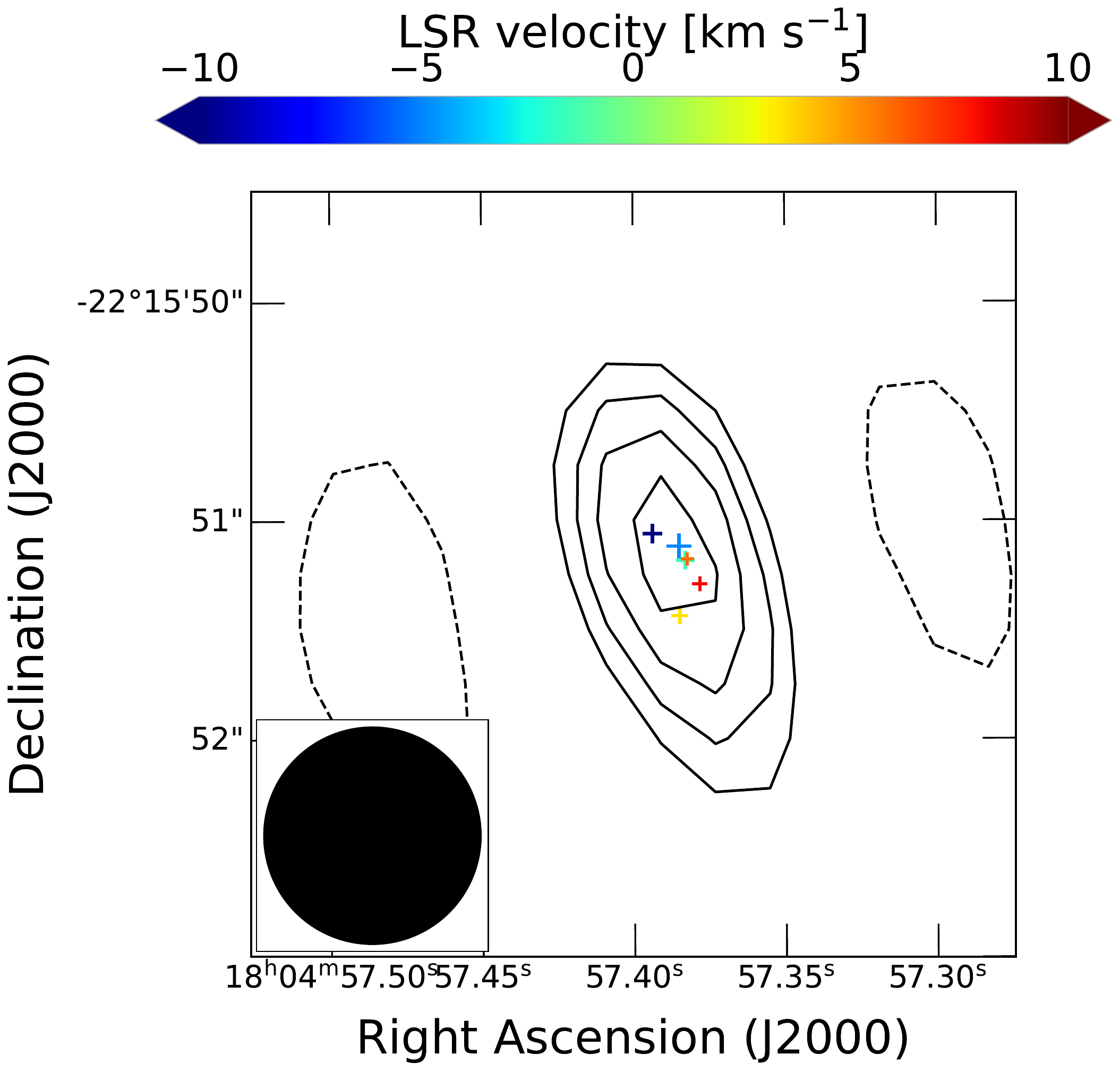}
\end{subfigure}
\caption{Radio continuum emission at 5.7 GHz of IRAS 18019$-$2216 {from GLOSTAR}. The contours are $-$3, 3, 4, 5, and 6\,$\times$\,1$\sigma$ (where $\sigma$\,=\,90 $\mu$Jy\,beam$^{-1}$ is the rms of the map). {The crosses are the positions of the H$_{2}$O masers, as shown in Fig.\,\ref{fig:maser_spa} (top left panel), assuming that the peak emission of the radio continuum emission at 22.22 GHz from our VLA observations lies at the same position as that of this map at 5.7 GHz. The color of the crosses represent the LSR velocity of the masers as indicated in the colorbar. The size of the crosses is proportional to the positional uncertainty of the masers.} The beam is shown in the bottom left corner.}
\label{fig:i18019_pv_6GHz}
\end{figure}

\subsubsection{IRAS 18443$-$0231}
\label{sec:i18443}

This optically obscured object has been classified as a PN by \citet{coop13}\footnote{Further characteristics of this object are available in \url{http://rms.leeds.ac.uk/}}  and \citet{kan15}, based on its near-infrared spectrum with prominent emission lines and weak continuum emission{, which can distinguish it from objects with similar infrared colors,} such as H\,{\sc ii} regions or Wolf-Rayet stars. Although these works do not give any specific information that may allow us to discern whether the source is a late post-AGB star or a young PN, its SED (Fig. \ref{fig:sed}), peaking at $\simeq 25$ $\mu$m, strongly support that it is an evolved star.

An interesting feature in the infrared spectrum is that the ionized emission shows the recombination line of He\,{\sc i} at 2.058\,$\mu$m stronger than the Br\,$\gamma$ emission \citep{coop13, kan15}. This is consistent with the photoionized gas of young PNe that host a central star with $T_{\rm eff}\leq 40000$\,K and present nebular electron densities larger than $\sim$10$^{4}$\,cm$^{-3}$ \citep[][]{dep94, lum01a, lum01b}. However, using the 3MdB, we have checked that this emission line ratio can also be reproduced by shock-induced ionization (Cala et al., in preparation){, and it is also seen in post-AGB stars showing shock-excited H$_{2}$ emission \citep[][]{gle15}. Hence, IRAS\,18443$-$0231 seems to be a post-AGB star for which more research is needed to confirm its possible nature as a PN.}

A spectrum of H$_{2}$O maser emission obtained with the 100\,m Green Bank Telescope \citep{urq11} shows a total velocity spread of maser components of 68 km\,s$^{-1}$, from $V_{\rm LSR}=-28$ to +40\,km\,s$^{-1}$, indicating that this source could be  a WF, as suggested by \citet{ort24}. However, given the presence of several H\,{\sc ii} regions in its neighborhood \citep{ort24}, interferometric observations {were} necessary to discern whether all these maser components actually arise from IRAS\,18443$-$0231. Actually, our VLA observations do interferometrically confirm the association between H$_{2}$O masers and the radio continuum. Moreover, our data also show H$_{2}$O maser components covering a wider velocity range than detected by \citet{urq11}, with components between $V_{\rm LSR}=-28.5$ and +48.6\,km\,s$^{-1}$ (Fig.\,\ref{fig:maser_spe} and Table\,\ref{tab:masers_obs4}). All these components, covering a total velocity spread of 77 km\,s$^{-1}$ are spatially associated with the source, finally confirming IRAS\,18443$-$0231 as a WF. We detected no OH maser emission in this source (rms $\simeq 75$\,mJy\,beam$^{-1}$). 

All H$_2$O maser components in this source are blueshifted with respect of the reported central velocity of the system, $V_{\rm LSR}=+110.3$ km\,s$^{-1}$ \citep{urq08}. Thus, the detected maser emission would be tracing outflow motions of $\simeq-60$ to $-138$ km\,s$^{-1}$ along the line of sight. The systemic LSR velocity was estimated from a single-dish spectrum of the $^{13}$CO(1-0) rotational line \citep{urq08}. We note, however, that those spectra show multiple spectral components, ranging from $V_{\rm LSR}=+50.4$ to +110.3 km\,s$^{-1}$. There is obvious interstellar confusion in this spectrum, which is normal for single-dish observations of molecular lines at the low galactic latitude of this source. The criterion used by \citet{urq08} to select the $^{13}$CO component associated with the source was to choose the brightest one in the spectrum. This criterion seems reasonable for high-mass star-forming regions, which were the main targets for these authors, within the RMS survey. However, this criterion is not necessarily applicable to evolved objects, such as WFs. As an illustration, the CO(1-0) single-dish observations of ten WFs carried out by \citet{riz13} show contamination from multiple interstellar components in most of the spectra, and in no case the most intense component corresponds to the velocity of the source. The emission associated with these WFs, when detectable, is extremely weak. \citet{and11} also reported radio recombination line emission at 9 cm toward IRAS 18443$-$0231, at $V_{\rm LSR} = +63.4$ and +98.5  km\,s$^{-1}$, from observations with the Green Bank Telescope (beam size $\simeq 82\arcsec$). These lines most likely arise from the neighboring H\,{\sc ii} regions. Thus, we can conclude that there is no solid estimate of the systemic velocity of the source so far, and we tentatively assume the intermediate LSR velocity between the two most extreme H$_2$O maser components ($\simeq +10$ km\,s$^{-1}$) in the graphical representation of the  maser velocities. If the systemic velocity of the source is confirmed to be close to this assumed one, then the weak CO emission around $+120$ km\,s$^{-1}$ that \citet{ort24} interpreted as tracing the redshifted lobe of molecular outflow from  IRAS 18443$-$0231 is unlikely to be associated with the source.

Fig.\,\ref{fig:maser_spa} (bottom left) shows the spatial distribution of all the H$_{2}$O masers we identified, with respect to the radio continuum emission peak at 22.22 GHz, from our VLA observations. The H$_{2}$O masers trace a jet-like structure with an S-shaped morphology with a general north-south orientation on a larger scale. The maser components farther away from the center are the weakest ones, and they do not show a clear velocity distribution as expected in a jet (a trend of red- and blueshifted components separated on either side of the central source). However, the strongest innermost components (Fig.\,\ref{fig:maser_spa}, bottom right) are arranged in a linear structure in the northeast (NE) to southwest (SW) direction, with a rough trend of more blueshifted components to the SW. The brightest maser components are blueshifted, and are located closer to {this radio continuum emission peak}. They seem to form a relatively dense cluster with a roughly circular or arc shape, suggesting the existence of bow-shocks \citep[as seen in, e.g., IRAS\,18113$-$2503;][]{oro19}, or the origin of a molecular jet as it emerges from the surrounding circumstellar material \citep[as seen in, e.g., W43A;][]{taf20}.

Another outstanding characteristic of IRAS 18443$-$0231 is its spectral index, as mentioned in Section\,\ref{sec:continuum}. Clearly negative spectral indices, characteristic of non-thermal radio emission, have been reported before in this source, with observations that indicate values $\alpha \simeq-1.15$ to $\alpha \simeq-0.36$ \citep[e.g.,][]{xu06, dzi23}. Some of these estimates should be taken with care, as the available information at radio frequencies of the source shows variability of a factor $>2$ \citep{bec94, xu06, pur13, med19, dzi23, ort24}. In this sense, the result of \cite{xu06} is more robust, as it was obtained with observations taken close in time.

Negative spectral indices, such as the one seen in IRAS\,18443$-$0231, are characteristic of jets in which free electrons move in a magnetized environment. These jets are traced by the WF, and the derived spectral index $\alpha$$\simeq$$-$0.42 is similar to the spectral index of synchrotron jets of active galactic nuclei. Moreover, it has also been suggested that the powerful jets in this object can produce X- and $\gamma$-ray emission \citep{ort24}, although the association of these high-energy emission with the source is still uncertain. Negative spectral index values have been reported in some WFs, such as the post-AGBs IRAS\,15445$-$5449 and IRAS\,18043$-$2116 \citep{ps13, ps17}, and the PN IRAS\,15103$-$5754 \citep{sua15}, as well as other post-AGB stars and PNe \citep{cer11, cer17}. 

This group of non-thermal radio continuum emitters may be ideal sources for investigating the role of magnetic shaping of circumstellar structures in these evolutionary phases \citep[e.g.][]{gar20}.   

Figure\,\ref{fig:i18443_24GHz} shows the radio continuum emission of IRAS\,18443$-$0231 at 24 GHz. It shows a compact, bright central component, plus some extended structure, with low-level extensions to the southwest, which are also observed in images at 3 and 5 GHz from VLASS and CORNISH, respectively \citep[see Fig.\,10 of][]{ort24}. Archival VLA continuum map at 5.7 GHz shows an unresolved source at a beam size of 1$\farcs$2$\times$0$\farcs$9. Similarly, \cite{xu06} reported an unresolved source (beam size  0$\farcs$9) at 15 GHz. We also show in Fig.\,\ref{fig:i18443_24GHz} the location of the H$_{2}$O masers, as presented in Fig.\,\ref{fig:maser_spa} (bottom panels). The central linear structure of masers is located near the bright central non-thermal component of the ionized gas, although the weakest masers extend over a larger region and could be related with mass-loss along the direction of the extended continuum emission.

\begin{figure}
    \centering
    \includegraphics[width=1\linewidth]{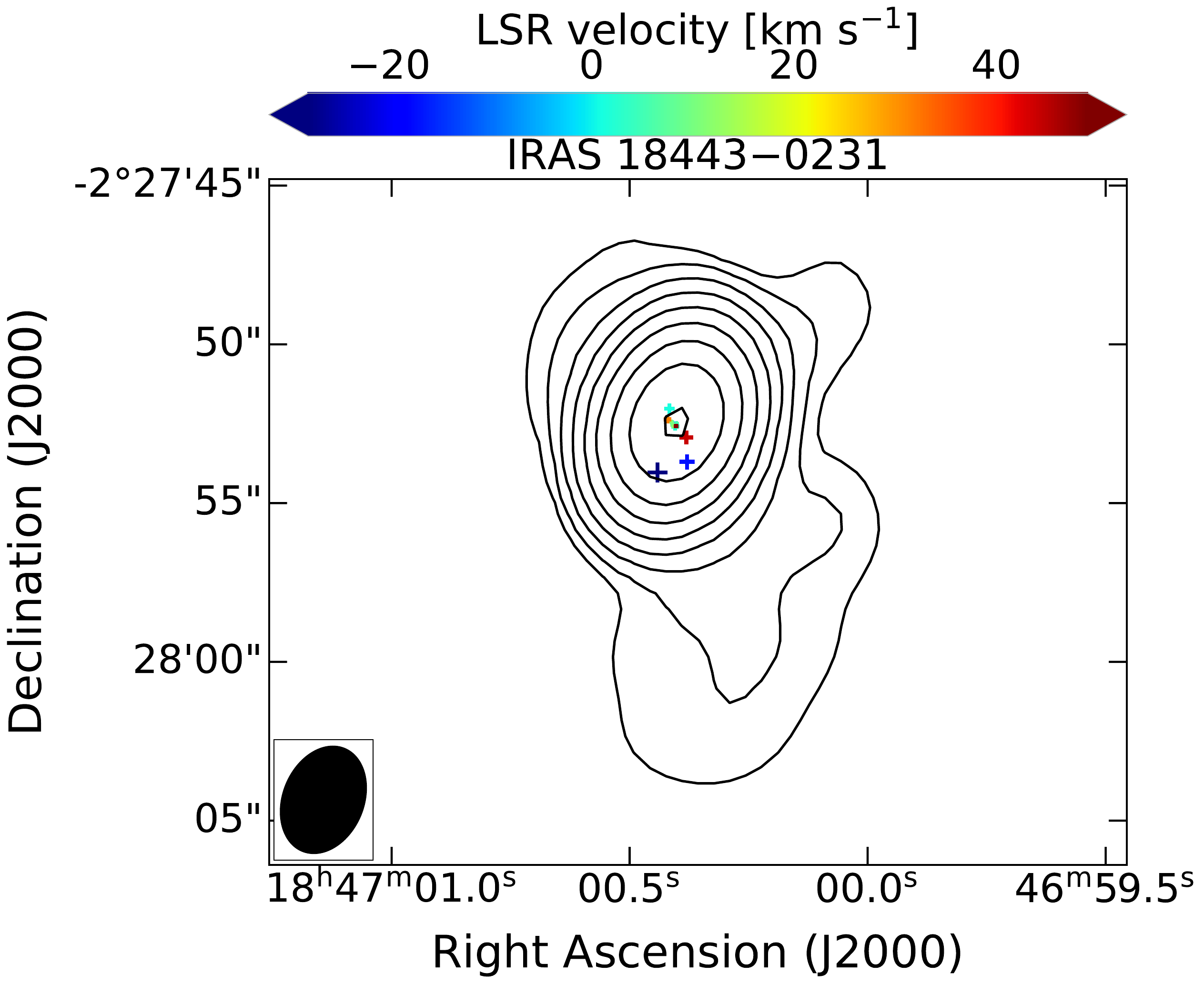}
    \caption{Contour map of the radio continuum emission of IRAS\,18443$-$0231 at 24 GHz. The contours increment in steps of 3$\sigma \times$2$^{n}$ starting from n=0 ($\sigma$= 0.23 mJy\,beam$^{-1}$ is the rms of the map). The crosses are the positions of the H$_{2}$O masers, as shown in Fig.\,\ref{fig:maser_spa} (bottom panels). The color of the crosses represent the LSR velocity of the masers as indicated in the colorbar. The size of the crosses is proportional to the positional uncertainty of the masers. The beam is shown in the bottom left corner.}
      \label{fig:i18443_24GHz}
\end{figure}

\subsubsection{IRAS 18480$+$0008}
\label{sec:i18480}

The VLA THOR survey reported a single peak 1665 MHz maser feature at $V_{\rm LSR}$=$+$12.0 km\,s$^{-1}$ from this source \citep{beu19}. VLA surveys of radio continuum have also reported this source with emission at 3 and 5 GHz \citep[][]{med19, gor21, dzi23}. From these observations we report here for the first time this source as a OHPN candidate. Furthermore, our VLA observations show a match between H$_{2}$O masers and radio continuum emission at 22.22 GHz. The latter is unresolved in our data (beam $\simeq 3\arcsec$). Moreover, archival continuum data performed with the VLA in the B configuration at 5.7 GHz (beam $\simeq$1$\arcsec$) also shows unresolved emission. The H$_{2}$O maser spectrum displays multiple spectral features emitted between $V_{\rm LSR}$=$+$14\,km\,s$^{-1}$ and $V_{\rm LSR}$=$+$28\,km\,s$^{-1}$ (Fig.\,\ref{fig:maser_spe} and Table\,\ref{tab:masers_obs3}). The positions of the H$_{2}$O masers (Fig.\,\ref{fig:maser_spa}, top right) trace a linear distribution from the northeast to the southwest direction, which is expected from either a collimated mass-loss event or a toroidal structure. Hence, IRAS 18480$+$0008 is a new H$_{2}$OPN and OHPN candidate, and could be a new WF. A confirmation of its nature as a WF would require further studies to determine the orientation of the nebula with respect to the water masers, or proper motion studies of the latter, to determine whether this maser emission actually traces a jet.

\subsubsection{IRAS 19035+0801}

This source has been classified as a candidate for PN based on its infrared colors and radio continuum emission \citep{van95, ira18}. Based on single-dish observations, \cite{eder88} reported single-peaked 1612 MHz OH maser emission at $V_{\rm LSR}$=$-$80.2 km\,s$^{-1}$. We have also detected this maser emission at around $-$78 km\,s$^{-1}$ in both ATCA and VLA observations and confirm that it has a positive positional coincidence with the source of radio continuum previously reported (see Fig.\,\ref{fig:maser_spe} and Table\,\ref{tab:match}). 
The positional accuracy of our data is not enough to determine the spatial distribution of the OH masers. We have not detected any H$_{2}$O maser emission. Hence, IRAS 19035+0801 is a new candidate for OHPN.

\subsubsection{IRAS 19194$+$1548}
\label{sec:i19194}

This source was spectroscopically classified as a likely PN G50.4802+0.7058 \citep{sab14}. OH maser emission at 6035 MHz has also been reported in this source, using ATCA interferometric observations \citep{avi16}. The 6035 MHz spectrum shows an asymmetric double peak at $V_{\rm LSR}$ of $+$48.12 and $+$48.22\,km\,s$^{-1}$ (see Table\,\ref{tab:match}). Data from the THOR survey report the presence of both OH maser emission at 1612 MHz \citep{beu19} and 1.5 GHz continuum emission \citep{wan18} at the source position. These maser and continuum detections were reported separately, {although they were} obtained in simultaneous observations. The 1612 MHz OH maser emission shows only one emission feature at $V_{\rm LSR}$=$+$43.5\,km\,s$^{-1}$. The match between OH maser and radio continuum emission, as well as their association with this likely PN is reported here for the first time. In our VLA observations, we did not detect any 22.235 GHz H$_{2}$O maser emission. 

In Figure\,\ref{fig:i19194_ukirt} we present an archival infrared image of IRAS\,19194+1548 at 2.1\,$\mu$m taken from the archive of the United Kingdom Infrared Telescope (UKIRT), which shows a point-symmetric bipolar nebula with a narrow waist. The position of the radio continuum emission \citep{wan18} is near the center of the nebula, while the OH emission may be associated with the eastern lobe (relative positional accuracy of 0$\farcs$5). A Gaussian fit to our radio continuum map at 24 GHz indicates that the emission peak [RA(J2000)=19h21m40.421, DEC(J2000) +15d53m54.83s] lies at a position closer to the waist than to the lobes (see black point in Fig.\,\ref{fig:i19194_ukirt}).

\begin{figure}
      \centering
            \includegraphics[width=0.95\hsize]{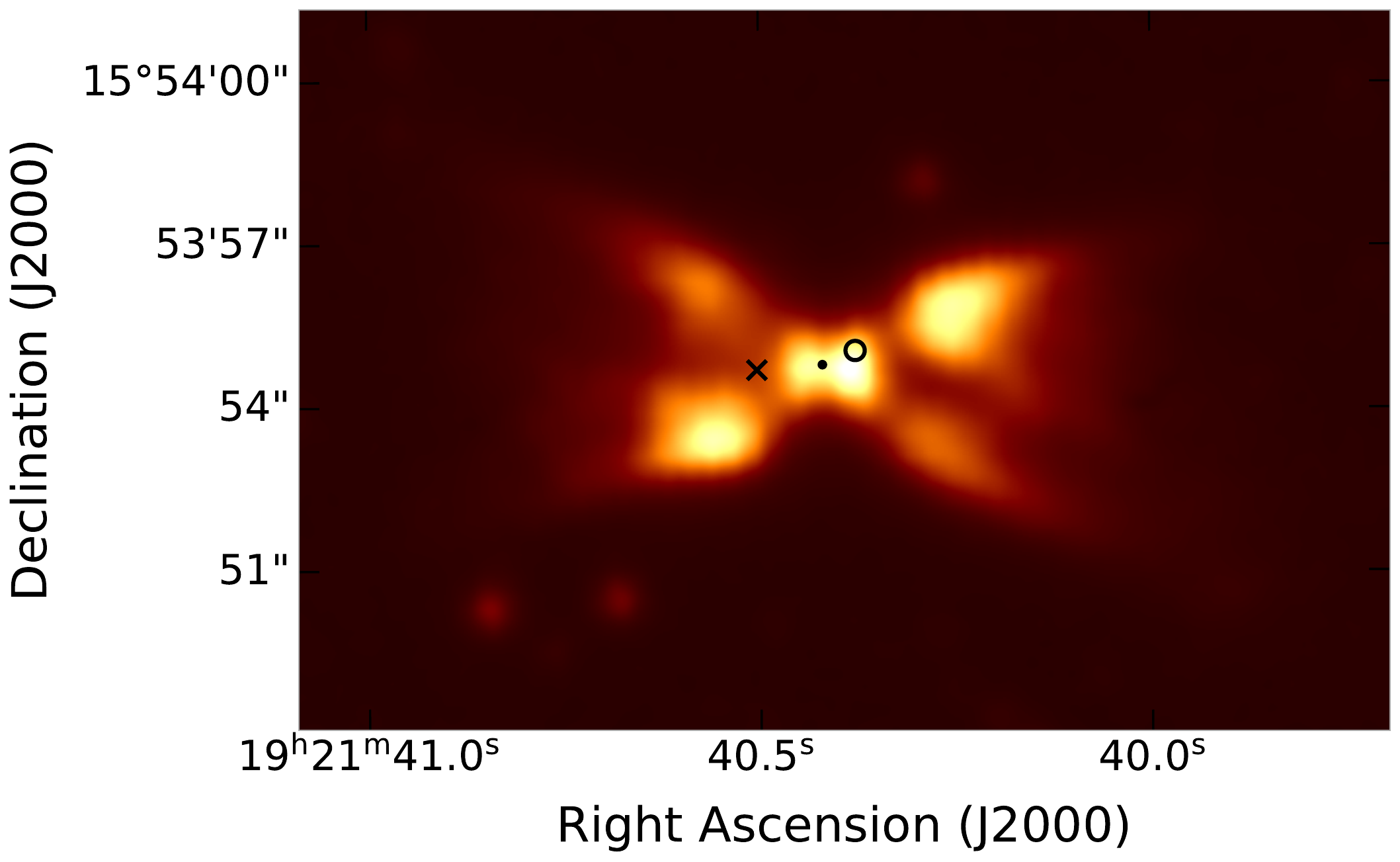}     
          \caption{Near-infrared broadband emission at 2.1\,$\mu$m of IRAS 19194$+$1548 by the UKIRT in logarithmic scale. The black empty circle and `x' lie at the reported positions of the continuum emission peak at 1.5 GHz and 1612 MHz OH maser, respectively. The sizes are proportional to their relative positional accuracy. The black point lies at the position of the continuum emission peak at 24 GHz from our VLA observations. Its size is proportional to its positional uncertainty.}
      \label{fig:i19194_ukirt}
\end{figure}

It should be noted that \citet{akr19} classified this object as a possible SySt, based on its SED, whose peak they reported to be at $\simeq 4$ $\mu$m. However, as shown in Fig.\, \ref{fig:sed}, our updated SED, including additional data, shows a peak between 30 and 70\,$\mu$m, which is compatible with an obscured PN, rather than a SySt, as discussed in Sec.\,\ref{sec:sed}. In addition, an inspection of the spectrum shown by \citet{sab14}, suggests the presence of an emission line at 6830\,$\AA$ that may be associated with the O\,{\sc vi} Ramman-scattered line. However, this emission feature line is observed in both SySt and young PNe or may be a blending of several emission lines \citep[see][and references therein]{akr19}. On the other hand, the near-infrared spectrum obtained from the RMS survey\footnote{The spectra and further characteristics of this object are available at \url{http://rms.leeds.ac.uk/}} shows CO(3--1) emission at 2.31\,$\mu$m \citep{coop13}, which is observed in stars that have already entered the post-AGB phase \citep[e.g.][]{hri94, oud95, gle11, gle12}. This emission is not expected in a SySt, where the main component is a red giant which only show CO absorption lines \citep{ray09}. {We also point out that, as mentioned before, this source has OH emission both at 1612 and 6032 MHz. The former is extremely rare in SySts \citep[e.g.,][]{nor84, sea95}, since the companion white dwarf ionizes the outer envelopes where these masers are produced, dissociating the OH molecules. Moreover, the 6035 MHz OH line has only been detected in young PNe \citep[in Vy\,2-2 and K\,3-35;][]{jew85, des02, des10}, but not in SySts.} 

{In summary, there are some pieces of indirect evidence suggesting that this object could be a PN. However, here we report IRAS 19194+1548 as a new OHPN candidate \citep[see also][]{mar24}, for which more research is needed to definitely confirm its nature as a PN.}

\subsection{New candidate maser-emitting PNe from the literature search}
\label{sec:new_IP_lit}

\subsubsection{IRAS 16029$-$5055}

This source, also called OH 331.158+00.781, has been classified as a post-AGB star due to its infrared colors and single peak OH spectrum at 1612 MHz, with $V_{\rm LSR} = -89.1$ km\,s$^{-1}$ \citep{sev97}. In the CORNISH South Survey, a weak radio continuum source has been reported, with a flux density of 3.20$\pm$0.39 mJy at 5.5 GHz \citep{ira23}. No other radio continuum nor maser line emission has been reported on this source. We reported for the first time the match between the interferometric positions of this single-peak OH maser and the radio continuum emission. Moreover, the analysis presented in Sec.\,\ref{sec:spitzer} of the archival \textit{Spitzer} spectrum shown in Fig.\,\ref{fig:spitzer} (top panel) allows us to spectroscopically confirm the presence of photoionized gas. Hence, we report IRAS\,16029$-$5055 as a new OHPN candidate, for which more observations are necessary before a final confirmation as bona fide OHPN (see Section\,\ref{sec:nascent_PNe}).

\subsubsection{IRAS 18213$-$1245A}

Based on its infrared and radio continuum emission, this source has been classified as the PN candidate G018.5242+00.1519 \citep{urq09,ira18}. Based on ATCA observations, \cite{walsh14} reported two H$_{2}$O maser features at $V_{\rm LSR}$ of $\simeq$+21.7\,km\,s$^{-1}$ and $\simeq$+23.6\,km\,s$^{-1}$ on this source. In addition, a wide 1665 MHz OH single peaked feature at $V_{\rm LSR}$=27\,km\,s$^{-1}$ covering a range of velocities of $\sim$15\,km\,s$^{-1}$ was reported by \cite{beu19} in THOR, based on VLA observations. Although these maser and continuum data were not obtained simultaneously, their spatial locations are sufficiently close (distance $\simeq 1''$), and therefore, here we report the match between the interferometric positions of the PN candidate and these masers. Hence, IRAS 18213$-$1245A is a new H$_{2}$OPN and OHPN candidate. 

\subsubsection{IRAS 18464$-$0140}
\label{sec:i18464}

Based on its infrared and radio continuum emission, this source has been classified as the PN candidate G031.2131$-$00.1803 \citep{ira18,med19}. This classification as PN is supported by the non-detection of thermal ammonia emission \citep{ang96}, which is {more commonly} detected in star-forming regions {than in evolved stars \citep[see][]{ang96, won18}}. VLA observations in 1984 detected H$_{2}$O maser emission spread between $V_{\rm LSR}$ of $-$6.3 and $-$39.2\,km\,s$^{-1}$ and a peak emission of 20.2\,Jy\,beam$^{-1}$ at $V_{\rm LSR}$=$-$37.1\,km\,s$^{-1}$ \citep{fc89, fc99}. In addition, single emission peaks of OH maser emission at 1612, 1665, and 1667 MHz have been detected at $V_{\rm LSR}$ of $-$26.6 to $-$30.2 km\,s$^{-1}$ based on VLA observations \citep{fc89, fc99, beu19}. {The 1665 MHz OH emission presents significant variability, with flux densities of $\simeq 0.50$ and 0.35 Jy (Table \ref{tab:masers_obs7}) reported by reported by \cite{fc89}, but it was not detected in THOR} \citep[rms=0.01\,Jy\,beam$^{-1}$;][]{beu19}. {Based on the spatial coincidence of maser and continuum emission, we report }IRAS\,18464$-$0140 as a new H$_{2}$OPN and OHPN candidate.

Fig.\,\ref{fig:i18464_10GHz_h2o} shows a radio continuum map at 10\,GHz of IRAS 18464$-$0140, which shows an elongated source with a {deconvolved size of (344$\pm$14)\,$\times$\,(53$\pm$36)\,mas$^2$ elongated at }position angle{ 56$\pm$2$\degr$}. We also show the location of the H$_{2}$O maser components from observations performed on 1984, and initially reported by \cite{fc99}, which we reprocessed completely, including calibration and self-calibration. The increased signal-to-noise ratio, of a factor $\sim$2 in the maser emission, due to the self-calibration process, allowed us to detect two new components (at $-28.68$ and $-6.29$ km\,s$^{-1}$) not reported by \cite{fc99}. The parameters of the detected maser components are shown in Table\,\ref{tab:masers_i18464}. The blueshifted and redshifted maser clusters are separated $\sim$0$\farcs$7, and distributed along the same direction as the elongated continuum emission. This suggests that the masers are tracing a jet, which is characteristic of WFs \citep[e.g.][]{sua08, gom17}. Hence, in addition to being a new H$_{2}$OPN and OHPN candidate, and considering this small spread of velocities of its H$_2$O maser emission, IRAS 18464$-$0140 is also a new `low-velocity' WF \citep[for this classification see][]{yung13}. 
We note that, while the association of maser and continuum emission, and their similar orientations are clear, it is not possible to exactly determine an accurate positional alignment between these two types of emission, since those data were not obtained simultaneously. It is likely that the masers actually lie at the tips of the bipolar structure traced by the continuum emission. Thus, new simultaneous H$_{2}$O maser and continuum observations are needed to accurately determine the location of the maser emission with respect to the nebula.

\begin{figure}
      \centering
      \includegraphics[width=0.93\hsize]{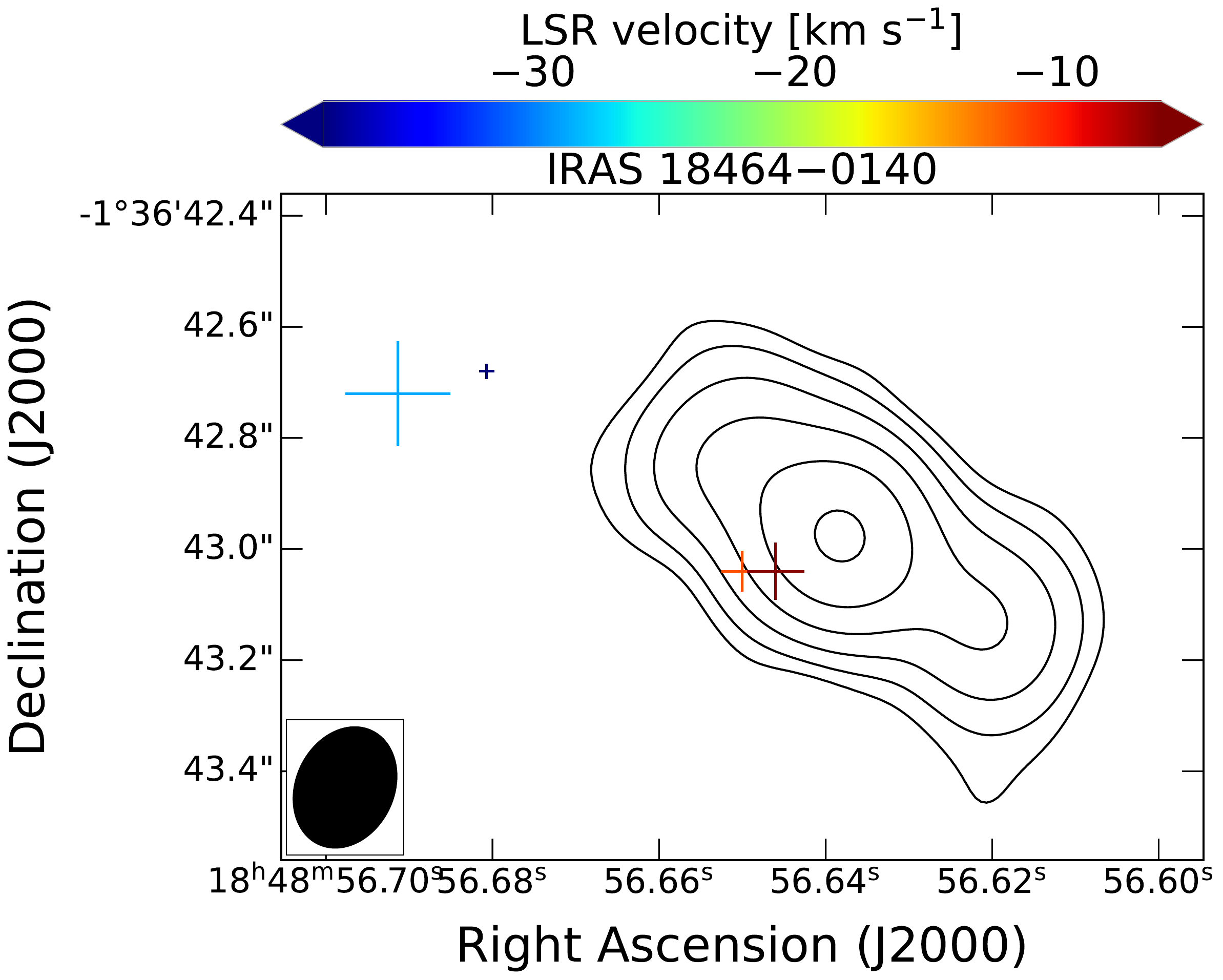}      
      \caption{Contours of the radio continuum emission at 10 GHz of IRAS 18464$-$0140 from archival VLA data from observations performed on 2022. The contours increment in steps of 3$\sigma \times$2$^{n}$ starting from n=0 ($\sigma$\,=\,31\,$\mu$Jy\,beam$^{-1}$). The crosses are the positions of the H$_{2}$O masers obtained from our reprocessing of archival VLA data from observations performed on 1984. The color of the crosses represent the LSR velocity of the masers as indicated in the colorbar. The size of the crosses is proportional to the positional uncertainty of the masers.}
      \label{fig:i18464_10GHz_h2o} 
\end{figure}

\begin{figure}
      \centering
      \includegraphics[width=0.77\hsize]{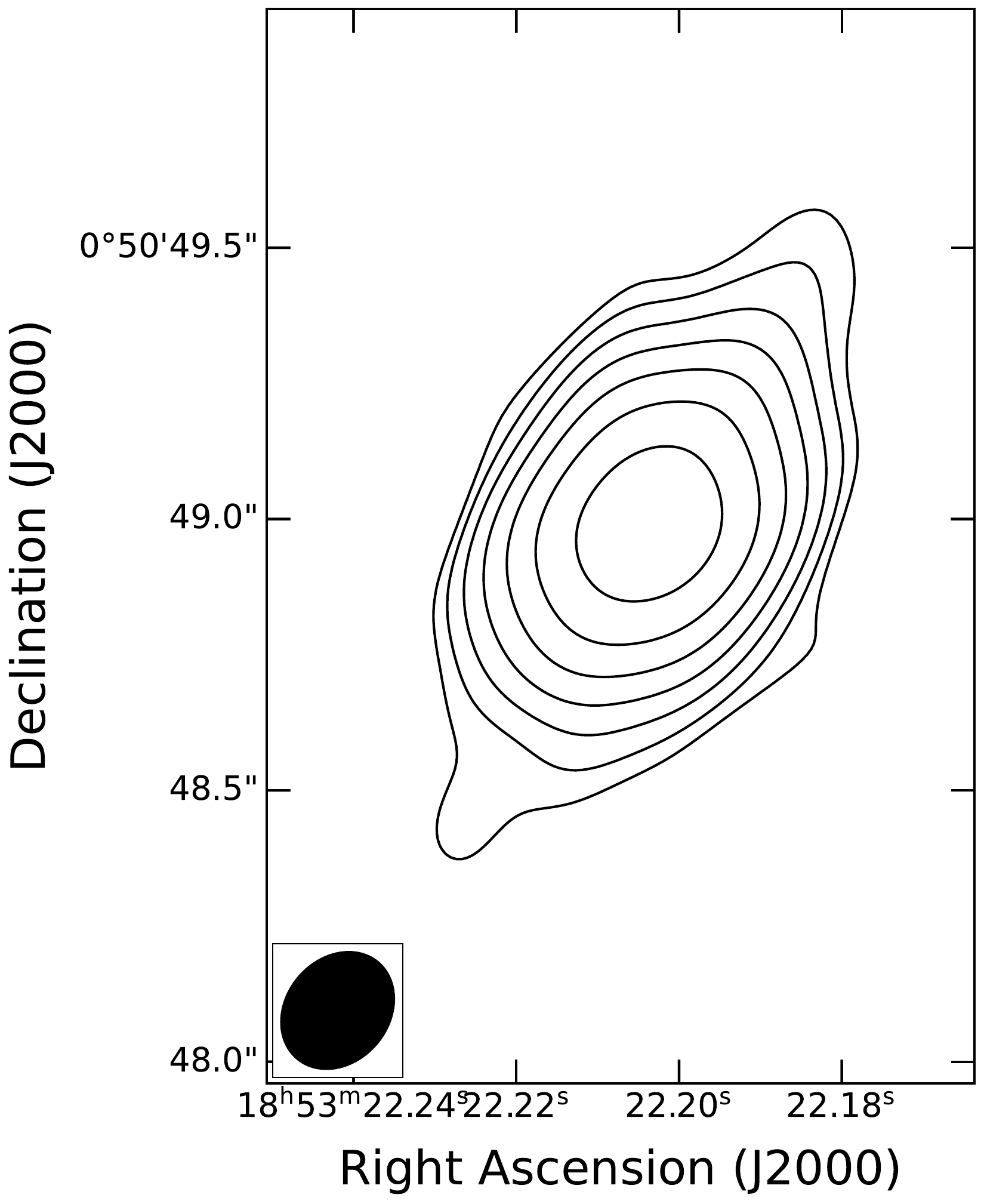}      
      \caption{Contours of the radio continuum emission at 10 GHz of IRAS 18508+0047. The contours increment in steps of 3$\sigma \times$2$^{n}$ starting from n=0 (where $\sigma$\,=\,44 $\mu$Jy\,beam$^{-1}$ is the rms of the map). The filled ellipse is the synthesized beam of the radio continuum emission map.}
      \label{fig:i18508} 
\end{figure}

\subsubsection{IRAS 18508$+$0047}

Based on its infrared and radio continuum emission, this source has been classified as the PN candidate G033.9059$-$00.0436 \citep{urq09,ira18}. A single 1612 MHz OH maser feature at +76.5 km\,s$^{-1}$ was reported at galactic position 33.906$-$0.044 by the THOR survey \citep{beu19}. In this work, we report the match between the interferometric positions of the PN candidate and the OH maser emission (Table\,\ref{tab:match}). Hence, IRAS 18508+0047 is a new OHPN candidate. Fig.\,\ref{fig:i18508} shows archival VLA continuum data from this source at 10\,GHz. It originates from a strong compact source and weak extended emission in the {northwest-southeast} direction. A Gaussian fit indicates a deconvolved size of 284$\pm$3$\times$200$\pm$2 mas$^{2}$, pa 147.9$\pm$1.2$\degr$. 

\subsubsection{IRAS 19176$+$1251}
\label{sec:i19176}

This source has been classified as the PN candidate G047.6884$-$00.3024, based on its infrared and radio continuum emission \citep{prei88, van95, scz07, ram12, ira18}. Observations of H$_{2}$O masers have not detected emission on the source \citep[rms=0.08\,Jy\,beam$^{-1}$;][]{gom15a}. Simultaneous observations of continuum and line emission detected both the 1.5 GHz radio continuum and the single peak 1720 MHz OH maser emission at $-$39.0\,km\,s$^{-1}$ (see Table\,\ref{tab:match}). These findings have been independently reported in the THOR survey for continuum and OH line emissions at the position of source G47.688$-$0.302 \citep{wan18,beu19}. The analysis presented in Sec.\,\,\ref{sec:spitzer} of the archival \textit{Spitzer} spectrum shown in Fig.\,\ref{fig:spitzer} (bottom right panel) is consistent with the presence of photoionized gas. Hence, we report IRAS 19176+1251 as a new OHPN candidate, for which more observational data are necessary before a final confirmation as bona fide OHPN (see Section\,\ref{sec:nascent_PNe}).

To our knowledge, IRAS 19176+1251 is the first evolved low- to intermediate-mass star known to have detected OH maser emission at 1720 MHz, and without any association with other OH and H$_{2}$O maser, providing an excellent opportunity for investigation of the physical conditions for this OH maser transition in evolved stars. Although OH emission at 1720 MHz has been reported in other PNe \citep[][]{mir01, qia16a, gom16}, it was always found together with other OH maser transitions. The exclusive presence of OH masers at 1720 MHz is typical of shock waves of supernova remnants encountering interstellar molecular clouds \citep[see][]{war02}{, some star forming regions \citep[][]{ogb20}, and in FU Ori-type outbursts \citep[see][and references therein]{sza25}.}

\subsubsection{OH 025.646$-$00.003}
\label{sec:oh25}

A symmetric double-peaked maser spectrum at 1612 MHz has been reported from OH 025.646$-$00.003 in the THOR survey \citep{beu19}. Radio continuum emission at 4.8 GHz has also been reported at this position \citep{whi05}. Here, the match between radio continuum and OH masers is reported for the first time. We did not detect H$_{2}$O maser nor radio continuum emission at 24 GHz from the source. Hence OH 025.646$-$00.003 is a new OHPN candidate.

\section{Discussion}

\label{sec:nascent_PNe}

As a summary after our results in this paper, in Table\,\ref{tab:PNe_masers} we report the total number of maser-emitting PNe and candidates identified in this work, together with those previously reported in other works. We have increased the number of confirmed maser-emitting PNe known to nine, while the number of candidates has increased {to 16}. In addition, we have increased the number of WFs {to 18} (Table\,\ref{tab:water_fountains}), including the first two `low-velocity' WFs radio continuum emitters. 

\begin{table*}[]
\setlength{\tabcolsep}{3.8pt}
      \caption[]{List of all known maser-emitting PNe and candidates.}
      \label{tab:PNe_masers}
      \centering                          
\begin{tabular}{lcccccccccc}     
\hline\hline                 
& & & \multicolumn{3}{c}{Maser emission$^{a}$} & & \multicolumn{4}{c}{References$^{b}$} \\
Name & Alt. name & Galactic Coordinates & SiO & H$_{2}$O & OH & Status$^{c}$ & SiO & H$_{2}$O & OH & Status \\
\hline 
IRAS 07027$-$7934 & Vo 1 & 291.375$-$26.294 & -- & No & Yes & IP & -- & 1 & TW & 2, 3 \\
IRAS 15103$-$5754 & GLMP 405 & 320.907$-$00.293 & No & Yes & -- & IP & 4 & 5 & -- & 5, 6 \\
IRAS 16029$-$5055 & {SCHD 68} & 331.158+00.781 & -- & -- & Yes & IC & --& -- & 7 & TW \\
IRAS 16333$-$4807 &  & 336.644$-$00.696 & No & Yes & Yes & IP & 4 & 8 & 9 & 8, 9 \\
IRAS 16372$-$4808 & {SCHD 117} & 337.064$-$01.173 & No & No & Yes & IC & 4 & 10 & 11 & 12 \\
IRAS 17150$-$3754 & {PBZ 1} & 349.351$-$00.211 & -- & No & Yes & IC & -- & -- & 13 & 13 \\
IRAS 17347$-$3139 & {RPZM 28} & 356.801$-$00.055 & No & Yes & Yes & IP & 4 & 14 & 15 & 14, 16, 15, TW \\
IRAS 17371$-$2747 & JaSt 23 & 000.340+01.574 & No & -- & Yes & IP & 4 & -- & 17 & 18, 19 \\ 
IRAS 17375$-$2759 & {PBOZ 26} & 000.208+01.412 & -- & -- & Yes & IC & -- & -- & {20} & 19, TW \\ 
IRAS 17393$-$2727 & {OH 0.9+1.3} & 000.892+01.341 & No & -- & Yes & IP & 4 & -- & 13 & 13, 21, TW \\ 
IRAS 17487$-$1922 & {OH 008.9+03.7} & 008.905+03.712 & -- & No & Yes & IC & -- & TW & TW & TW \\ 
IRAS 17494$-$2645 & {SCHB 220} & 002.640$-$00.191 & -- & No & Yes & IC & -- & TW & 22 & 12, TW \\ 
IRAS 18019$-$2216 & {SCHB 281} & 007.958$-$00.393 & -- & Yes & Yes & IC & -- & TW & 23 & 12, TW \\
IRAS 18061$-$2505 & {MaC 1-10} & 005.974$-$02.611 & No & Yes & No & IP & 4 & 24 & 24 & 24, 25 \\
IRAS 18213$-$1245A &  & 018.526+00.150 & -- & Yes & Yes & IC & -- & 26 & 27 & TW \\
IRAS 18443$-$0231 & {RFS 505} & 030.234$-$00.138 & -- & Yes & No & IC & -- & TW & TW & TW \\
IRAS 18464$-$0140 & OH\,31.21$-$0.18 & 031.213$-$00.180 & -- & Yes & Yes & IC & -- & 28 & 28 & TW \\
IRAS 18480+0008 &  & 033.023+00.279 & -- & Yes & Yes & IC & -- & TW & 27 & TW \\
IRAS 18508+0047 & {GPSR5 33.906-0.044} & 033.906$-$00.043 & -- & No & Yes & IC & -- & TW & 27 & TW  \\
IRAS 19035+0801 & {PM 1-279} & 041.787+00.488 & -- & No & Yes & IC & -- & TW & TW & TW  \\
IRAS 19176+1251 & {GLMP 898} & 047.688$-$00.302 & -- & No & Yes & IC & -- & 29 & 27 & TW  \\
IRAS 19194+1548 &   & 050.480+00.705 & -- & No & Yes & {IC} & -- & TW & 27 & 30, TW  \\
IRAS 19219+0947 & Vy 2-2 & 045.498$-$02.703 & No & No & Yes & IP & 4 & 31 & 32 & 33, 32  \\
IRAS 19255+2123 & K 3-35 & 056.096+02.094 & No & Yes & Yes & IP & 4 & 34 & 34 & 35, 34, TW \\
OH 25.646$-$0.003  &  & 025.646$-$00.003 & -- & No & Yes & IC & -- & 27 & TW & TW  \\
OH 341.681+00.264 &  & 341.681+00.263 & No & -- & Yes & IC & 4 & -- & 11 & 12 \\
\hline
\end{tabular}

Notes. $^{a}$ `--': no interferometric observations available. $^{b}$ {Reference for the interferometric maser emission detected. In the status column, we provide references that have contributed to classifying the source as a PN candidate or confirmed PN.} TW: This work. 1. \citet{sua09}. 2. \citet{men90}. 3. \citet{sur02}. 4. \citet{cal24}. 5. \citet{gom15b}. 6. \citet{sua15}. 7. \citet{sev97}. 8. \citet{usc14}. 9. \citet{qia16a}. 10. \citet{dea07}. 11. \citet{qia16b}. 12. \citet{cal22}. 13. \citet{pot87}. 14. \citet{deG04}. 15. \citet{taf09}. 16. \citet{gom05}. 17. \citet{sev02}. 18. \citet{jast04}. 19. \cite{usc12}. {20. \citet{zij89}}. 21. \citet{gar07}. 22. \citet{qia18}. 23. \citet{qia20}. 24. \cite{gom08}. 25. \citet{mir21}. 26. \citet{walsh14}. 27. \citet{beu19}. 28. \citet{fc99}. 29. \citet{gom15a}. 30. \citet{sab14}. 31. \citet{gom90}. 32. \cite{sea83}. 33. \citet{vy45}. 34. \cite{mir01}. 35. \cite{mir00}. $^{c}$ {As defined in Table\,\ref{tab:sample}. IC: interferometrically confirmed maser-emitting PN candidate. IP: interferometrically confirmed maser-emitting PN.}

\end{table*}

This increase in candidate maser-emitting PNe is an important step forward, although eventually confirming their nature as bona-fide PNe would be crucial to better understand the nature and properties of this category of sources.

To confirm the identified PN candidates as bona fide maser-emitting PNe, it is first necessary to ascertain the presence of photoionized gas. Considering that these candidates are, in general, optically obscured objects, we suggest that this confirmation can be achieved by means of near- and mid-infrared spectroscopic observations, complemented with a comparison of the observed emission line ratios with diagnostic diagrams that represent photoionization and shock ionization models. A first attempt has been made for some sources in Section\,\ref{sec:spitzer}. Using several diagnostic diagrams involving different line rations would strongly support whether or not photoionized gas is present in our objects. Furthermore, high-resolution imaging of the emission lines that are used to create diagnostic diagrams, such as those of [Ne\,{\rm II}] and [Ne\,{\rm III}] shown in Fig.\,\ref{fig:3mdb_dd3}, should be obtained in order to create spatially resolved maps of emission line ratios across the ionized circumstellar nebula, pinpointing the distribution of the photoionized gas from the PN. In addition, high-resolution imaging of radio and millimeter/submillimeter recombination lines can also reveal whether their kinematics is more consistent with an ionized wind, or with the expansion of a photoionized nebula.

Even if we could confirm as PNe a significant number of the identified candidates, their numbers may still be scarce. to draw firm conclusions {about} important issues such as their evolutionary status, or whether they all evolve from stars with similar characteristics (e.g., initial masses, metallicity). It is then important to continue working on increasing the number of known members of this type of sources, for instance, confirming the nature of the identified candidates as PNe (by means of optical and infrared spectroscopy) or undergoing new sensitive mappings of masers in our Galaxy, such as the future Galactic Australian Square Kilometre Array Pathfinder (GASKAP)-OH project \citep{daw24}.

\begin{table}[]
    \setlength{\tabcolsep}{4.2pt}
      \caption[]{List of all confirmed `water fountains'.}
      \label{tab:water_fountains}
      \centering                          
\begin{tabular}{lcccccc}     
\hline\hline                 
Name & Alt. name & $V_{\rm range}$$^{a}$ & Reference$^{b}$\\
 &  &  km\,s$^{-1}$ & \\

\hline 
IRAS 15103$-$5754 & GLMP 405 & 70 &  1 \\
IRAS 15445$-$5449 & OH 326.5$-$0.4 & 90 &  2 \\
IRAS 16342$-$3814 & OH 344.1+5.8 & 260 &  3 \\
IRAS 16552$-$3050 & GLMP 498 & 170 &  4 \\
IRAS 17291$-$2147 & GLMP 498 & 70 &  5 \\
IRAS 18019$-$2216 & {SCHB 281} & 20 & This work\\
IRAS 18043$-$2116 & OH 9.1$-$0.4 &  800 & 6, 7\\
IRAS 18113$-$2503 & PM 1$-$221 &  500 & 8\\
IRAS 18139$-$1816 & OH 12.8$-$0.9 & 50 & 9\\
IRAS 18286$-$0959 & OH 21.80$-$0.13 & 200 & 10\\
IRAS 18443$-$0231 & {RFS 505} & 80 & This work \\
IRAS 18450$-$0148 & W43A & 180 & 11\\
IRAS 18455+0448  &  &  40 &  12  \\
IRAS 18460$-$0151 &  OH 31.0$-$0.2 & 310 &  13   \\
IRAS 18464$-$0140 & OH 31.21$-$0.18  & 30 & This work  \\
IRAS 18596+0315 &  OH 37.1$-$0.8 & 60 & 5  \\
IRAS 19134+2131 &    & 100 &  14 \\
IRAS 19190+1102 &  PM 1$-$298  & 100 &  15 \\

\hline
\end{tabular}

Notes. $^{a}$ Velocity range of the H$_{2}$O masers to date. $^{b}$ Interferometric confirmation as WF and velocity range. 1. \cite{gom15b}. 2. \cite{ps11}. 3. \cite{cla09}. 4. \cite{sua08}. 5. \cite{gom17}. 6. \cite{wal09} 7. \cite{usc23}. 8. \cite{gom11}. 9. \cite{bob05}. 10. \cite{yung11}. 11. \cite{ima02}. 12. \cite{vle14}. 13. \cite{ima13}. 14. \cite{ima04}. 15. \cite{day10}.

\end{table}

 In addition to the scarcity of known maser-emitting PNe, observational biases are also a potential problem to understand properties such as the nature, timescales, or prevalence of maser emission in PNe. Many searches for masers in evolved stars have selected their target samples based on particular infrared properties \citep[e.g.][]{dav93, tel96}, which may leave out many PNe. These biases are being mitigated by the increasing number of published maser surveys, which performed a complete mapping of wide regions throughout the Galactic center and plane \citep[e.g.][]{walsh14, beu19, qia16b, qia18, qia20}. Keeping in mind these caveats, the current list of sources can start to provide some interesting patterns, although they will have to be taken with care. 

The infrared properties of these maser-emitting sources, as shown in the \textit{MSX} (Fig.\,\ref{fig:msx}) and \textit{WISE} (Fig.\,\ref{fig:wise}) color-color diagrams, seem very diverse, as they populate all regions of those diagrams. In principle, this suggests that they do not constitute a homogeneous group and that different types of PNe can give rise to maser emission. However, there are hints of a higher density of sources in certain regions of the color-color diagrams, such as the top right quadrant in Fig.\,\ref{fig:msx} and 9$\leq$[3.4]$-$[22]$\leq$13 in Fig.\,\ref{fig:wise}, probably indicating that PNe with particular properties have a higher tendency to pump masers. 

In line with the suggestion of \cite{cal24b}, we found, apart from the known case of IRAS\,17347$-$3139, one more OHPN (IRAS\,17393$-$2727) and three more OHPN candidates (IRAS\,16029$-$5055, IRAS\,17375$-$2759, and IRAS\,19176+1251)} that could have mixed chemistry, with O-rich outer envelopes (where OH and H$_{2}$O masers are emitted), but with C-rich inner nebulae and/or central stars (see Sect.\,\ref{sec:spitzer}). Thus, they could potentially be similar to IRAS\,07027$-$7934 and IRAS 18061$-$2505, which harbor C-rich central stars of the type [WC] \citep[see][and Sec.\,\ref{sec:i07027}]{zij89,mir21}. In general, PNe harboring [WC] have been found surrounded by both O- and C-rich dust \citep[e.g.,][]{dem02}, but only in these PNe the presence of masers links the [WC] with an O-rich star progenitor in the AGB phase. Hence, these PNe are potentially key in characterizing the formation mechanisms of [WC], which may involve a recent thermal pulse and/or a common envelope \citep[see][and references therein]{iva13, zou20}. A common envelope has previously been suggested in the formation and evolution of other maser-emitting PNe \citep[][]{gom18a, mir21}. In any case, the formation mechanisms of [WC] are still not completely understood \citep[e.g.][]{dem02, gar18, haj20}. 

In this paper we have also identified two objects (IRAS\,18019$-$2216 and IRAS\,18464$-$0140) which could be classified as `low-velocity' WFs because they show bipolar H$_{2}$O maser distributions, a velocity coverage of the H$_{2}$O masers larger or in a range different from that of the OH masers, but a H$_{2}$O maser velocity coverage of only tens of km\,s$^{-1}$ \citep[][]{yung13, fan24}, in contrast to the usual velocity spreads of hundreds of km\,s$^{-1}$ of classical WFs (Table\,\ref{tab:water_fountains}). \citet{yung13} found that WFs with low velocity masers showed infrared colors suggesting that they are in the late AGB or early post-AGB phases. If this is correct, IRAS\,18019$-$2216 and IRAS\,18464$-$0140 would probably not be PNe, and their radio continuum emission would arise from shock-ionized gas. 

\section{Conclusions}

We have undergone an extensive search for new maser-emitting PNe. Our main conclusions are as follows.

\begin{itemize}
    \item We have increased the number of confirmed maser-emitting PNe known to nine and increased the number of candidates to 16. Furthermore, we have increased the number of known WFs to 18, and reported the first two `low-velocity' WFs emitters of radio continuum. 

    \item The wide distribution of maser-emitting PNe in the \textit{MSX} and \textit{WISE} color-color diagrams suggests that they do not constitute a homogeneous group and that PNe of different properties may host maser emission during its evolution. However, there is some evidence that PNe with particular infrared properties may have a higher tendency to pump maser emission.
    
    \item In addition to the OHPN IRAS 07027$-$7934 and the H$_{2}$OPN IRAS\,18061$-$2505, which had already been shown to host a [WC] central star with an O-rich envelope, we report the presence of PAH emission in one more OHPN (IRAS\,17393$-$2727) and three more OHPN candidates (IRAS\,16029$-$5055, IRAS\,17375$-$2759, and IRAS\,19176+1251), suggesting that they may also show mixed chemistry, with C-rich central star or inner circumstellar regions. Together with the OHPN and H$_{2}$OPN IRAS\,17347$-$3139, previously reported with PAHs, they could represent an emergent group of PNe that underwent a recent thermal pulse and/or common envelope evolution.
    
    \item The new {interferometrically confirmed} WF IRAS\,18443$-$0231 shows strong and variable non-thermal radio continuum emission, as in magnetized outflows, and near infrared spectroscopy indicates ionized emission compatible with both shocks and photoionization. More research is needed on this object to confirm it as a PN.
        
\end{itemize} 

\begin{acknowledgements}
The Australia Telescope Compact Array is part of the Australia Telescope National Facility (grid.421683.a) which is funded by the Australian Government for operation as a National Facility managed by CSIRO. We acknowledge the Gomeroi people as the traditional owners of the Observatory site. The National Radio Astronomy Observatory is a facility of the National Science Foundation operated under cooperative agreement by Associated Universities, Inc. We used continuum images from the VLA Sky Survey, downloaded from the Canadian Initiative for Radio Astronomy Data Analysis (CIRADA), which is funded by a grant from the Canada Foundation for Innovation 2017 Innovation Fund (Project 35999), as well as by the Provinces of Ontario, British Columbia, Alberta, Manitoba and Quebec. This work has made use of the SIMBAD database, operated at the CDS, Strasbourg, France, and the NASA/IPAC Infrared Science Archive, which is operated by the Jet Propulsion Laboratory, California Institute of Technology, under contract with the National Aeronautics and Space Administration. It also makes use of data products from 2MASS (a joint project of the University of Massachusetts and the Infrared Processing and Analysis Center/California Institute of Technology, funded by NASA and the NSF), AKARI (a JAXA project with the participation of ESA), DENIS (partly funded by the SCIENCE and the HCM plans of the European Commission under grants CT920791 and CT940627), HERSCHEL (Herschel is an ESA space observatory with science instruments provided by European-led Principal Investigator consortia and with important participation from NASA), IRAS (was a joint project of the US, UK and the Netherlands), Pan-STARRS (have been made possible through contributions by the Institute for Astronomy, the University of Hawaii, the Pan-STARRS Project Office, the Max-Planck Society and its participating institutes), \textit{Spitzer} Space Telescope (operated by the Jet Propulsion Laboratory, California Institute of Technology under a contract with NASA), MSX (funded by the Ballistic Missile Defense Organization with additional support from NASA Office of Space Science), UKIDSS (The project is defined in \cite{law07} and uses the UKIRT Wide Field Camera \citep[WFCAM;][]{cas07} and a photometric system described in \cite{hew06}. The pipeline processing and science archive are described in \cite{ham08}), USNO-B catalog \citep[the construction and contents of the catalog can be found in][]{mon03}, VVV survey (is supported by the European Southern Observatory, by BASAL Center for Astrophysics and Associated Technologies PFB-06, by FONDAP Center for Astrophysics 15010003, by the Chilean Ministry for the Economy, Development, and Tourism’s Programa Iniciativa Científica Milenio through grant P07-021-F, awarded to The Milky Way Millennium Nucleus), and WISE (a joint project of the University of California, Los Angeles, and the Jet Propulsion Laboratory/California Institute of Technology, funded by the NASA). We also used the VLASS QLimage cutout server at URL cutouts.cirada.ca, operated by the Canadian Initiative for Radio Astronomy Data Analysis (CIRADA). CIRADA is funded by a grant from the Canada Foundation for Innovation 2017 Innovation Fund (Project 35999), as well as by the Provinces of Ontario, British Columbia, Alberta, Manitoba and Quebec, in collaboration with the National Research Council of Canada, the US National Radio Astronomy Observatory and Australia’s Commonwealth Scientific and Industrial Research Organisation. This work made use of Astropy:\footnote{http://www.astropy.org} a community-developed core Python package and an ecosystem of tools and resources for astronomy \citep{ast13, ast18, ast22}. RAC, JFG, LFM, GA are financially supported by grants PID2020-114461GB-I00, PID2023-146295NB-I00, and CEX2021-001131-S, funded by MCIN/AEI /10.13039/501100011033. RAC also acknowledges support by the predoctoral grant PRE2018-085518, funded by MCIN/AEI/ 10.13039/501100011033 and by ESF Investing in your Future. K.O. acknowledges the support of the Agencia Nacional de Investigación Cient{\'i}fica y Desarrollo (ANID) through the FONDECYT Regular grant 1240301.

\end{acknowledgements}

\begin{appendix}

\section{Details of individual maser components}
\label{app:a}
In this appendix we present tables with the parameters of the spectral components of masers, in the sources where there is more than one such component. In sources with a single maser component, relevant information is already given in Table\,\ref{tab:match}. For each maser component, we indicate its position (right ascension and declination), their positional uncertainties ($\delta_{\rm R.A.}$,  $\delta_{\rm dec}$), the observed transition (H$_{2}$O at 22235 MHz, or OH  at 1612, 1665, 1667, and 1720 MHz), LSR velocity ($V_{\rm LSR}$) in km\,s$^{-1}$, and peak flux density ($S_{\nu}$) in Jy. In most cases, the masers components in these tables were obtained from the observations reported in this paper, but we have also added values from the literature, when relevant. 

We note that the reported positional uncertainties are those obtained with a fit of an elliptical Gaussian to the image of the channel with the peak emission of each component. These are to be understood as relative errors among the components of an individual maser transition and observational epoch. To compare positions between different epochs and transitions, one should consider the uncertainties in absolute astrometry, which are typically on the order of 10\% of the synthesized beam.

\begin{table*}
    \setlength{\tabcolsep}{4pt}
      \caption[]{OH maser emission in our ATCA observations of IRAS 07027$-$7934.}
      \label{tab:masers_obs}
      \centering                          
\begin{tabular}{llccccccccccrr}
\hline\hline                 
 \multicolumn{2}{c}{Maser coordinates} & $\delta_{\rm R.A.}$ & $\delta_{\rm dec}$ & Transition & $V_{\rm LSR}$ & $S_{\nu}$   \\    
	   	R.A.(J2000) & Dec(J2000) & (arcsec) & (arcsec) & (MHz) & (km\,s$^{-1}$) & (Jy)   \\
     \hline 
06:59:26.273 & $-$79:38:46.99 & 0.04 & 0.05 &  1612 & $-$46.45 &  0.646$\pm$0.027   \\
 06:59:26.780 & $-$79:38:47.54 & 0.22 & 0.21 & 1612 & $-$40.28 &  0.30$\pm$0.04   \\
06:59:25.218 & $-$79:38:49.04 & 0.24 & 0.24 & 1612 & $-$41.37 &  0.173$\pm$0.031   \\
\hline
\end{tabular}
\end{table*}

\begin{table*}
    \setlength{\tabcolsep}{4pt}
      \caption[]{OH maser emission in our ATCA and VLA observations of IRAS 17487$-$1922 }
      \label{tab:masers_obs1}
      \centering                          
\begin{tabular}{llccccccccrr}     
\hline\hline                 
 \multicolumn{2}{c}{Maser coordinates} & $\delta_{\rm R.A.}$ & $\delta_{\rm dec}$ & Transition & $V_{\rm LSR}$ & $S_{\nu}$   \\    
	   	R.A.(J2000) & Dec(J2000) & (arcsec) & (arcsec) & (MHz) & (km\,s$^{-1}$) & (Jy)   \\
\hline 
 17:51:44.89120 & $-$19:23:46.8 & 0.0020 & 0.4 & 1612 & $-$39.18 & 0.39$\pm$0.07   \\
 17:51:44.79420 & $-$19:23:46.66 & 0.0013 &  0.20 &  1612 & $-$39.00 & 0.41$\pm$0.04   \\
 17:51:44.9810 & $-$19:23:41.91 & 0.015 &  0.24 & 1612 & $-$38.82 & 0.44$\pm$0.05    \\
 17:51:44.91960 & $-$19:23:40.9 & 0.0019 &  0.4 & 1612 & $-$38.64 &  0.44$\pm$0.09   \\
 17:51:44.8501 & $-$19:23:46.8 & 0.004 &  0.6 & 1612 & $-$38.37 & 0.345$\pm$0.011    \\
 17:51:44.98140 & $-$19:23:40.21 & 0.0020 &  0.31 & 1612 & $-$38.10 & 0.44$\pm$0.07   \\
 17:51:45.12 & $-$19:23:42.9 & 0.4 &  0.4 &  \textbf{1612} & $-$38.15 & 0.58$\pm$0.05  \\
 17:51:44.711 & $-$19:23:43.41 & 0.21 &  0.21 & \textbf{1612} & $-$39.06 & 0.48$\pm$0.08   \\
 17:41:45.36 & $-$19:23:42.0 & 0.3 &  0.4 & \textbf{1612} & $-$39.60 & 0.38$\pm$0.04  \\
\hline
\end{tabular}

\tablefoot{The VLA observations are marked with the frequency of the transition in boldface. The ones with normal fonts, represent the ATCA observations}

\end{table*}

\begin{table*}
    \setlength{\tabcolsep}{4pt}
      \caption[]{H$_{2}$O maser emission in our VLA observations of IRAS 18019$-$2216.}
      \label{tab:masers_obs2}
      \centering                          
\begin{tabular}{llrccccccccrr}     
\hline\hline                 
 \multicolumn{2}{c}{Maser coordinates} & $\delta_{\rm R.A.}$ & $\delta_{\rm dec}$ & Transition & $V_{\rm LSR}$ & $S_{\nu}$   \\    
	   	R.A.(J2000) & Dec(J2000) & (arcsec) & (arcsec) & (MHz) & (km\,s$^{-1}$) & (Jy)   \\
\hline 
 18:34:57.3719 & $-$22:15:50.65 & 0.005 &  0.24 & 22235 & $-$10.11 & 0.194$\pm$0.017 \\
 18:34:57.3520 & $-$22:15:50.8 & 0.013 &  0.4 & 22235 & $-$4.63 &   0.111$\pm$0.020   \\
 18:34:57.3472 & $-$22:15:50.92 & 0.006 &  0.21 & 22235 & $-$1.47 &   0.222$\pm$0.013  \\
 18:34:57.3511 & $-$22:15:50.23 & 0.005 &  0.18 & 22235 & +3.37 &  0.167$\pm$0.013  \\
 18:34:57.3468 & $-$22:15:50.91 & 0.003 &  0.11 & 22235 & +5.90 &   0.332$\pm$0.010  \\
 18:34:57.3365 & $-$22:15:51.17 & 0.004 &  0.14 & 22235 & +8.01 &   0.247$\pm$0.008  \\

\hline
\end{tabular}

\end{table*}

\begin{table*}
    \setlength{\tabcolsep}{4pt}
      \caption[]{H$_{2}$O maser emission (our VLA observations) and OH (from the literature) maser emission of IRAS 18480+0008.}
      \label{tab:masers_obs3}
      \centering                          
\begin{tabular}{llccccccccrr}     
\hline\hline                 
 \multicolumn{2}{c}{Maser coordinates} & $\delta_{\rm R.A.}$ & $\delta_{\rm dec}$ & Transition & $V_{\rm LSR}$ & $S_{\nu}$   \\    
	   	R.A.(J2000) & Dec(J2000) & (arcsec) & (arcsec) & (MHz) & (km\,s$^{-1}$) & (Jy)   \\
\hline 

 18:50:36.685 & +00:12:28.34 & 0.10 &  0.10 & 1665 & +12.00 & 0.345$\pm$0.009   \\
 18:50:36.66107 & +00:12:28.140 & 0.0009 &  0.020 & 22235 & +14.53 & 1.026$\pm$0.011   \\
 18:50:36.6635 & +00:12:28.21 & 0.003 &  0.05 & 22235 & +18.75 &  0.244$\pm$0.007  \\
 18:50:36.6720 & +00:12:28.41 & 0.017 &  0.17 & 22235 & +20.22 & 0.110$\pm$0.012  \\
 18:50:36.66141 & +00:12:28.191 & 0.0004 &  0.007 & 22235 & +22.54 &  1.9844$\pm$0.0008  \\
 18:50:36.66040 & +00:12:28.07 & 0.0013 &  0.03 & 22235 & +25.07 &  0.690$\pm$0.010  \\
 18:50:36.66330 & +00:12:28.147 & 0.0010 &  0.020 & 22235 & +26.54 &  0.782$\pm$0.009  \\

\hline
\end{tabular}

\tablefoot{The parameters of the OH maser transition at 1665 MHz were reported by \cite{beu19}.}

\end{table*}

\begin{table*}
    \setlength{\tabcolsep}{4pt}
      \caption[]{H$_{2}$O maser emission in our VLA observations of IRAS 18443$-$0231.}
      \label{tab:masers_obs4}
      \centering                          
\begin{tabular}{llccccccccrr}     
\hline\hline                 
 \multicolumn{2}{c}{Maser coordinates} & $\delta_{\rm R.A.}$ & $\delta_{\rm dec}$ & Transition & $V_{\rm LSR}$ & $S_{\nu}$   \\    
	   	R.A.(J2000) & Dec(J2000) & (arcsec) & (arcsec) & (MHz) & (km\,s$^{-1}$) & (Jy)   \\
\hline 
18:47:00.4421 & $-$02:27:54.0 & 0.014 &  0.8 & 22235 & $-$28.49 & 0.048$\pm$0.009  \\
18:47:00.380 & $-$02:27:53.7 & 0.03 &  0.5 & 22235 & $-$17.11 & 0.055$\pm$0.012  \\
18:47:00.40411 & $-$02:27:52.552 & 0.0005 &  0.010 & 22235 & $-$10.37 & 2.458$\pm$0.013  \\
18:47:00.40336 & $-$02:27:52.591 & 0.0003 &  0.007 & 22235 & $-$9.32 & 3.875$\pm$0.014    \\
18:47:00.4049 & $-$02:27:52.54 & 0.005 & 0.09 & 22235 & $-$6.37 &  0.234$\pm$0.012  \\
18:47:00.40182 & $-$02:27:52.587 & 0.0005 &  0.011 & 22235 & $-$4.90 & 1.796$\pm$0.010  \\
18:47:00.40120 & $-$02:27:52.552 & 0.00016 &  0.004 & 22235 & $-$3.42 & 9.826$\pm$0.018 \\
18:47:00.4069 & $-$02:27:52.56 & 0.003 &  0.06 & 22235 & +0.16 & 0.315$\pm$0.009   \\
18:47:00.4165 & $-$02:27:52.03 & 0.009 &  0.23 & 22235 & +1.42 & 0.098$\pm$0.012    \\
18:47:00.4044 & $-$02:27:52.60 & 0.006 &  0.12 & 22235 & +2.47 & 0.14$\pm$0.08   \\
18:47:00.4004 & $-$02:27:52.56 & 0.003 &  0.05 & 22235 & +5.05 & 0.390$\pm$0.012   \\
18:47:00.40080 & $-$02:27:52.472 & 0.0004 &  0.009 & 22235 & +5.84 & 2.370$\pm$0.011    \\
18:47:00.41120 & $-$02:27:52.46 & 0.0017 &  0.04 & 22235 & +7.74 & 0.543$\pm$0.011   \\
18:47:00.41020 & $-$02:27:52.53 & 0.0018 &  0.04 & 22235 & +9.64 & 0.572$\pm$0.011    \\
18:47:00.41332 & $-$02:27:52.43 & 0.0015 &  0.03 & 22235 & +11.54  & 0.653$\pm$0.010    \\
18:47:00.40773 & $-$02:27:52.58 & 0.0013 &  0.03 & 22235 & +12.80 & 0.681$\pm$0.010   \\
18:47:00.4208 & $-$02:27:52.37 & 0.007 &  0.12 & 22235 & +31.76 & 0.144$\pm$0.010    \\
18:47:00.3810 & $-$02:27:52.9 & 0.011 &  0.4 & 22235 & +43.98 & 0.103$\pm$0.011   \\
18:47:00.40251 & $-$02:27:52.58 & 0.0017 &  0.04 & 22235 & +48.61 & 0.488$\pm$0.010    \\

\hline
\end{tabular}
\end{table*}

\begin{table*}
    \setlength{\tabcolsep}{4pt}
      \caption[]{H$_{2}$O and OH maser emission of IRAS 18213$-$1245A (from the literature).}
      \label{tab:masers_obs7}
      \centering                          
\begin{tabular}{llccccccccrr}     
\hline\hline                 
 \multicolumn{2}{c}{Maser coordinates} & $\delta_{\rm R.A.}$ & $\delta_{\rm dec}$ & Transition & $V_{\rm LSR}$ & $S_{\nu}$   \\    
	   	R.A.(J2000) & Dec(J2000) & (arcsec) & (arcsec) & (MHz) & (km\,s$^{-1}$) & (Jy)   \\
\hline 

18:24:09.73 & $-$12:43:23.727 & 0.20 &  0.10 & 1665 & +27.00 & 0.139$\pm$0.005    \\
18:24:09.71 & $-$12:43:23.96 & 8.7 &  1.36 & 22235 & +23.60 & 3.70$\pm$0.05    \\
18:24:09.71 & $-$12:43:24.04 & 9.6 &  1.36 & 22235 & +21.70 & 0.75$\pm$0.05     \\
\hline
\end{tabular}

\tablefoot{The H$_{2}$O masers were reported in HOPS \citep{walsh14}, and the OH masers in THOR \citep{beu19}}.

\end{table*}

\begin{table*}
    \setlength{\tabcolsep}{4pt}
          \caption[]{H$_{2}$O and OH masers in IRAS 18464$-$0140.}
      \label{tab:masers_i18464}
      \centering                          
\begin{tabular}{llccccccccrr}     
\hline\hline                 
 \multicolumn{2}{c}{Maser coordinates} & $\delta_{\rm R.A.}$ & $\delta_{\rm dec}$ & Transition & $V_{\rm LSR}$ & $S_{\nu}$   \\    
	   	R.A.(J2000) & Dec(J2000) & (arcsec) & (arcsec) & (MHz) & (km\,s$^{-1}$) & (Jy)   \\
\hline 

 18:48:56.523 & $-$01:36:43.32 & 0.20 &  0.20 & 1612 & $-$28.10 & 1.749$\pm$0.021   \\
 18:48:58.62 & $-$01:36:44.3 & 0.1 &  0.1 & 1665 & $-$28.90 & 0.350$\pm$0.020     \\
 18:48:52.72 & $-$01:36:44.3 & 0.1 &  0.1 & 1665 & $-$29.45 & 0.500$\pm$0.010     \\
 18:48:56.52 & $-$01:36:43.0 & 0.4 &  0.4 & 1667 & $-$29.00 & 0.377$\pm$0.010    \\
 18:48:56.680580 & $-$01:36:42.6796 & 0.00020 &  0.0025 & 22235 & $-$37.89 & 34.85$\pm$0.17 \\
 18:48:56.6914 & $-$01:36:42.72 & 0.014 &  0.13 & 22235 & $-$28.68 & 1.01$\pm$0.20   \\
 18:48:56.65030 & $-$01:36:43.040 & 0.0013  &  0.013 & 22235 & $-$11.56 & 5.71$\pm$0.22   \\
 18:48:56.6462 & $-$01:36:43.04 & 0.004 & 0.03 & 22235 & $-$6.29 & 2.56$\pm$0.21  \\
                 
\hline
\end{tabular}

\tablefoot{The 1612 and 1667 MHz OH maser emission were reported in THOR \citep{beu19}, and the 1665 MHz OH maser in \cite{fc99}. The H$_{2}$O maser emission was initially reported by \cite{fc99}, but we reprocessed it completely, including normal calibration and self-calibration. The values given in this table correspond to those we obtained after this reprocessing.}

\end{table*}

\section{Targets without positional association between maser and radio continuum in our ATCA/VLA observations}
\label{app:non_match}

In Table\,\ref{tab:non_match_discarded} we report the cases for which maser and continuum emission were both detected in our field of view, but their position did not match, and a careful analysis indicates that the emission of each type arises from different sources, and thus we can discard them as possible maser-emitting PNe. Each of these sources are described in the subsections of this appendix.

Table\,\ref{tab:non_match} shows the sources in which continuum and/or maser emission was not detected. Given the possible variability of the sources, in those cases we cannot discard that the previously reported emission of these types was actually associated, but one or both of them fell below our sensitivity threshold. We note that we also included the source OH~25.646$-$0.003 in that table, although it is a confirmed OHPN candidate, but we show there our non-detection of H$_2$O masers at 22 GHz, and of continuum emission around that frequency.

\begin{sidewaystable*}
    \setlength{\tabcolsep}{3.5pt}
      \caption[]{Sources with discarded association in our ATCA/VLA observations.}
      \label{tab:non_match_discarded}
      \centering                          
\begin{tabular}{lccc|cccccr}     
\hline\hline                 
 \multicolumn{4}{c}{Radio continuum emission} & \multicolumn{4}{c}{Maser emission}  \\
 Source &  Position & Flux density$^a$  & Frequency & Source & Position & Flux density$^b$  &  Transition  \\    
Name & RA, DEC (J2000) & (mJy) & (GHz) & Name & RA, DEC (J2000) & (Jy) &  &     \\
\hline 
RACS-DR1 J141249.8$-$69210 & 14:12:50.756 $-$69:21:03.56 & 48.56$\pm$0.69 & 2.1 & IRAS\,14086$-$6907 & 14:12:50.393 $-$69:21:09.75 & 5.612$\pm$0.035 & OH\,1612\,MHz   \\
IRAS\,15559$-$5546 & 15:59:57.644 $-$55:55:33.24 & 69.27$\pm$1.81 & 2.1 & IRAS\,15557$-$5546 & 15:59:41.667 $-$55:54:51.21 & 1.49$\pm$0.06 &  OH\,1612\,MHz   \\
IRAS\,17385$-$2211 & 17:41:36.590 $-$22:13:03.90 & 0.68$\pm$0.09 & 2.1 & IRAS 17388$-$2203 & 17:41:48.979 $-$22:05:17.03 & 2.56$\pm$0.05 &   OH\,1612\,MHz   \\
&  & $<$1.01 & 1.5 & & 17:41:48.963 $-$22:05:15.27 & 2.529$\pm$0.031 & OH\,1612\,MHz  \\
& 17:41:36.610 $-$22:13:03.37 & 0.70$\pm$0.05 & 24 & & & $<$14 &  H$_2$O\,22\,GHz  \\
GPSR\,020.590+0.425 & 18:27:07.018 $-$10:46:08.70 & 20.22$\pm$1.05 & 2.1 & 
IRAS\,18243$-$1048 & 18:27:08.273 $-$10:46:06.93 & 0.915$\pm$0.038 & OH\,1612\,MHz   \\
NGC\,6644 & 18:32:34.682 $-$25:07:44.15 & 13.04$\pm$0.28 & 2.1 & IRC\,$-$20491 & 18:32:07.675 $-$24:57:18.95 & 3.084$\pm$0.05 & OH\,1612\,MHz   \\
& 18:32:34.687 $-$25:07:44.27 & 79.22$\pm$1.56 & 1.5 & & 18:32:07.675 $-$24:57:18.95 & 3.089$\pm$0.037 & OH\,1612\,MHz  \\
& 18:32:34.689 $-$25:07:44.18 & 77.98$\pm$0.35 & 24 &  & & $<$14  &  H$_2$O\,22\,GHz \\
K\,3-33 & 19:22:26.694 +10:41:23.19 & 13.04$\pm$0.28 & 2.1 & IRAS\,19201+1040 & 19:22:29.168 +10:46:21.62 & 1.23$\pm$0.05 &  OH\,1612\,MHz  \\
& 19:22:26.669 +10:41:21.06 & 10.72$\pm$0.51 & 1.5 &  & 19:22:29.172 +10:46:21.00 & 2.56$\pm$0.05 & OH\,1612\,MHz \\
& 19:22:26.645 +10:41:20.30 & 15.34$\pm$0.08 & 24 &  & & $<$12 &  H$_2$O\,22\,GHz \\

\hline
\end{tabular}

Notes: $^a$ For sources in which radio continuum emission is not detected, we report the 3\,$\sigma$ upper limit to the intensity (in mJy\,beam$^{-1}$). $^b$ For sources in which maser emission is not detected, we report the noise level (in mJy\,beam$^{-1}$), measured at the position and channel map with the $V_{\rm LSR}$ of the reported OH maser from single-dish observations.

\end{sidewaystable*}

\begin{table*}
      \caption[]{Sources with undetected continuum and/or maser emission in our ATCA/VLA observations.}
      \label{tab:non_match}
      \centering                          
\begin{tabular}{ccccccc}     
\hline\hline                 
Target name	   & Date & \multicolumn{2}{c}{Continuum$^a$}  & Maser$^b$ & Telescope & Association$^c$ \\
  &  &  Frequency & Flux density  & Flux density \\
 &  &  (GHz) & (mJy)  & (Jy) \\
\hline     
IRAS 14079$-$6402 & 2021/10/19  & 2.1 &  1.81$\pm$0.16 & $<$13 & ATCA & UM\\
IRAS 16280$-$4008 & 2021/10/19  & 2.1 &  615.28$\pm$1.46 & $<$29 &  ATCA & UM\\
IRAS 17180$-$2708 & 2021/10/20 & 2.1 & 265.01$\pm$0.33 &  $<$59 &  ATCA & UM \\
 & 2021/04/03 & 1.5 &  268.59$\pm$1.84 & $<$39 &  VLA & UM\\
 & 2021/04/01  & 24 &  174.87$\pm$1.62 & $<$8.7 & VLA & UM\\
IRAS 17221$-$3038 & 2021/10/20 & 2.1 & 8.60$\pm$0.13 & $<$38 &  ATCA & UM\\
IRAS 17253$-$2824 & 2021/10/20 & 2.1 &  $<$0.19 & 0.171$\pm$0.020 & ATCA & UC\\
 & 2021/04/03 & 1.5 &  $<$1.98 & 0.224$\pm$0.014  & VLA & UC\\
 & 2021/04/01  & 24 &  $<$0.04 & $<$6.7 & VLA & UC, UM\\
IRAS 17374$-$2700  & 2021/10/20 & 2.1 & $<$0.18 & $<$33  & ATCA & UC, UM \\
& 2021/04/03 & 1.5 & $<$1.12 & $<$24 & VLA & UC, UM \\
&  2021/04/01 & 24 &  $<$0.04 & $<$8.4 & VLA & UC, UM \\\
IRAS 17403$-$2107 & 2021/10/20 & 2.1 &  11.26$\pm$0.16 & $<$40 & ATCA & UM \\
& 2021/04/03 & 1.5 &  11.82$\pm$0.47 &  $<$38 & VLA & UM\\
& 2021/04/01 & 24 & 7.79$\pm$0.09 & $<$11 & VLA & UM \\
IRAS 18271$-$1014 & 2021/10/19 & 2.1 &  1.53$\pm$0.37 & $<$11 &  ATCA & UM \\
 & 2021/04/03 & 1.5 &  $<$2.89 & $<$20 &  VLA & UM\\
&  2021/04/01 & 24 &  1.95$\pm$0.09 & $<$8.8 & VLA & UM\\
OH 25.646$-$0.003$^d$ & 2021/03/25 & 24 &  $<$0.08 & $<$12 & VLA & UC, UM\\
IRAS 18551+0159 & 2021/04/03 & 1.5 & 142.64$\pm$4.64  &  $<$70 & VLA & UM\\
& 2021/03/25 & 24 &  189.64$\pm$0.36  & $<$14 & VLA & UM\\
IRAS 19123+1139 & 2021/10/19 & 2.1 &  29.50$\pm$1.05 & $<$29 & ATCA & UM\\
& 2021/03/25 & 1.5 &  33.14$\pm$3.34 & $<$79 & VLA & UM\\
&  2021/03/25 & 24 &  17.09$\pm$0.11 & $<$12 & VLA & UM\\
IRAS 19127+1717 & 2021/10/19 & 2.1 &  0.79$\pm$0.16  & $<$20 &  ATCA & UM\\
 & 2021/03/25 & 1.5 & 2.24$\pm$0.37 & $<$30 & VLA & UM\\
&  2021/03/25 & 24 &  5.94$\pm$0.05 &  $<$11 & VLA & UM\\
IRAS 19508+2659 & 2021/03/25 & 24 &  $<$0.06 & 0.565$\pm$0.011 & VLA & UC\\
IRAS 20266+3856 & 2021/04/03 & 1.5 & $<$3.54 & 13.36$\pm$0.18 & VLA & UC\\
&  2021/03/25 & 24 & $<$0.06 & $<$15 & VLA & UC, UM \\
\hline 
\end{tabular}

Notes.$^a$ For sources in which radio continuum emission is not detected, we report the 3\,$\sigma$ upper limit to the intensity (in mJy\,beam$^{-1}$). $^b$ For sources in which maser emission is not detected, we report the noise level (in mJy\,beam$^{-1}$), measured at the position and channel map with the $V_{\rm LSR}$ of the reported OH maser from single-dish observations. $^c$ UC: undetected continuum emission. UM: undetected maser emission. $^d$ This object is a new OHPN candidate, but we have included it in this table because we did not detect any H$_{2}$O maser nor radio continuum emission in our VLA observations (see Sec\,\ref{sec:oh25}).\\

\end{table*} 

\subsection{The field of IRAS 14086$-$6907}

OH maser emission was  fist reported toward IRAS 14086$-$6907  from single-dish observations by \citet{tel91}. Radio continuum emission (source RACS-DR1 J141249.8-692104) was detected with the ASKAP interferometer near this object \citep{hale21}. The radio continuum position is $\sim$7\arcsec\,away from the source 2MASS J14125041-6921092, which is the near-infrared counterpart of IRAS 14086$-$6907. We have detected both OH masers and radio continuum emission in our interferometer observations. We confirm that the OH maser emission is associated with the AGB star IRAS 14086$-$6907 \citep{lebz23}, and is indeed separated from the continuum emission by $\sim$7\arcsec. The parallax of the Gaia source associated with IRAS 14086$-$6907 (Gaia DR3 5846602400091413632) indicates it is at a distance of $\simeq 1$ kpc. The separation of the radio continuum emission would correspond to 7000 au at this distance. We can confidently discard that the presence of this radio continuum emission could indicate that IRAS 14086$-$6907 is a PN, since we would not expect unresolved radio continuum emission that far away from the central star without any hint of emission closer to it. The nature of the radio continuum emission is unknown, but it is probably associated to an extragalactic object, since it is brighter at 887.5 MHz \citep{hale21} than at 1367.5 MHz \citep[][]{duc24}, suggesting non-thermal emission. A possible optical counterpart for the radio continuum source is Gaia DR3 5846602404392109312.

\subsection{The field of IRAS 15559$-$5546}

IRAS 15559$-$5546 (Hen 2-142) has been spectroscopically classified as a PN \citep{frew13}. OH maser emission with a double-peaked spectral profile has been reported in its neighborhood with single-dish observations \citep{tel91}. We have detected both radio continuum and OH maser emission within the observed field, with a separation between them of $\sim$2.4\,arcmin. The radio continuum emission is clearly associated with the PN IRAS 15559$-$5546, 
while The OH masers are emitted by IRAS 15557-5546 \citep{sev97}, which has been previously classified as an AGB star \citep{sev02}. We did not detect any radio continuum emission from IRAS 15557-5546.

\subsection{The field of IRAS 17385$-$2211}

IRAS\,17385$-$2211 has been spectroscopically classified as PN \citep{frew13}. Double-peaked 1612 MHz OH maser emission was reported in its neighborhood based on single-dish observations \citep{tel91}. We have detected the radio continuum emission from the PN. However, the position of the OH maser emission coincides with that of IRAS\,17388$-$2203, which has been spectroscopically classified as a post-AGB star \citep{sua06}, and it is $\sim$8.3\,arcmin away from IRAS\,17385$-$2211. We confirm the association of OH maser emission from this post-AGB star for the first time. The maser spectrum displays multiple emission features, whose velocities coincide with those seen in the single-dish spectrum reported by  \cite{tel91}. 
 We did not detect any radio continuum emission at 2.1\,GHz from IRAS 17388$-$2203. 

\subsection{The field of IRAS 18243$-$1048}

Double-peaked 1612 MHz OH maser emission was reported by single-dish observations towards IRAS 18243$-$1048 \citep{tel91}. There is a radio continuum source located $\sim 20"$ away from this infrared source, whose emission has been reported at different frequencies \citep[e.g.,][]{con98, pur13}, and that lies within the beam of the OH single-dish observations. Our interferometric observations detect both the OH maser and radio continuum emission in the field, and we can confirm that the OH maser emission is indeed associated with IRAS 18243$-$1048 (2MASS J18270824-1046099), $\sim 20"$ away from the radio continuum source. The position of the latter is located next ($\sim 2"$ away) to a different infrared source in 2MASS (source 2MASS J18270688-1046087) and Spitzer (source SSTGLMC G020.5893+00.4247) images, with no extended emission connecting it with IRAS 18243$-$1048.

\subsection{The field of IRAS 18295$-$2510}

IRAS 18295$-$2510 (NGC\,6644) is a spectroscopically classified PN \citep[e.g.,][]{frew13}. 
The possible association of this PN with  OH maser emission at 1612 MHz had been previously suggested by \cite{zij89}, based on single-dish observations. We have detected the radio continuum emission from the the PN. We also found double-peaked OH maser emission in the field, but it arises from a different source, IRC\,20491, located $\simeq 2'$ away from NGC\,6644.

\subsection{The field of IRAS 19200+1035}

Double-peaked 1612 MHz OH maser-emission has been reported towards IRAS\,19200+1035 based on single-dish observations \citep{eder88}. The position of this source (associated to 2MASS 19222483+1040506) lies $\sim$41\arcsec\,away from the spectroscopically confirmed PN K\,3-33, so both objects fall within the beam of the OH maser observations. In our data, we have detected the radio continuum emission from the PN. We have also detected OH maser emission in the field, but it is not associated with either PN K\,3-33 or IRAS\,19200+1035, but with the source  IRAS\,19201+1040 (2MASS J19222916+1046200) instead, an object which is $\sim$5\,arcmin away from PN K\,3-33. This same double-peaked OH maser emission has also been reported in IRAS\,19201+1040 \citep[e.g.,][]{lew94}, and ours is the first interferometric detection of this emission, which allows us to properly associate it to IRAS\,19201+1040, rather than to IRAS\,19200+1035. We have not detected any radio continuum emission from IRAS\,19201+1040.

\section{Parameters of the shock and photoionization models}
\label{ap:models}

Taking into account the mid-infrared spectra of newly and previously identified objects (Fig.\,\ref{fig:spitzer}), we have used the fine-structure spectral lines of [Ne\,{\sc ii}], [Ne\,{\sc iii}], and [Ne\,{\sc v}] seen at 12.81, 15.55, and 14.32\,$\mu$m, respectively, to create diagnostic plots in the 3MdB \citep{mor15, ala19} that represent shock ionization and photoionization models, which we describe below.

\subsection{Shock models}

We have used the 3MdB \citep{mor15, ala19} to obtain the shock models of \cite{sut17}, which follow the same grid definition as \citet{all08}, for a total of 15968 models. These models were then applied to generate the diagnostic diagram shown in Fig.\,\ref{fig:3mdb_dd3} (left). The same models and parameters were used in \citet{cal24c}, where a more detailed explanation can be found.

\subsection{Photoionization models}

We have used the 3MdB \citep{mor15, ala19} to obtain the PN photoionization models of \citet{del14}. These models were computed using Cloudy v.17 \citep[]{fer17}, resulting in a total of 68041 models. These models were then applied to generate the diagnostic diagram shown in Fig.\,\ref{fig:3mdb_dd3} (right). The same models and parameters were used in \citet{cal24c}, where a more detailed explanation can be found.

\end{appendix}

\end{document}